\documentclass[a4paper,11pt]{article}
\pdfoutput=1 

\usepackage{jcappub} 

\usepackage[T1]{fontenc} 

\usepackage{newtxtext,newtxmath}
\usepackage{orcidlink}

\newcommand{\lya}{Ly$\alpha$ }
\newcommand{\siii}{\mathrm{Si_{\textsc{ii}}}}
\newcommand{\siiii}{\mathrm{Si_{\textsc{iii}}}}
\newcommand{\pk}{P^{1\mathrm{D}}}
\newcommand{\bk}{B^{1\mathrm{D}}}
\newcommand{\bksq}{B^{1\mathrm{D}}_{SQ}}

\usepackage{graphicx}	
\usepackage{subfigure}
\usepackage{multirow}
\usepackage{booktabs}
\usepackage{fontawesome}
\usepackage[toc,page]{appendix}
\usepackage{hyperref}
\usepackage{amsmath}	
\usepackage{amssymb}	
\usepackage{cuted}
\usepackage[normalem]{ulem}

\title{\boldmath A title with some math: $x=1$}

\title{First \lya 1D Bispectrum Measurement in eBOSS}

\author[a]{Rodrigo de la Cruz\orcidlink{0000-0001-9908-9129},}
\author[a]{Gustavo Niz\orcidlink{0000-0002-1544-8946},} 
\author[b,c]{Vid Ir\v{s}i\v{c}\orcidlink{0000-0002-5445-461X},} 
\author[d,e]{Corentin Ravoux\orcidlink{0000-0002-3500-6635},}
\author[f]{César Ramírez,}
\author[a,g,h]{Hiram K. Herrera-Alcantar\orcidlink{0000-0002-9136-9609}}

\affiliation[a]{Departamento de Física, División de Ciencias e Ingenierías, Universidad de Guanajuato,
Campus León, C.P. 37150, León, Mexico.}
\affiliation[b]{Kavli Institute for Cosmology, University of Cambridge, Madingley Road, Cambridge CB3 0HA, UK.}
\affiliation[c]{Cavendish Laboratory, University of Cambridge, 19 J. J. Thomson Ave., Cambridge CB3 0HE, UK.}
\affiliation[d]{Aix Marseille Univ, CNRS/IN2P3, CPPM, Marseille, France.}
\affiliation[e]{Université Clermont-Auvergne, CNRS, LPCA, 63000 Clermont-Ferrand, France.}
\affiliation[f]{Institut de F{\'i}sica d'Altes Energies (IFAE), The Barcelona Institute of Science and Technology, 08193 Bellaterra (Barcelona), Spain.}
\affiliation[g]{Université Paris-Saclay, CEA, IRFU, 91191, Gif-sur-Yvette, France.}
\affiliation[h]{Sorbonne Université, CNRS, UMR 7095, Institut d’Astrophysique de Paris, 98 bis bd Arago, 75014 Paris, France.}

\emailAdd{rodrigo.cruz.noriega93@gmail.com}
\emailAdd{g.niz@ugto.mx}

\abstract{We present the first robust measurement of the one-dimensional Lyman alpha (Ly$\alpha$) forest bispectrum using the complete extended Baryon Oscillation Spectroscopic Survey (eBOSS) quasar sample, corresponding to the sixteenth data release (DR16) of the Sloan Digital Sky Survey (SDSS). The measurement employs an FFT estimator over 12 redshift bins, ranging from $z=2.2$ to $z=4.4$, and extends to scales of $0.02 ~ (\mathrm{km/s})^{-1}$. The sample consists of 122,066 quasar spectra, although only the first six redshift bins contain sufficient data to extract a physical bispectrum. To validate and correct the bispectrum measurement, we use synthetic datasets generated from lognormal and 2LPT mocks. Additionally, we detect clear evidence of correlations between Si$_\mathrm{III}$ absorption lines and the \lya forest within the bispectrum signal, which we describe with an extension of the model used for the analogue of 1d power spectrum signal. In this context, the pipeline developed for this study addresses the impact of instrumental and methodological systematics and is ready for application to larger spectroscopic datasets, such as those from the first year of DESI observations. Finally, A simple perturbation theory model provides a reasonable explanation of the eBOSS bispectrum, suggesting that higher-order one-dimensional statistics in the \lya forest can complement cosmological model inference based on the power spectrum in future analyses.
}

\begin{document}
\maketitle
\flushbottom

\section{Introduction and summary of main results} \label{Intro}
The intergalactic medium (IGM) resides in the low-density regions between galaxies and is important to various astrophysical disciplines. Understanding the IGM has been essential since the pioneering work by \cite{Gunn:1965hd} in the 1960s, where it was first used to test models of structure formation on the smallest comoving scales (see, for example, \cite{PhysRevD.71.103515}, \cite{Viel_2005}, \cite{Palanque-Delabrouille:2019iyz}, or \cite{Goldstein:2023gnw}). Variations in the IGM's density are observed in quasar (QSO) spectra as a continuous series of absorption features. Although various atomic transitions serve as biased tracers of this density field, the strongest signal arises from the Lyman-alpha (\lya) transition of neutral hydrogen atoms within the IGM. These absorption features in QSO spectra are referred to as the \lya forest, optimally studied at redshifts between 2 and 5 (see \cite{doi:10.1146/annurev-astro-082214-122355} for a review). 

The small-scale IGM distribution is captured in the fluctuations of the \lya forest superimposed on the quasar continuum, defined as the QSO unabsorbed spectrum. These fluctuations can be analyzed through different summary statistics, either in Fourier or real space. One approach in Fourier space is through the three-dimensional power spectrum ($P^{3D}$), a correlation between QSO spectrum's absorption of multiple line-of-sights separated by small angles on the sky (see for example \cite{McDonald_2003, Font-Ribera_2018,debelsunce20243d,Abdul_Karim_2024}). Due to the observational nature of the \lya forest, there is high resolution dataset over line of sights but a limited sampling on the perpendicular directions, representing a technical challenge for measuring the $P^{3D}$. The advent of new \lya forest surveys with greater cover of lines of sights over the sky, such as Dark Energy Spectroscopic Instrument (DESI) (\cite{desi_collaboration_desi_2016}, \cite{abareshi_overview_2022}), opens a realistic perspective into measuring the $P^{3D}$. 

A more straightforward approach is to study the \lya forest power spectrum averaged over the line of sight, known as the one-dimensional power spectrum ($\pk$). Early measurements of $\pk$ can be found in \cite{Kim_2004,Croft_1998, Croft_2002, McDonald_2000}. Medium-resolution surveys have measured $\pk$ using data from the Sloan Digital Sky Survey (SDSS) (\cite{SDSS:2000hjo}), the Baryon Oscillation Spectroscopic Survey (BOSS), and the extended BOSS (eBOSS). Initial measurements, such as \cite{McDonald_2006} ($\mathcal{O}$($10^{3}$ spectra)), were followed by \cite{Palanque-Delabrouille:2013gaa} ($\mathcal{O}$($10^{4}$ spectra)) and finally by \cite{Chabanier_2019}. More recently, $\pk$ was measured using the first DESI data with a Fast Fourier Transform (FFT) approach (\cite{ravoux2023dark}) and compared with the quadratic maximum likelihood estimator (QMLE) approach from \cite{karacayli2023optimal} (or \cite{Palanque-Delabrouille:2013gaa} for BOSS), with both methods yielding consistent results. High-resolution $\pk$ measurements have also been made by SQUAD (\cite{Murphy_2018}, $\mathcal{O}$($4\times10^{2}$ spectra)), KODIAK (\cite{Meara_2015}, $\mathcal{O}$($3\times10^{2}$ spectra)), and XQ-100 (\cite{refId0}, $\mathcal{O}$($1\times10^{2}$ spectra)). Further examples of $\pk$ analysis are found in \cite{Walther_2017}, \cite{10.1093/mnras/stab3201}, \cite{10.1093/mnras/stw3372}, \cite{PhysRevD.88.043502}, \cite{10.1093/mnras/stz2214}, \cite{10.1093/mnras/stz344}, \cite{Boera_2019}, and \cite{10.1093/mnras/stab2017}.

The $\pk$ analysis provides valuable information on structure formation, as seen in \cite{LinaresCedeno:2020dte}, \cite{refId1}, \cite{Bose:2018juc}, and \cite{Borde:2014xsa}. It is also a tool for studying neutrino masses, as demonstrated by \cite{Palanque-Delabrouille:2015pga}, \cite{Yeche:2017upn}, and \cite{Ivanov:2024jtl}. We hope to achieve similar results using the one-dimensional bispectrum ($\bk$) in future studies. To extract more information from \lya forest data, one can either use higher-order Lyman series (references for Wilson+22; Dijkstra+10) or higher-order statistics such as the one-dimensional bispectrum (e.g., \cite{Zaldarriaga:2000rg}; \cite{Mandelbaum:2003km}).

In this work, we focus on higher-order statistics, specifically the bispectrum, which provides a complementary description of the non-Gaussian features in the \lya forest. The bispectrum quantifies correlations between three Fourier modes, offering insight into the non-linear processes shaping the large-scale structure of the universe (\cite{Chongchitnan_2014}, \cite{Liguori:2010hx}, \cite{Scoccimarro:2000sn}, \cite{Ivanov:2023qzb}). \cite{Zaldarriaga:2000rg} mention that gravitational growth induces correlations between large-scale modes and small-scale power, which can be demonstrated through the bispectrum. These three-point correlations can also distinguish between fluctuations caused by large-scale structure in the matter distribution and those arising from non-gravitational processes, such as variations in the continuum emission of quasars. Other physically interesting phenomena can be studied using higher-order statistics, including the recent parity study by \cite{Adari:2024vkf} employing the four-point correlation function.

Several studies in the literature have explored the use of the \lya bispectrum. Notably, \cite{10.1111/j.1365-2966.2004.07404.x} measured the $\bk$ using 27 high-resolution quasar spectra in the redshift range of 2.0 to 2.4. They compared the $\bk$ results with non-Gaussian hydrodynamical simulations to investigate the potential for constraining primordial non-Gaussianity (\cite{Viel:2008jj}). In contrast, \cite{Zaldarriaga:2000rg} and \cite{Mandelbaum:2003km} measured a related quantity and studied the cross-correlation coefficient. In \cite{Maitra:2020jrk}, the three-point correlation function was measured using hydrodynamical simulations, decomposing the \lya forest into multiple Voigt profile components (clouds). Another interesting work by \cite{Chiang_2017} measured the squeezed-limit cross-bispectrum between the \lya power spectrum and the long-wavelength quasar overdensity using simulations. This is similar to the position-dependent power spectrum approach (\cite{Chiang:2014oga}). Additionally, \cite{PhysRevLett.109.121301} attempted to place stringent constraints on primordial non-Gaussianity by analyzing the measured 3D bispectrum of the \lya forest. In \cite{Sarkar:2012mq}, they further pursued non-Gaussianity testing, this time incorporating the three-dimensional 21-cm bispectrum and examining the cross-bispectrum between 21-cm brightness temperature and \lya transmitted flux fields. With this background on the \lya bispectrum ($\bk$) established, we can now describe the main results of this work.

\begin{figure}
\begin{center}
\includegraphics[width=8.1cm]{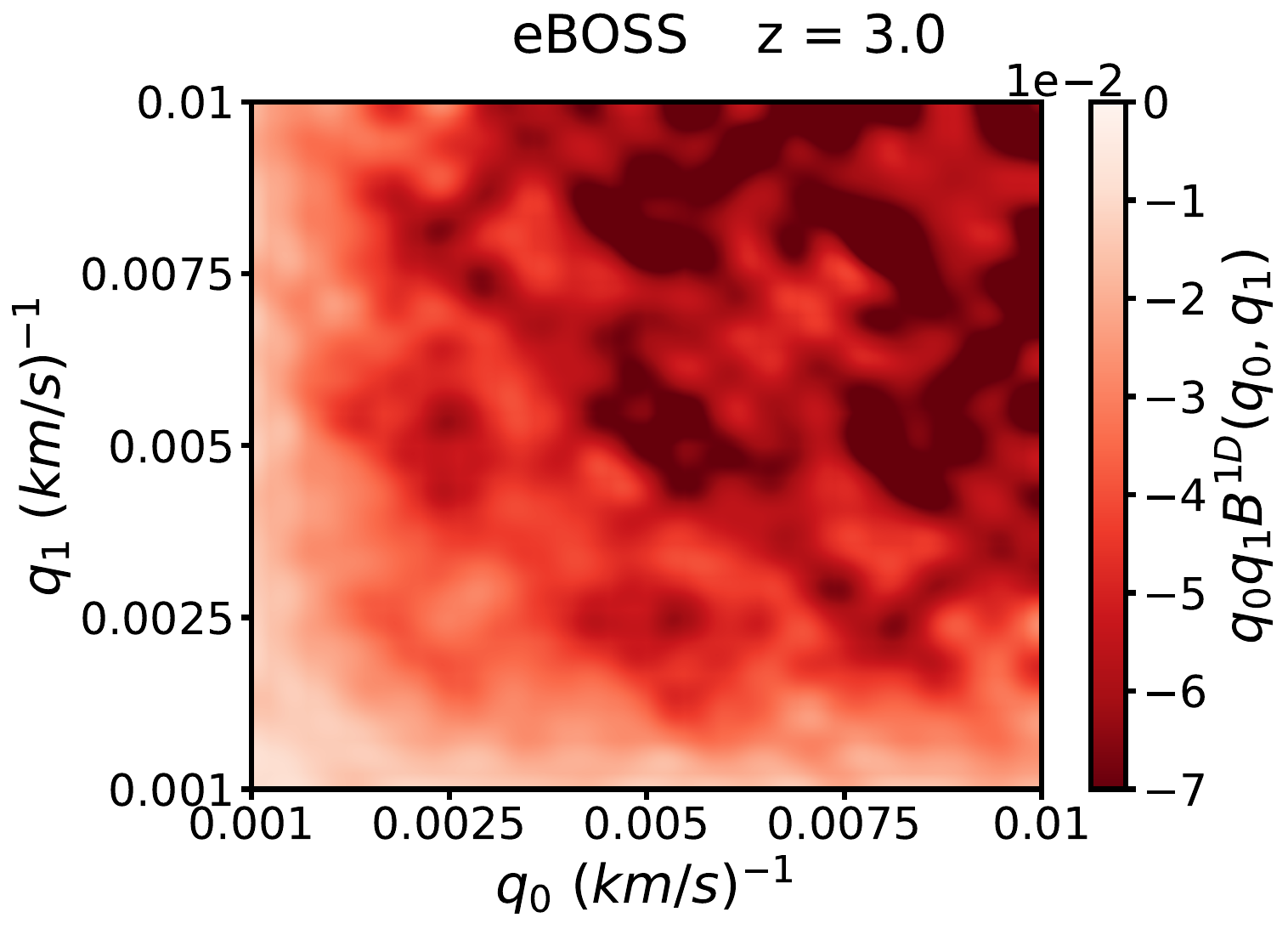}
\includegraphics[width=7.2cm]{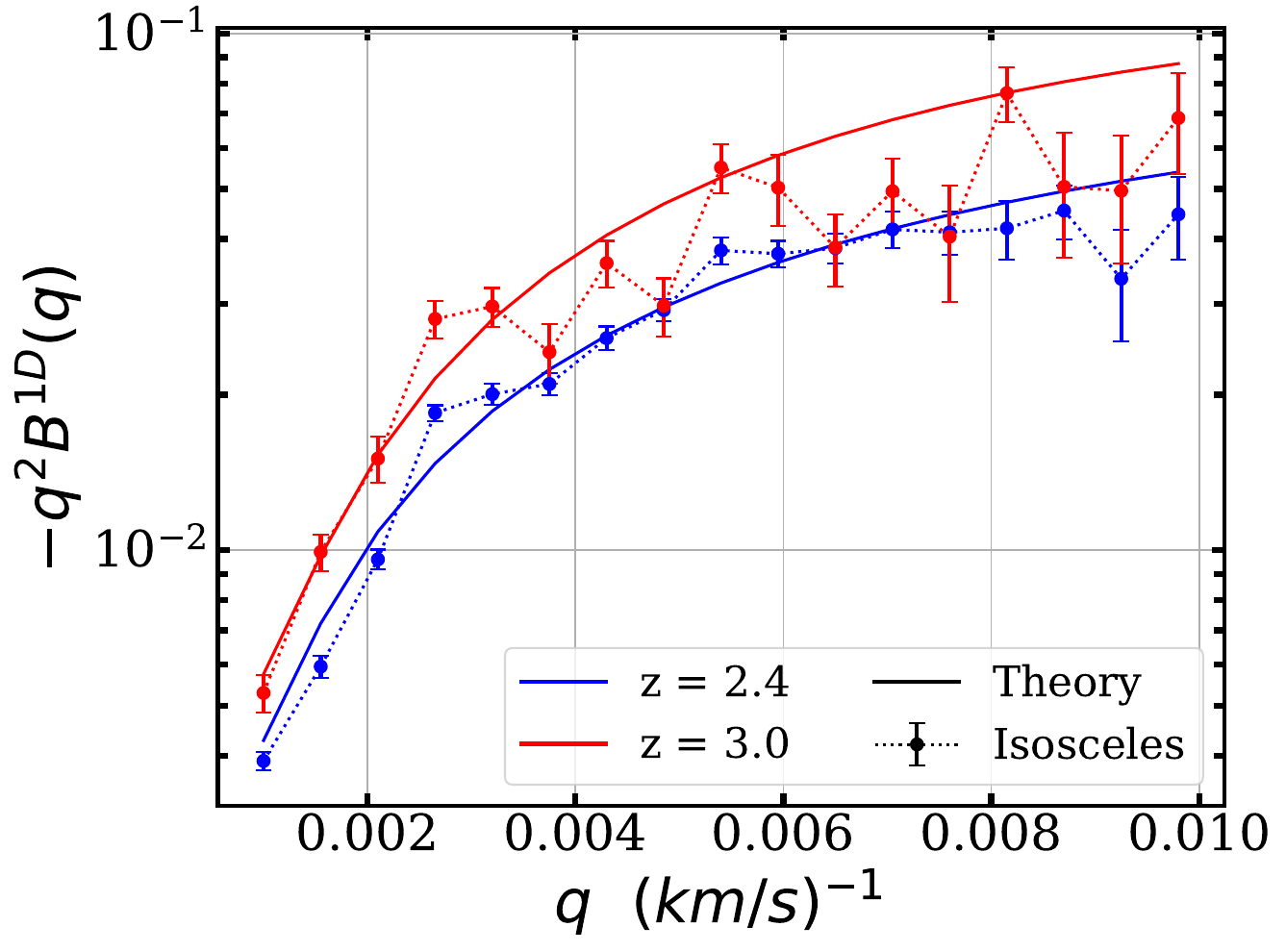}
\caption{One-dimensional bispectrum in the \lya forest using eBOSS DR16 data. \textit{Left:} 2D $\bk$ measurement for all triangle configurations along the line of sight at a redshift of $z = 2.4$. The dark areas are caused by the cross-correlations between \lya and $\siiii$. \textit{Right:} the 1D signal is obtained by extracting data from the diagonal of the 2D signal. The data points with error bars despict the $\bk$ measurements for isosceles triangles $(q_{0}=q_{1})$. Results are presented for two specific redshift bins: one at $z=2.4$ (blue) and the other at $z=3.2$ (red). The variation in amplitude between these signals is attributed to the evolution of the $\bk$ with redshift. The solid lines, corresponding to the same redshifts as the data points, represent the theoretical modeling of the signal based on second-order perturbation theory. These measurements were made by applying the Eq.~(\ref{eq08}). The error bars are obtained from the diagonal of the bootstrap covariance matrix and systematic uncertainties added in quadrature.}
\label{fig:plot_t0}
\end{center}
\end{figure}

Our primary result is the first measurement of the one-dimensional bispectrum in the \lya forest using data from a large galaxy spectroscopic survey, specifically the eBOSS DR16 dataset. We estimate the $\bk$ signal over 12 redshift bins, though only six of these bins provide sufficient statistical significance. After carefully analyzing the noise and other systematics, we show that the signal is compatible with simple models based on either second-order perturbation theory (2OPT) or synthetic mocks using LogNormal/2LPT simulation methods. Figure \ref{fig:plot_t0} summarizes our main results on the $\bk$ measurement from eBOSS.

Three-point correlations are defined over triangular configurations in real or Fourier space. In the case of $\bk$, the function is supported over co-linear triangles, defined by two momentum scales ($q_0, q_1$). The left panel of Figure \ref{fig:plot_t0} shows the full $\bk$ for all triangular configurations, while the right panel focuses on isosceles triangles. These isosceles configurations, corresponding to the diagonal in the $q_0q_1$ plane, exhibit the strongest bispectrum signal. Although a more detailed discussion will follow, there is clear redshift evolution in the signal, consistent with the 2OPT model. Moreover, as seen in the density plot in the left panel, periodic darker regions are observed, which are caused by the cross-correlation between the absorption of the \lya and the $\siiii$ quasar emission lines. Another notable result is the first measurement of the one-dimensional bispectrum in the sidebands, caused by metals on the red side of the \lya peak. Finally, the pipeline presented in this work can be directly applied to the next generation of large-scale spectroscopic \lya surveys, such as DESI (\cite{abareshi_overview_2022}), or high-resolution quasar data like those from SQUAD (\cite{Murphy_2018}).

The rest of the paper is structured as follows: In Sec. \ref{Sec02}, we present the formalism of the three-point statistic and the pipeline for performing the $\bk$ measurement. Sec. \ref{Sec03} focuses on the data analysis. In the following section, synthetic spectra are computed using Gaussian random field methods and Lagrangian perturbation theory, extending to mildly non-linear scales. The treatment of statistical and systematic uncertainties for the $\bk$ measurement is detailed in Sec. \ref{Sec05}. In Sec. \ref{Sec06}, we present our measurement using eBOSS DR16 data, along with the theoretical model of the bispectrum based on second-order perturbation theory. Finally, we offer conclusions in the last section.

\section{Overview of the data and the bispectrum formalism} \label{Sec02}
\subsection{Observational dataset} \label{sec021}
The DR16 quasar catalog includes 750,132 confirmed quasars, covering wavelengths from 3,600~\AA~ to 10,400~\AA~, with a spectral resolution of $\Delta \lambda = 0.11$~\AA. The dataset spans a redshift range of $0.0 < z < 6.0$. For this paper, we focus only on quasars with a \lya forest visible in their spectra, applying a redshift cut of $z_{Ly\alpha} \geq 2.1$. This subset consists of 193,346 quasar spectra with \lya forests, visually confirmed redshifts, and covering the range $2.1 < z < 4.8$. Furthermore, we limit the wavelength range to $3,650$~\AA~$< \lambda < 7,235$~\AA~ to avoid the CCD camera edges. For quasars with multiple observations, we use the combined spectra from all exposures on a given plate. To study redshift dependence, we consider a sub-sample with $2.2 < z < 4.4$, divided into 12 bins of size $\Delta z = 0.2$, with magnitudes in the r-band between 16.25 and 22.75.

Additionally, 117,458 spectra containing Damped Lyman Alpha systems (DLAs) were identified by \cite{Chabanier:2021dis} within the same redshift range. To detect more DLAs, we apply a simple DLA finder, which first calculates the observed flux power spectrum (fPk) and, using the continuum flux, computes the transmission power spectrum (TPk). If $\langle TPk \rangle/\langle fPk \rangle \geq 2.5$, the spectrum is flagged as containing a DLA, followed by manual visual inspection. In some cases, these visually inspected spectra exhibit significant portions of the \lya forest being cut, resulting in the exclusion of those quasars from the sample.

Moreover, 27,117 quasar spectra with Broad Absorption Lines (BALs) have been identified (\cite{Lyke_2020}). We exclude these BAL quasars, identified by a non-zero BI$\_$CIV flag, reducing the final sample to 166,229 \lya quasars.

\subsection{Methodology}\label{SubSec1-2}
The starting point consists of cleaning the initial sample. To that end, we first remove spectra with fewer than 50 pixels in the forest or where the mean spectral resolution is larger than 85 km/s. Moreover, since sky lines affect the $\pk$ and $\bk$ by increasing the pixel noise, we mask several sky lines in each forest \href{https://github.com/igmhub/picca/tree/master/etc}{\faGithub}\footnote{Formally, sky removal implies a convolution of the masks with the field in Fourier space. However, we assume it acts locally, so we only remove the sky lines before extracting the delta field and applying the Fourier transform. Sky lines are emission lines produced by various processes in Earth's atmosphere, which emit light at specific wavelengths. We use the list of sky lines from \url{https://github.com/igmhub/picca/tree/master/etc}.}, and instead of removing the entire pixel, we interpolate between neighboring pixels. This interpolation is performed before computing the continuum fit to avoid affecting it. However, once we calculate the delta field from the fitted continuum, we set the corresponding sky-line pixels to zero before performing the Fourier transform. We assess the impact of this masking scheme when discussing systematics. Additional steps for cleaning the initial dataset include accounting for galactic extinction and discarding bad spectra, which we define as those with a negative mean flux or no pixels in the \lya forest region. Finally, for the initial dataset, we mask DLAs using the catalog from \cite{Chabanier_2022}, at the redshift $z_{\mathrm{abs}}$, in the range $[\lambda_{C}-W/2, \lambda_{C}+W/2]$, where $\lambda_{C}=(1+z_{\mathrm{abs}})\lambda_{Ly\alpha}$ and $W$ is the equivalent width given by equation 20 of \cite{10.1093/mnras/stab3201}, $W=7.3(1+z_{\mathrm{abs}}) \sqrt{N_{H_{I}}/10^{20}cm^{-2}}$~\AA. Additional DLA systems identified by our simple DLA finder are removed from the catalog.

For the $\bk$ analysis, we restrict the \lya forest region to the range $1050$~\AA~$ < \lambda_{R.F.} < 1180$~\AA~ to avoid contamination of the correlation functions by astrophysical effects near the quasars \cite{Chabanier_2019}. We use a pixel size in velocity units of $\Delta v(\lambda) = 69$ km/s. This forest region is shown for an average quasar in Fig.~\ref{fig:plot0001}. We assume an average signal-to-noise ratio $\overline{SNR}$\footnote{We define $\overline{SNR}$ as $\langle f(\lambda)/\sigma_{p}(\lambda)\rangle$, where $f(\lambda)$ is the observed flux and $\sigma_{p}(\lambda)$ is the estimate of the standard deviation of the flux.} greater than 1.0.

The \lya forest region spans a redshift range $\Delta z\sim 0.4$ for quasars at $z_{Ly\alpha} = 2.4$, $\Delta z\sim 0.5$ at $z_{Ly\alpha} = 3.2$, and $\Delta z\sim 0.6$ at $z_{Ly\alpha} = 4.4$. To reduce redshift correlations, it is common to split the forest into three subregions (each containing about 170 pixels per QSO spectrum for the eBOSS sample) and assign them to different redshift bins when calculating $\pk$ (\cite{Chabanier_2019}, \cite{Palanque-Delabrouille:2013gaa}, \cite{ravoux2023dark}). However, in this work, we divide the forest into chunks for each line of sight and assign each chunk to its corresponding redshift bin. These chunks correspond to consecutive, non-overlapping sub-forests of different lengths. Each chunk spans at most $\Delta z = 0.2$. In addition to the previous cuts, each chunk must pass two further cuts. To improve the quality of the $\pk$ and $\bk$ estimations, we select chunks with $\overline{SNR}$ greater than 2 and exclude chunks with fewer than 50 pixels, a different approach from \cite{Chabanier_2019}. We are now ready to calculate the quasar continuum for each individual chunk.

\subsubsection{Continuum fitting}
The observable quantity in the \lya forest is the transmission, $F = e^{-\tau}$, where $\tau$ is the optical depth. By definition, it is also expressed as the transmitted flux fraction $F(\lambda) = f(\lambda)/C(\lambda)$, where $f(\lambda)$ is the observed flux and $C(\lambda)$ is the unabsorbed quasar flux. One can define the transmission fluctuation, or "delta" field, $\delta_{F}(\lambda)$, by
\begin{equation}
    \delta_{F}(\lambda) = \frac{F(\lambda)-\overline{F}(\lambda)}{\overline{F}(\lambda)} = \frac{f(\lambda)}{C(\lambda)\overline{F}(\lambda)} - 1, \label{eq00}
\end{equation}
where $\overline{F}(\lambda) = \langle f(\lambda)/C(\lambda) \rangle$ is the transmission ensemble average and $\langle\cdot\rangle$ denotes the ensemble average.

Different strategies can be used to compute the $\delta_{F}$ field. For example, one may choose to estimate the product $C(\lambda)\overline{F}(\lambda)$, as done with the public code PICCA \href{https://github.com/igmhub/picca}{\faGithub}\footnote{The Package for IGM Cosmological-Correlations Analyses (PICCA) is available at \url{https://github.com/igmhub/picca}.}, or independently estimate the $C(\lambda)$ and $\overline{F}(\lambda)$ functions, as is done when fitting the continuum using Principal Component Analysis (PCA) \cite{P_ris_2011}. In this work, we test both methods for measuring the $\delta_{F}(\lambda)$ field using mock spectra and compare them with the true continuum. For the PCA method, we employ the \emph{empca} code from \cite{Bailey_2012}. Interestingly, we find that with this PCA code, we are not directly measuring $C(\lambda)$, but rather the product $C(\lambda)\overline{F}(\lambda)$, similarly to the PICCA method. We observe that calculating $\left\langle f(\lambda_{\mathrm{obs}})/PCA(\lambda_{\mathrm{obs}})\right\rangle_{\lambda_{\mathrm{obs}}}$, where $PCA(\lambda_{\mathrm{obs}})$ is the PCA model result, closely approximates a unity value for each wavelength in the forest. This indicates that $\left\langle f(\lambda_{\mathrm{obs}})/PCA(\lambda_{\mathrm{obs}}) - 1 \right\rangle_{\lambda_{\mathrm{obs}}} \sim \langle \delta(\lambda_{\mathrm{obs}})\rangle$, confirming that PCA measures $C(\lambda)\overline{F}(\lambda)$. To capture as much variance as possible with the PCA, we normalize each quasar spectrum so that its integral over the forest is equal to a constant, where the value of the constant is chosen to have an integrate flux of one. This step is repeated independently for each line of sight.

\begin{figure}
\centering
\includegraphics[width=12.5cm]{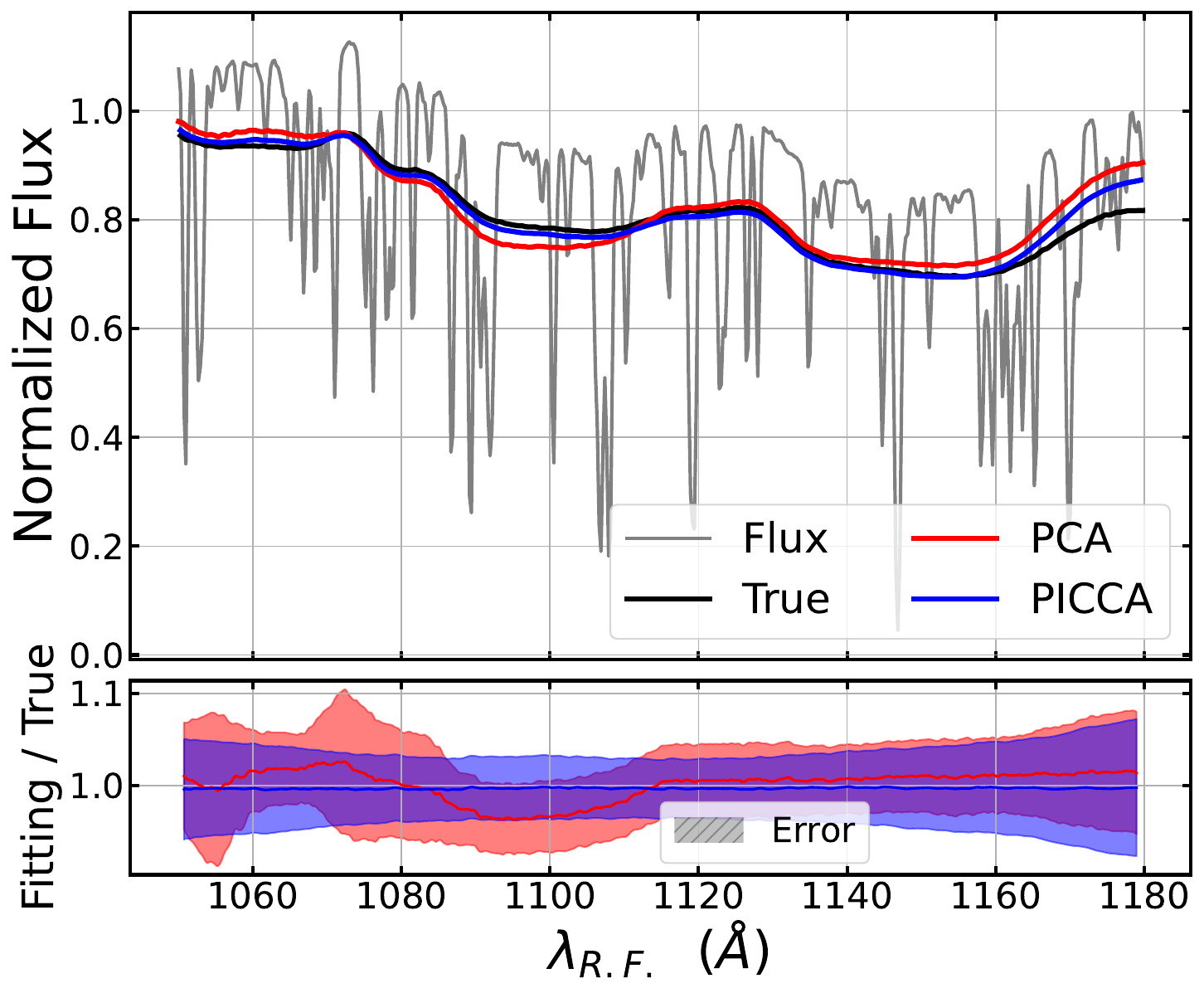}
\caption{Example of a synthetic quasar spectrum at z = 2.64 (gray line) in the rest-frame. The color lines cover our defined forest region, $\lambda_{R.F.}$ $\in$ [1050, 1180]~\AA, which lies between the quasar's Ly$\beta$ (1025.72~\AA\enspace) and \lya (1215.67~\AA\enspace) emission lines. On the top panel, the blue, red and black lines represent, respectively, the quantity $C(\lambda)\overline{F}(\lambda)$ measured using PICCA, PCA and the true continuum in the mocks.  The bottom panel shows the average ratio of the continuum fitting and the true continuum (solid lines), as well as the dispersion from the average (shaded bands).
}
\label{fig:plot0001}
\end{figure}
Fig. \ref{fig:plot0001} compares the $C(\lambda)\overline{F}(\lambda)$ quantity obtained using both the PICCA and PCA methods for a simulated quasar, compared to its true continuum. In both cases, the errors are below $10\%$, with smaller discrepancies in the central regions of the spectra and increasing towards the edges, particularly near the \lya and Ly$\beta$ peaks. Generally, the method employing PICCA, as described in detail in \cite{ravoux2023dark}, \cite{du_Mas_des_Bourboux_2020}, and \cite{10.1093/mnras/stad3781}, appears to provide greater homogeneity over the forest region compared to PCA. This stability is evident as PICCA does not exhibit the long-mode features seen in the PCA results, especially below $\sim$1,120~\AA. These features in the PCA method arise from the selection of eigenvectors used for spectrum reconstruction, where higher principal components correct the mean spectrum of each quasar. As a result, using more principal components introduces greater features in the continuum relative to the true continuum. The emergence of features before $\lambda = 1,120$~\AA~suggests that, on average, absorption lines around $\lambda = 1,070$~\AA~are shallower, while those around $\lambda = 1,100$~\AA~are deeper compared to absorption lines at $\lambda \geq 1,120$~\AA, where the depth remains generally constant. Regardless of the continuum fitting choice, it introduces a bias to both the $\pk$ and $\bk$ signals. Continuum fitting generally suppresses the $\pk$ signal relative to the true continuum, particularly on large scales. With PICCA, this suppression is approximately $3\%$, while with PCA, it is about $5\%$. We further discuss these biases in the continuum fitting methods when addressing systematics in the bispectrum. The suppression of the $\bk$ signal varies depending on the triangle configuration used, but generally exceeds that of the $\pk$ signal, as discussed in Sec. \ref{Sec04}. Despite PICCA demonstrating a smaller deviation from the true continuum, we opt to use the PCA method for continuum extraction, as it provides a clearer redshift evolution of the $\bk$ signal compared to PICCA. In fact, we observe little redshift evolution when using PICCA. With the continuum fitting method established, we can proceed to calculate the delta field, accounting for the contributions of various absorbers involved in the \lya forest.

\subsubsection{FFT 1D power spectrum} \label{subsubsec222}
As will become evident in the next section, where we introduce an estimator for the 1D bispectrum of the "delta field," $\delta_{F}(\lambda)$ from Eq.~(\ref{eq00}), it is also necessary to estimate the 1D power spectrum. Therefore, we begin by deriving the $\pk$. As discussed in \cite{McDonald_2006}, \cite{ravoux2023dark}, and \cite{Chabanier_2019}, the $\delta_{F}$ field in the \lya region can be decomposed into several contributions: one component from the \lya peak absorption and another originating from metal absorption. Metal contributions in the \lya forest analysis can be further classified into two types: those redward (i.e., at larger wavelengths) of the \lya peak, known as side-bands, and those within the \lya region itself. The absorptions redward of the \lya peak are independent of the \lya peak absorption and are know as side-bands. We define side band one (SB1) in the range $1270$~\AA~$ < \lambda_{\mathrm{R.F.}} < 1380$~\AA~ and side band two (SB2) for $1410$~\AA~$ < \lambda_{\mathrm{R.F.}} < 1520$~\AA. Metal absorptions within the \lya region are primarily due to the $\mathrm{Si_{\textsc{ii}}}$ ($\lambda_{\mathrm{Si_{\textsc{ii}}}} = 1,190$~\AA) and $\mathrm{Si_{\textsc{iii}}}$ ($\lambda_{\mathrm{Si_{\textsc{iii}}}} = 1,206.50$~\AA) emission lines. These metal features are correlated with \lya absorption lines and generate oscillations in both the power spectrum and the bispectrum. In summary, the delta field in the \lya region can be expressed as a sum of the following contributions:
\begin{align}
    \delta_{F}(\lambda) = \left(\delta^{Ly\alpha}(\lambda) + \delta^{\siii/\siiii}(\lambda) + \delta^{\mathrm{SB1/SB2}}(\lambda)\right)  \mathrm{W}(\lambda,R,\Delta \lambda) + \delta^{n}(\lambda), \label{eq01}
\end{align}
where $\delta^{n}$ represents noise fluctuations, and the function $W$ accounts for the spectrograph resolution, which we will characterize later. The 1D \lya power spectrum can be estimated from this decomposition by applying a fast Fourier transform (FFT) to $\delta_{F}$ for each chunk separately. The delta pixels are assumed to be equally spaced to allow the use of a simple FFT algorithm. The eBOSS quasar catalog provides a constant pixel width of $\Delta [\log(\lambda)] = 10^{-4}$, corresponding to $\Delta v (\lambda) = c \ln(10)\Delta[\log(\lambda)] = 69$ $km/s$ in velocity units. As a result, the Fourier space wave-vector, $k = 2\pi/\Delta v$, is measured in $(km/s)^{-1}$.

We now define an estimator for the $\pk$. In the absence of instrumental effects, the $\pk$ can be written as the product of two Fourier-transformed $\delta_{F}(\lambda)$, namely $(2\pi)\delta_{\mathrm{D}}(k-k')P^{\mathrm{raw}} = \langle \delta_{F}(k)\delta_{F}(k') \rangle$, where $\delta_{\mathrm{D}}(k)$ is the 1D Dirac delta function. However, when all contributions to the delta field from Eq.~(\ref{eq01}) are included, the raw power spectrum becomes
\begin{align}
    P^{\mathrm{raw}}(k) = \left(P^{Ly\alpha}(k)+P^{Ly\alpha-\siii/\siiii}(k)+P^{\mathrm{SB1/SB2}}(k)\right) W^{2}(k,R,\Delta v) + P^{n}(k). \label{eq02}
\end{align}
In this decomposition, $P^{Ly\alpha}$, $P^{Ly\alpha-\siii/\siiii}$, $P^{\mathrm{SB1/SB2}}$, and $P^{n}$ represent the power spectra corresponding to the components $\delta^{Ly\alpha}$, $\delta^{\siii/\siiii}$, $\delta^{\mathrm{SB1/SB2}}$, and $\delta^{n}$, respectively, as defined in Eq.~(\ref{eq01}). The estimation of the power spectrum, accounting for the different absorbers in the \lya forest, was first studied in \cite{McDonald_2006}. The window function $W$, which characterizes the spectrograph’s spectral response and pixel width, is defined in \cite{Palanque-Delabrouille:2013gaa} as
\begin{equation}
W(k,R,\Delta v) = \exp\Big(-\frac{(kR)^{2}}{2}\Big) \times \frac{\sin(k\Delta v / 2)}{k\Delta v / 2}, \label{window}
\end{equation}
where $R$ is the spectrograph resolution measure in $(km/s)^{-1}$. This estimator for the $\pk$ is our starting point to construct the 1D bispectrum.

\subsection{\texorpdfstring{\lya}{lya} Bispectrum Formalism and Estimator}\label{Subsec2-2-1}
Before constructing an estimator for the \lya $\bk$, let us first review some general properties of three-point statistics in cosmology. The 3D matter three-point correlation function (3PCF),
$$\xi^{(3)}(\mathbf{x}_0,\mathbf{x}_1,\mathbf{x}_2)\equiv \langle \delta(\mathbf{x}_0)\delta(\mathbf{x}_1)\delta(\mathbf{x}_2) \rangle ,$$
measures the correlation between overdensity triplets in a sample. It depends on 9 degrees of freedom (dofs), which naturally define the vertex coordinates of a triangle in real space. Statistical homogeneity (or equivalently, translation invariance, which allows us to choose one vertex as the origin) reduces this dependence to 6, which can be interpreted as the two vector sides of the triangle ($\mathbf{r}_1=\mathbf{x}_1-\mathbf{x}_0$, $\mathbf{r}_2=\mathbf{x}_2-\mathbf{x}_0$). Statistical isotropy further reduces the number of variables to 3, which we choose to be the lengths of the two sides of the triangle ($|\mathbf{r}_1|$ and $|\mathbf{r}_2|$) and their corresponding opening angle ($\mu=\mathbf{r}_1\cdot\mathbf{r}_2$). 

The bispectrum, $B(\mathbf{k}_0,\mathbf{k}_1,\mathbf{k}_2)$, is simply the Fourier transform of the 3PCF. In terms of an ensemble average over the $\delta$ fields, it reads
\begin{eqnarray}
\langle \delta(\mathbf{k}_0) \delta(\mathbf{k}_{1}) \delta(\mathbf{k}_{2}) \rangle &=&\int d^3\mathbf{x}_0 d^3\mathbf{x}_1 d^3\mathbf{x}_2\,
e^{i(\mathbf{x}_0\cdot\mathbf{k}_0+\mathbf{x}_1\cdot\mathbf{k}_1+\mathbf{x}_2\cdot\mathbf{k}_2)}
\langle \delta(\mathbf{x}_0) \delta(\mathbf{x}_{1}) \delta(\mathbf{x}_{2}) \rangle 
\nonumber \\&=&\int d^3\mathbf{x}_0 \,
e^{i\mathbf{x}_0\cdot(\mathbf{k}_0+\mathbf{k}_1+\mathbf{k}_2)} \int d^3\mathbf{r}_1 d^3\mathbf{r}_2\,
e^{i(\mathbf{r}_1\cdot\mathbf{k}_1+\mathbf{r}_2\cdot\mathbf{k}_2)}
\xi^{(3)}(\mathbf{r}_1,\mathbf{r}_2)
\nonumber \\&=&\left(2\pi\right)^3\delta^{D}(\mathbf{k}_0+\mathbf{k}_{1}+\mathbf{k}_{2})B(\mathbf{k}_0,\mathbf{k}_{1},\mathbf{k}_{2}), \label{Beqn}
\end{eqnarray}
where we have used the homogeneity of the 3PCF. In the last equality, we applied the definition of the Dirac delta function, $\delta^{D}$, whose consequence is that the k-vectors must form a triangle. Therefore, the bispectrum also has 6 dof when statistical homogeneity is assumed, and it is further reduced to 3 variables when isotropy is taken into account. As in real space, the resulting isotropic bispectrum can be expressed in terms of the two side lengths of triangles in k-space ($k_1=|\mathbf{k}_1|$, $k_2=|\mathbf{k}_2|$) and their corresponding opening angle ($\alpha=\mathbf{k}_1\cdot\mathbf{k}_2$). From symmetry arguments, one may expect the signal in Fourier or configuration space to peak for isosceles triangles in random data, with three notable configurations: equilateral, squeezed (when the opening angle approaches zero), and spread (when one side equals the sum of the other two). These configurations are also prominent in \lya analysis, as we will demonstrate later.

When restricted to a one-dimensional distribution (e.g., the \lya forest along the line of sight), the triangle configurations in both configuration and Fourier space collapse into a single line. In other words, the cosine of the opening angle ($\alpha$) reduces to $\pm1$, leaving only two independent variables. Formally, we split the $\mathbf{k}$ space into parallel ($q\equiv\mathbf{k}_\parallel$) and perpendicular ($\mathbf{k}_\bot$) directions. Using the 3D-to-1D Fourier space relation (see, e.g., \cite{kaiser91,Desjacques:2004sq}) $$ f_q^{(1d)}= \int \frac{d\mathbf{k}^2_\bot}{(2\pi)^2} f_{q=\mathbf{k}_\parallel, \mathbf{k}_\bot}^{(3d)},$$
the 1D Bispectrum from equation (\ref{Beqn}) reduces to
\begin{equation}
\langle \delta(q_0) \delta(q_{1}) \delta({q}_{2}) \rangle =
\left(2\pi\right)\delta^{D}(q_0+{q}_{1}+{q}_{2})B({q}_{0},{q}_{1},{q}_{2}), \label{B1deqn}
\end{equation}
where Dirac's delta function imposes the linear restriction $q_2=-q_0-q_1$. 

\subsubsection{FFT bispectrum estimator for \texorpdfstring{\lya}{lya}}\label{Subsec2-3}
As we proceeded with the $\pk$, the 1D bispectrum can be estimated by applying the FFT algorithm on $\delta_{F}$ from Eq.~(\ref{eq01}) for each chunk, and by estimating the ensemble average of the third moment of $\delta(\Delta v)$ values. In the absence of instrumental effects (such as noise and spectrograph resolution), the $\bk$ can be simply written as the real part of Eq.~(\ref{B1deqn}), namely:
$$B^{\mathrm{raw}}(q_{0},q_{1},q_{2}) = \langle \mathrm{Re}[\mathcal{F}(\delta(\Delta v_{0})) \cdot \mathcal{F}(\delta(\Delta v_{1})) \cdot \mathcal{F}(\delta(\Delta v_{2}))] \rangle,$$
where $\mathcal{F}(\delta(\Delta v))$ represents the Fourier transform of $\delta_{F}(\Delta v)$, binned into pixels of width $\Delta v$ over the \lya forest region. As a reminder, momentum conservation, or equivalently the presence of Dirac's delta function, enforces the closure of the momentum variables over the line of sight, restricting the 1D bispectrum to depend on two variables. We select $q_{0}$ and $q_{1}$ as these variables. As a result, the noiseless raw bispectrum is given by:
$$B^{\mathrm{raw}}(q_{0},q_{1}) = \langle \mathrm{Re}[\delta_{F}(q_{0})\delta_{F}(q_{1})\delta_{F}(-q_{0}-q_{1})] \rangle.$$
Notice that it is not necessary to compute the bispectrum over the entire two-dimensional plane because there are several symmetries in the signal. Since $\delta(\Delta v)$ is real, it follows that $\delta(-q) = \delta^{*}(q)$. Consequently, the bispectrum exhibits the following symmetries
\begin{align}
    &B(q_{0},q_{1}) = B(q_{1},q_{0}) = B^{*}(-q_{0},-q_{1}), \nonumber \\ &
    B(q_{0},q_{1}) = B^{*}(-q_{1},q_{0}+q_{1}) = B^{*}(-q_{0},q_{0}+q_{1}), \nonumber \\ & 
    B(q_{0},q_{1}) = B(-q_{0}-q_{1},q_{1}) = B(q_{0},-q_{0}-q_{1}). 
    \label{eq06}
\end{align}
Thus, we only need to compute $B^{\mathrm{raw}}(q_{0},q_{1})$ for $q_{0},q_{1} \geq 0$ and can use the first of the above symmetries to construct the full $q$-dependence shown in Fig. \ref{fig:plot_t0}. 

When considering the other effects in the delta field that lead to the decomposition in Eq.~(\ref{eq01}), the associated third moment in Fourier space simplifies to
\begin{eqnarray}
    B^{\mathrm{raw}}(q_{0},q_{1}) & = &\left(B^{Ly\alpha}(q_{0},q_{1})+B^{Ly\alpha-\siii/\siiii}(q_{0},q_{1})+B^{\mathrm{SB1/SB2}}(q_{0},q_{1})\right) \nonumber \\ && \cdot  W(q_{0},R,\Delta v)W(q_{1},R,\Delta v)W(q_{0}+q_{1},R,\Delta v) \nonumber \\ && +  P^{n}(q)\left[P^{\mathrm{raw}}(q_{0})+P^{\mathrm{raw}}(q_{1})+P^{\mathrm{raw}}(q_{0}+q_{1})\right]-2P_{n}^{2}(q). 
    \label{eq07}
\end{eqnarray}
In this decomposition, $B^{Ly\alpha}$ and $B^{\mathrm{SB1/SB2}}$ are the auto-bispectra associated with the transmission fluctuation fields $\delta^{Ly\alpha}$ and $\delta^{\mathrm{SB1/SB2}}$, respectively, with the same definition as the real part of Eq.~(\ref{B1deqn}) for the corresponding delta field. The transmission fluctuation field attributed to noise fluctuations, $\delta^{n}$, introduces two contributions to the bispectrum signal, as previously discussed in the literature (see, for example, \cite{Matarrese:1997sk} and \cite{PhysRevD.96.023528}). The first contribution arises from the cross-term between the noise power spectrum and the raw power spectrum, while the second term comes from the quadratic power of the noise power spectrum in Eq.~(\ref{eq07}). Moreover, the auto and cross-correlations between $\siii$ and $\siiii$ are subdominant and therefore neglected. All cross terms between two delta fields of $\siii/\siiii$ and one delta field of \lya, as well as cross terms with noise fluctuations and side-band metals, are also neglected. The only non-negligible cross terms are
\begin{eqnarray*}
    B^{Ly\alpha-\siii/\siiii}(q_{0},q_{1})&\approx & 2Re[\delta^{Ly\alpha}(q_{0})\delta^{\siii/\siiii}(q_{1})\delta^{Ly\alpha}(-q_{0}-q_{1})] \\ && + Re[\delta^{Ly\alpha}(q_{0})\delta^{Ly\alpha}(q_{1})\delta^{\siii/\siiii}(-q_{0}-q_{1})],
\end{eqnarray*}
which corresponds to the correlated absorptions of the \lya with either $\siii$ or $\siiii$.

The final FFT estimator for the 1D bispectrum is computed as an average over all available \lya forests within a given chunk. Since, at present, we are unable to independently extract the $B^{Ly\alpha}(q_{0},q_{1})$ and $B^{Ly\alpha-\siii/\siiii}(q_{0},q_{1})$ components of Eq.~(\ref{eq07}), we combine them by defining $\bk(q_{0},q_{1}) = \left\langle B^{Ly\alpha}(q_{0},q_{1}) + B^{Ly\alpha-\siii/\siiii}(q_{0},q_{1}) \right\rangle$. Thus, from Eq.~(\ref{eq07}), we conclude that the estimator for $\bk$ is
\begin{equation}
    \bk(q_{0},q_{1}) = \left\langle \frac{B^{\mathrm{raw}}_{q_{0},q_{1}} - P^{n}(q)\left[P^{\mathrm{r}}(q_{0})+P^{\mathrm{r}}(q_{1})+P^{\mathrm{r}}(q_{0}+q_{1})\right]+2P_{n}^{2}(q)}{W(q_{0},R,\Delta v)W(q_{1},R,\Delta v)W(q_{0}+q_{1},R,\Delta v)} \right \rangle - B^{\mathrm{SB1/SB2}}(q_{0},q_{1}), \label{eq08}
\end{equation}
where we use the shorter notation $P^{r} \equiv P^{\mathrm{raw}}$. The Eq.~(\ref{eq08}) is our final estimator to measure the 1D bispectrum.

The bispectrum, $\bk$, is divided into different redshift bins to account for its evolution, as discussed in Sec. \ref{Sec06}. The negative sign in the bispectrum is consistent with the findings of \cite{10.1111/j.1365-2966.2004.07404.x}, and we demonstrate that this negative sign persists for other combinations of $q_{0}$ and $q_{1}$. Furthermore, \cite{Zaldarriaga:2000rg} suggested that the negative sign might result from higher-order correlations arising due to gravitational growth.

In this work, we focus on two specific triangle configurations: the isosceles and the special case of the squeezed limit. As mentioned by \cite{Matarrese:1997sk} and \cite{Gil-Marin:2014sta}, the triangle configurations with the most prominent signals are those that share a common $q$-vector $(q_{1} = n\cdot q_{0})$, where $n$ is an integer. In appendix \ref{appendix:apex1}, we will verify that these two configurations yield the strongest signals. Isosceles triangles are characterized by $q_{1} = q_{0}$, and we have adopted the definition of the squeezed limit from \cite{Viel:2008jj}.
\begin{itemize}
    \item \emph{Configuration 1 (isosceles)}: $q_{1}= q_{0}$
    \begin{equation*}
        B(q)=Re[\langle \delta(q)\delta(q)\delta^{*}(2q)\rangle]
    \end{equation*}
    \item\emph{Configuration 2 (squeezed)}: $q_{0}=q - q_{\mathrm{min}}$ and $q_{1}=-q - q_{\mathrm{min}}$
    \begin{equation}
        B(q)=Re[\langle \delta(q - q_{\mathrm{min}})\delta^{*}(q + q_{\mathrm{min}})\delta(2q_{\mathrm{min}})\rangle]. \label{configs}
    \end{equation}
\end{itemize}
Formally the squeezed limit is when $q_{min}\rightarrow 0$, but in practise one cannot take that limit given the forest finite range. Therefore, one may use the first bin of the delta field, however and in order to avoid unphysical signals, we take $q_{min}$ as the second bin.
In the following, we present the flux bispectrum as a function of the wavenumber $q = q_{0}$. Hereafter, we will refer to the bispectrum of isosceles triangles as $\bk$, while for the squeezed limit bispectrum, we will use $\bksq$. In the case of the \lya forest, configurations close to the squeezed limit may offer an interesting approach for studying primordial non-Gaussianities (e.g., \cite{Viel:2008jj,PhysRevLett.109.121301,Chiang_2017}) or consistency conditions when compared with the power spectrum (see, for example, \cite{Peloso:2013zw, Mooij:2015yka,Goldstein:2022hgr}).

\section{Data analysis and \texorpdfstring{$\pk$}{P1D} assessment }\label{Sec03}
The instrumental effects in eBOSS have been extensively investigated in previous studies related to the measurement of $\pk$ (\cite{Palanque-Delabrouille:2013gaa}, \cite{Chabanier_2019}, and \cite{ravoux2023dark}). However, the impact of these effects on $\bk$ estimation remains unexplored. For this reason, we begin by analyzing these effects in the context of our study. Specifically, we focus on the impact of spectral resolution, instrumental noise, and the metal (side-bands) bispectrum. In addition, we provide a brief overview of our final sample selection and present the $\pk$ estimate within the \lya forest, comparing it with previous measurements to assess the effectiveness of our pipeline.

\subsection{Summary of the \texorpdfstring{\lya}{lya} sample} \label{subsec031}
As mentioned in Sec. \ref{sec021}, our \lya forest sample consists of 166,229 quasars. After removing BALs, missing DLAs, or bad spectra using our DLA finder, \lya forests with less than 50 pixels, and applying the $\overline{SNR}$ cut, the sample is reduced to 122,066 quasars. Since we work with chunks to measure the $\bk$, we improve the quality of the low/high-redshift chunks by considering CCD-observed wavelengths in the range 3770~\AA~$\leq \lambda_{\mathrm{obs}} \leq$ 6687~\AA. After removing chunks with $\overline{SNR} < 2$ and discarding those where the mean spectral resolution exceeds 85 km/s, our final \lya forest chunk sample contains 151,621 chunks. For SB1, we have 256,751 chunks.

\subsection{Spectrograph resolution} \label{subsec032}
The window function correction, as expressed in Eq.~(\ref{window}), encompasses two effects on the measurement of $\pk$ and $\bk$: one due to pixelization (spectrum binning or pixel width) and the other due to spectrograph resolution. While this correction has been thoroughly investigated for SDSS $\pk$ measurements (\cite{Palanque-Delabrouille:2013gaa}, \cite{Chabanier_2019}, and \cite{McDonald_2006}), its application to $\bk$ measurements remains unexplored. Eq.~(\ref{window}) shows that the influence of the spectrograph resolution $R$ is modeled by a Gaussian function. We determine $R$ following the same analysis performed in \cite{Chabanier_2019} and described in detail in \cite{Palanque-Delabrouille:2013gaa}, accounting for corrections to $R$ as a function of CCD position, fiber number, and wavelength. We then apply this correction to each pixel of every spectrum.
\begin{figure}
\centering
\includegraphics[width=10.5cm]{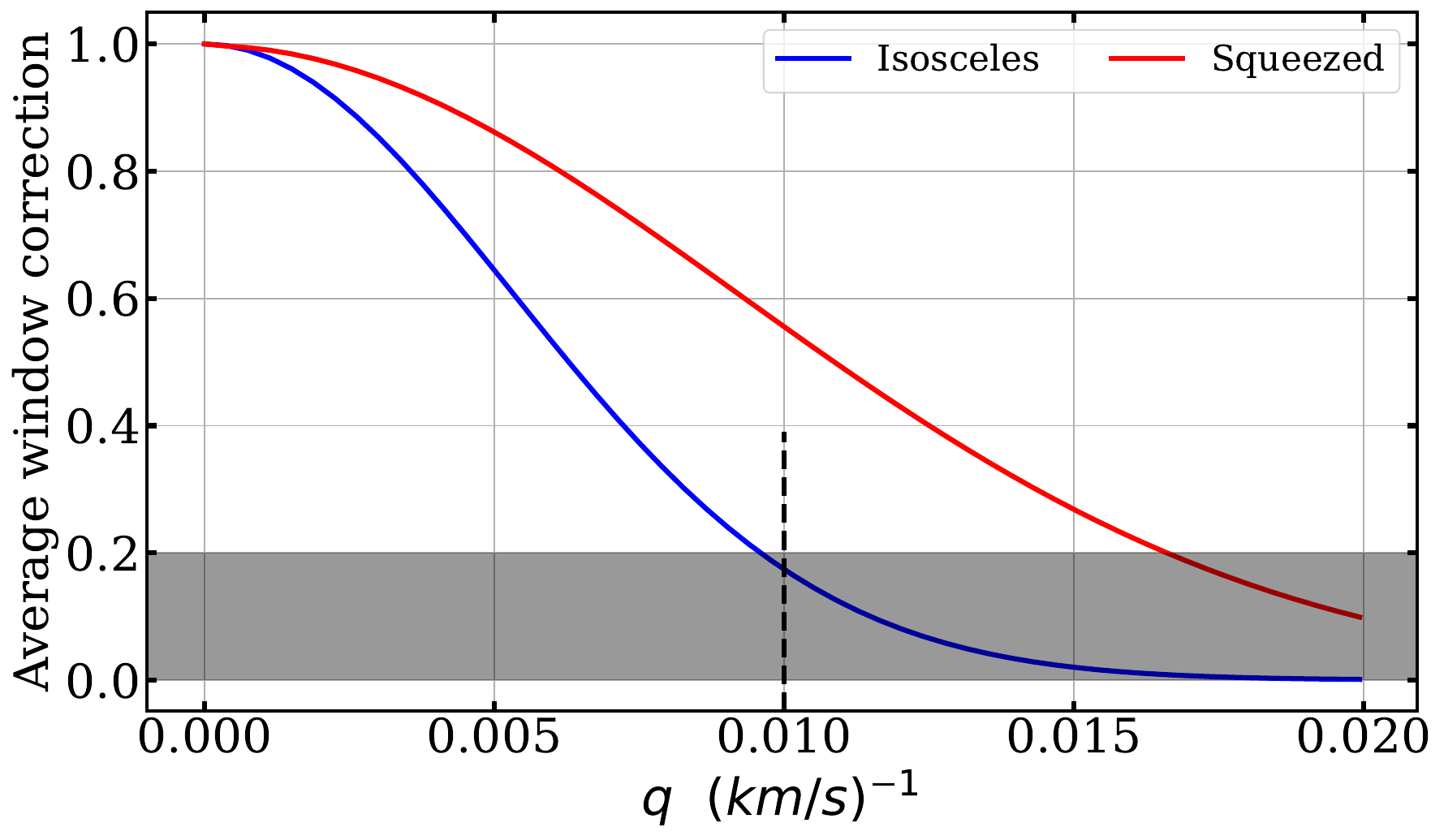}
\caption{Window function for $\bk$ measurements, either isosceles triangles or squeezed limit triangles. In thew case of isosceles triangles the suppression of the $\bk$ signal given by spectrograph resolution is close to 80 per cent, while $\bksq$ signal is 90 per cent. }
\label{fig:plot02}
\end{figure}
Fig. \ref{fig:plot02} shows the average window correction for the two configurations defined in \ref{configs}. These corrections reveal that both resolution and pixelization suppress the isosceles bispectrum signal by just over $80\%$ at $q = 0.01~(km/s)^{-1}$. For the squeezed limit bispectrum, the suppression exceeds $90\%$, similar to that of $\pk$. The maximum value of $q$ varies depending on the triangle configurations and the Nyquist-Shannon limit: $k_{\mathrm{Nyquist}} = \pi/\Delta v = 0.045~(km/s)^{-1}$. However, we choose $k_{\mathrm{max}} = 0.02~(km/s)^{-1}$ for $\pk$ and $\bksq$ ($q_{\mathrm{max}} = 0.02~(km/s)^{-1}$), as the impact of the spectrograph resolution becomes very strong beyond this value. For \emph{Configuration 1}, we set $q_{\mathrm{max}} = 0.01~(km/s)^{-1}$ due to a factor of two dependence.

\subsection{Noise Bispectrum measurement} \label{subsec033}
\subsubsection{Estimator of noise power spectrum}
One of the systematics that significantly impacts the $\pk$ measurement signal at small scales is the noise power spectrum $P^{n}$. Therefore, it is essential to accurately estimate it. Over the years, two methods have been employed for calculating $P^{n}$: the first involves directly estimating it from the pipeline noise, while the second utilizes the exposure difference method. For our bispectrum estimation, we opted for the exposure difference method, which is extensively detailed in \cite{ravoux2023dark} and has been applied in \cite{Chabanier_2019} and \cite{Palanque-Delabrouille:2013gaa}. To tune the value of $P^{n}$, we follow the procedure described in \cite{Chabanier_2019}, where, for each redshift bin, we fit $P^{\mathrm{raw}}$ on $q$-modes above $q = 0.02$ $(km/s)^{-1}$ with an exponential decay plus a constant $P^{\mathrm{raw}}_{lim}$. We then compute the ratio $\beta$ between $P^{\mathrm{pipeline}}_{n}$ and $P^{\mathrm{raw}}_{lim}$. The pipeline noise is defined as $P^{n} = P^{\mathrm{pipeline}}_{n}$ when $\beta < 1$, and $P^{n} = P^{\mathrm{pipeline}}_{n}/\beta$ otherwise.

\subsubsection{Characterization of noise in the bispectrum signal} 
\begin{figure}
\centering
\includegraphics[width=12.5cm]{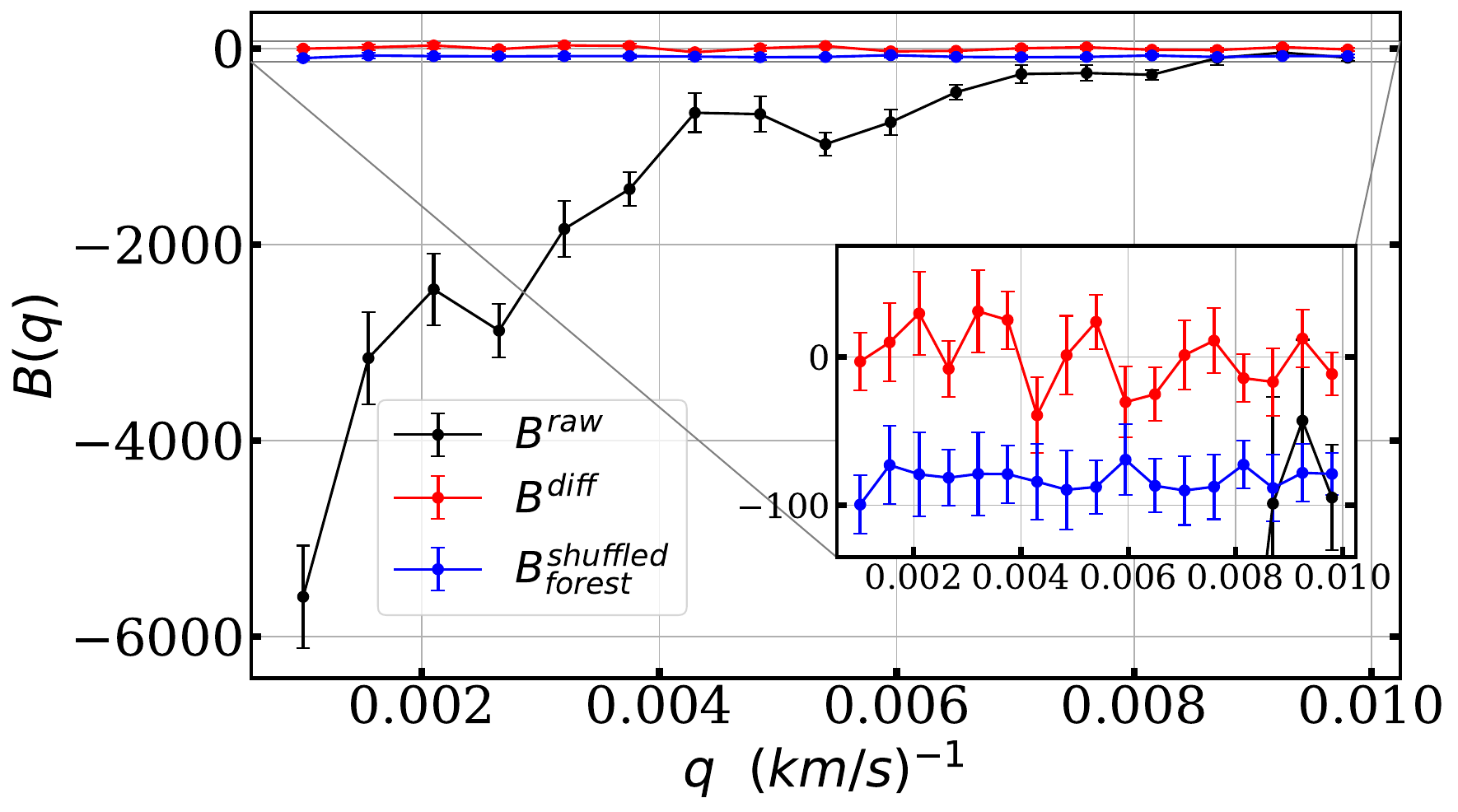}
\caption{Comparison between the flux bispectrum of observed fluxes (circles blacks), "randomized" absorption lines in the forest spectra (circles blues) and the bispectrum from the exposure difference noise (circles reds). The plot represent \emph{Configuration 1}. The bias present in the randomized bispectrum is given by the background power of the metals in the side-bands. The error bars come from the diagonal of the bootstrap covariance matrix.
}
\label{fig:plot03}
\end{figure}

To demonstrate that the raw bispectrum signal originates from a non-trivial source, we compare it with two distinct, trivial signals. The first comparison is with noise fluctuations as defined by Eq.~(\ref{eq01}), with its power spectrum computed using the exposure difference method. This time, our aim is to determine the bispectrum associated with the noise fluctuations, again using the exposure difference method. Specifically, we seek to measure $B^{\mathrm{diff}}(q_{0},q_{1},q_{2}) = Re[\mathcal{F}(n(\Delta v_{0}))\cdot \mathcal{F}(n(\Delta v_{1}))\cdot\mathcal{F}(n(\Delta v_{2}))] + \cdots$, assuming a Gaussian random field with zero mean, though its dispersion may explain some of the power in the raw data signal. However, as seen in Fig. \ref{fig:plot03}, this is not the case. Here, $\mathcal{F}(n(\Delta v_{i}))$ represents the Fourier transform of the normalized exposure difference spectrum, whose power spectrum is expected to be an accurate estimator of $P^{n}$ in the coadded spectrum. Furthermore, $P^{n}$ is found to be scale-independent, as expected for white noise. In the case of white noise, its bispectrum is expected to be zero, as shown by the red line in Fig. \ref{fig:plot03}. 
The second comparison is with a bispectrum signal obtained by shuffling the delta field at the pixel level ($B^{\mathrm{shuffled}}_{\mathrm{forest}}$). We shuffle each pixel of the delta field randomly, similar to what was done in \cite{10.1111/j.1365-2966.2004.07404.x}, with the main difference being that they shuffled absorption features instead of pixels, thus preserving the smallest structures. Since we aim to distinguish the raw signal from this other delta field configuration on large scales, we perform the shuffling at the pixel level. Again, the raw signal is not consistent with this shuffled configuration on scales below $q \lesssim 0.01$, as shown in Fig. \ref{fig:plot03}. To identify the source of bias in the $B^{\mathrm{shuffled}}_{\mathrm{forest}}$ signal, we computed the same for the side-bands. In this case, we found no bias, as the $B^{\mathrm{shuffled}}_{\mathrm{forest, SB1}}$ signal was consistent with zero. Thus, we conclude that the bias arises from the background power associated with the side-bands.

\begin{figure}
\centering
\includegraphics[width=12.5cm]{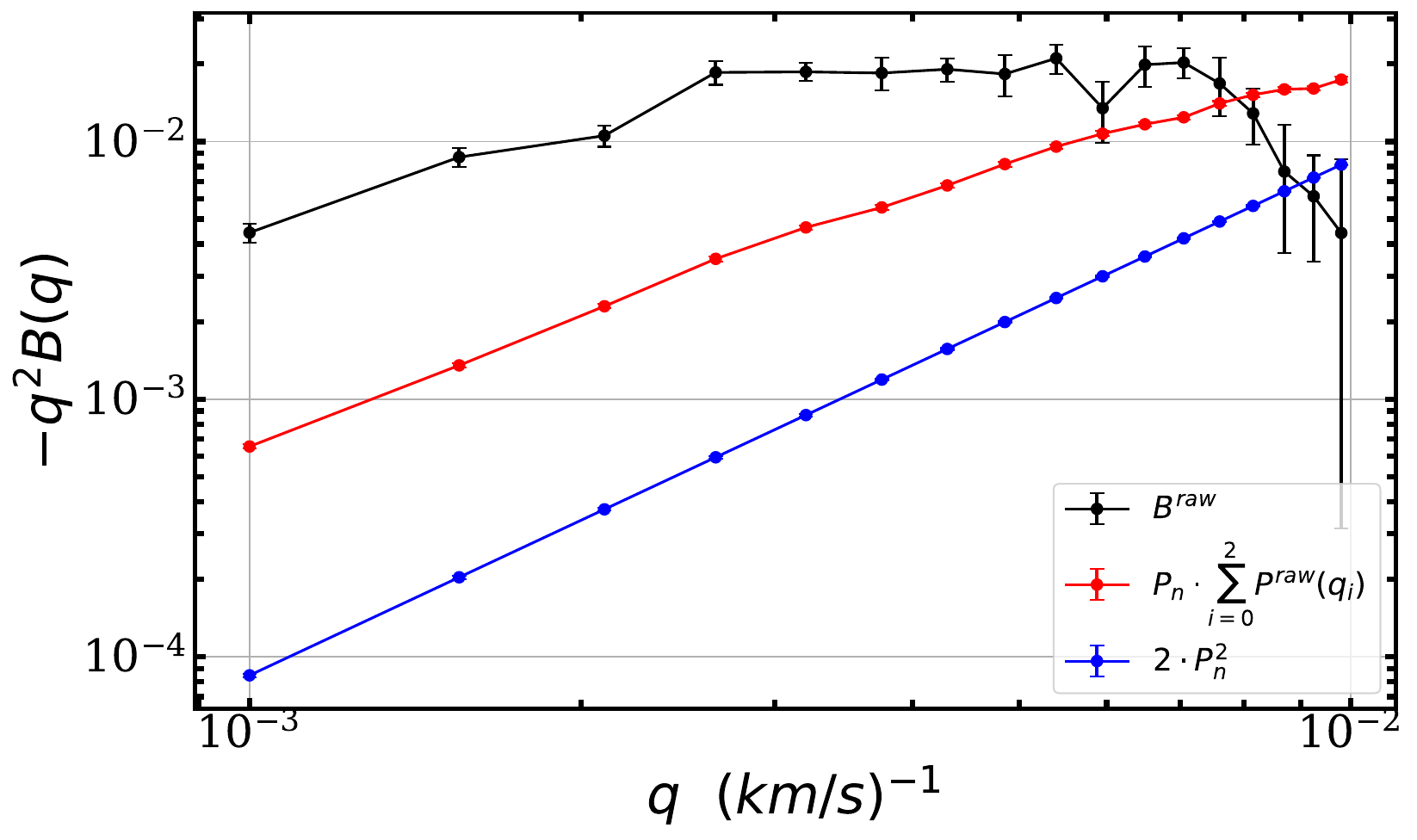}
\caption{Average bispectrum of the raw signal (black dots), the noise power spectrum times the power spectrum for different q-modes (ref dots) and twice the square of the noise power spectrum signal (blue dots). The blue line is a asymptote of the raw bispectrum signal. The red line contribute significantly of the raw bispectrum signal at small scales. The error bars come from the diagonal of the bootstrap covariance matrix.}
\label{fig:plot04}
\end{figure}

One might wonder which term in Eq.~(\ref{eq08}) dominates the signal (excluding $B^{\mathrm{metals}}$). Fig. \ref{fig:plot04} illustrates the contributions from the terms in Eq.~(\ref{eq08}). The largest contribution to $\bk$ comes from the term $B^{\mathrm{raw}}$, while the second contribution comes from the term $P^{n}\cdot \sum_{i=0}^{2}P_{\mathrm{raw}}(q_{i})$, as the value of $P^{n}$ is high, amplifying the sum of the $P^{\mathrm{raw}}(q_{i})$ signal. This second term enhances the $B^{\mathrm{raw}}$ signal, while the third term represents an asymptote to the combined signal of the first two terms. In the case of mocks, the $P^{n}$ is even smaller than in eBOSS, so the $\bk$ is dominated entirely by the $B^{\mathrm{raw}}$ signal. We plot the average of $B^{\mathrm{raw}}$, $P^{n}\cdot \sum_{i=0}^{2}P^{\mathrm{raw}}(q_{i})$, and $2P_{n}^{2}$, computed over the \lya forest. As expected, the noise is white, and the term $2P_{n}^{2}$ is scale-independent. The term $P^{n}\cdot \sum_{i=0}^{2}P^{\mathrm{raw}}(q_{i})$ is important, as it amplifies the $\bk$ signal, while in the mocks, it is irrelevant given the low noise level. For comparison, Fig. \ref{fig:plot04} also shows the raw bispectrum. We select $q_{\mathrm{max}} = 0.01~(km/s)^{-1}$ because at higher modes, the effect of the spectrograph resolution becomes significant, suppressing almost $80\%$ of the signal, although this percentage depends on redshift, with a larger effect at lower redshifts.

\subsection{Side-band bispectrum} \label{subsec034}
\begin{figure}
\centering
\subfigure{\includegraphics[width=7.5cm,height=6.cm]{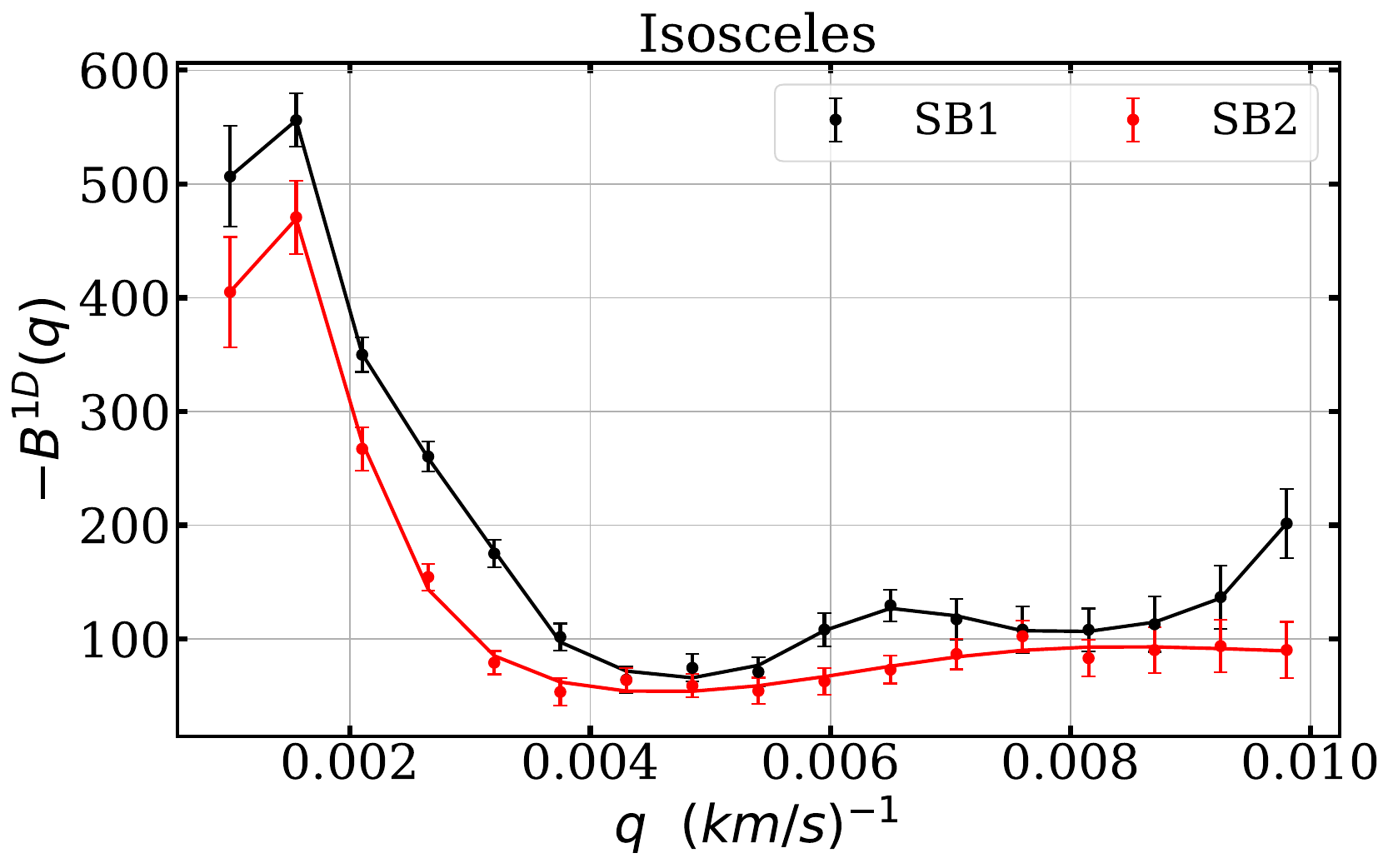}}
\subfigure{\includegraphics[width=7.5cm,height=5.68cm]{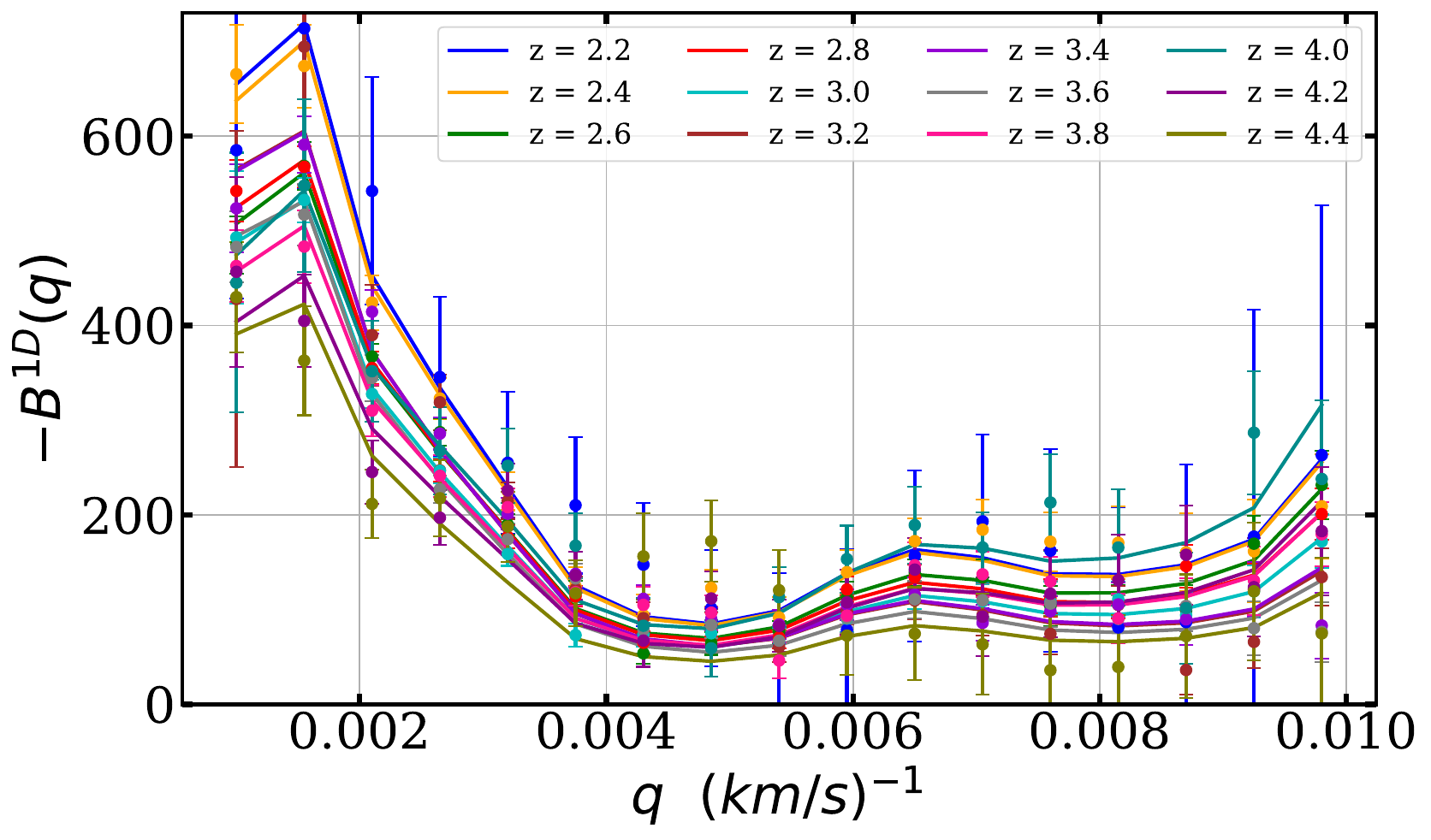}}
\caption{
1D bispectrum measured in the side-bands regions SB1 and SB2 for isosceles triangles. \textit{Top}: Average of the $\bk_{SB}$ over all redshift bins. The solid lines represent the fitted model given by Eq.~(\ref{eq11}). \textit{Bottom}: $\bk_{SB1}$ as a function of redshift. Each redshift bin is fitted using the product of the fitting in the top panel and a first-degree polynomial in which the two parameters are free. The error bars come from the
diagonal of the bootstrap covariance matrix.
}
\label{fig:plot05}
\end{figure}

\begin{figure}
\centering
\subfigure{\includegraphics[width=7.5cm,height=6cm]{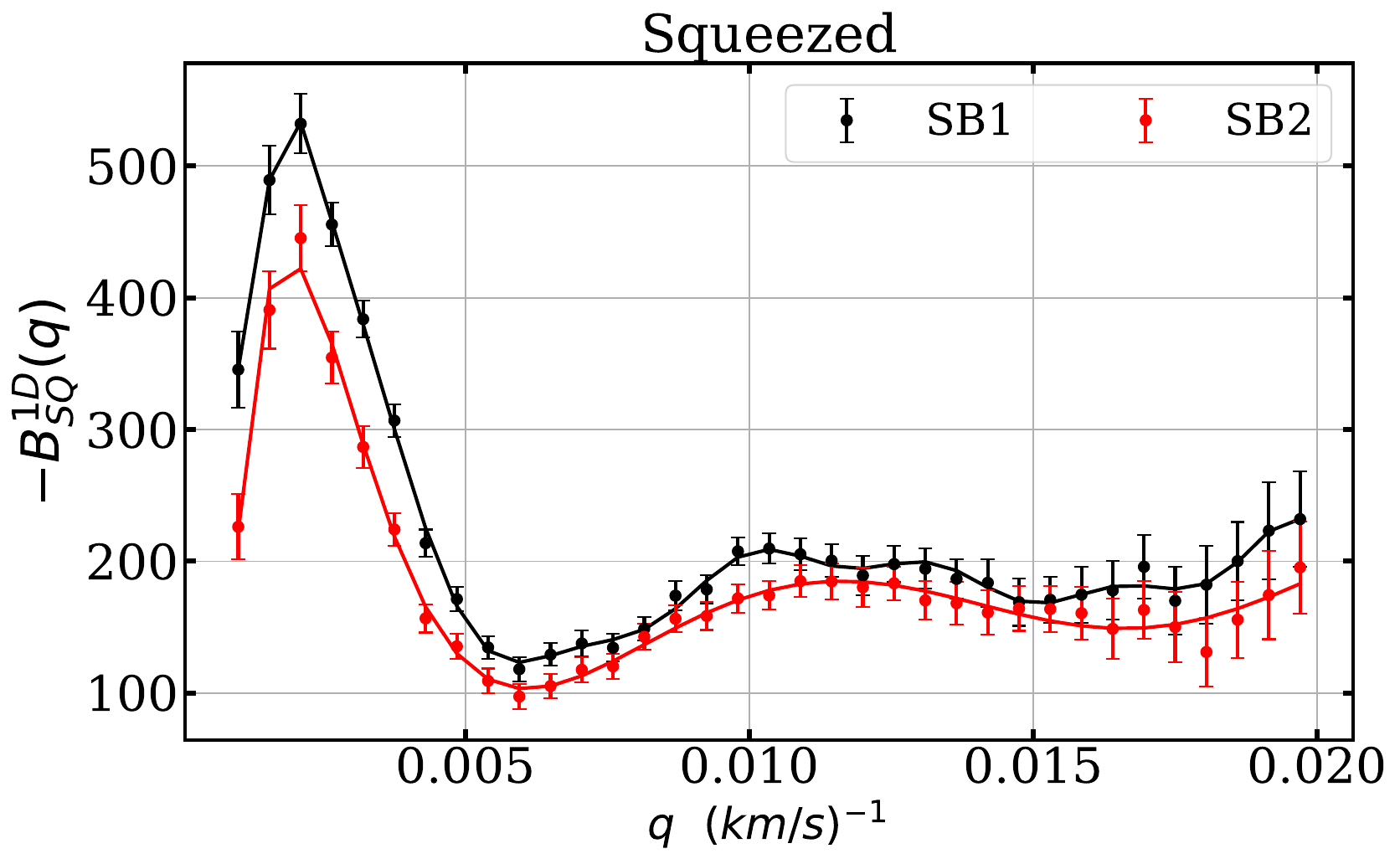}}
\subfigure{\includegraphics[width=7.5cm,height=5.68cm]{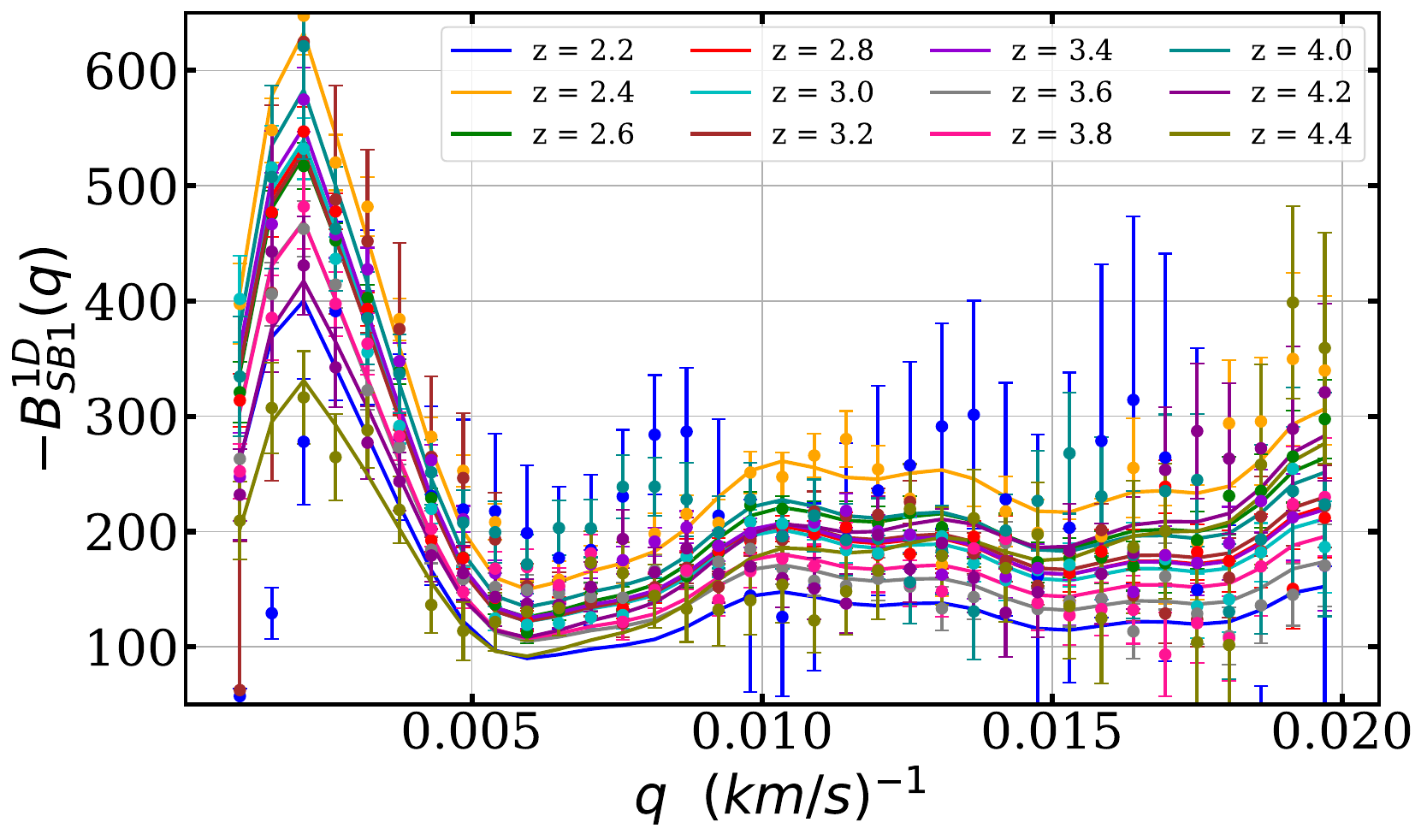}}
\caption{
1D bispectrum measured in the side-bands regions SB1 and SB2 for squeezed limit triangles. \textit{Top}: Average of the $\bk_{SQ~SB}$ over all redshift bins. The solid lines represent the fitted model given by Eq.~(\ref{eq11}). \textit{Bottom}: $\bk_{SQ~SB1}$ as a function of redshift. Each redshift bin is fitted using the product of the fitting in the top panel and a first-degree polynomial in which the two parameters are free. The error bars come from the
diagonal of the bootstrap covariance matrix.
}
\label{fig:plot06}
\end{figure}

In Sec. \ref{subsubsec222}, we discussed the absorption lines redward of the \lya emission line caused by metals. These absorptions produce a background power in the 1D \lya power spectrum. A similar effect occurs in the case of the \lya bispectrum. We assume that the cross-correlation between \lya absorption and metals in the side-bands is negligible. In both side-bands, the delta field $\left(\delta_{F}(\lambda)|_{SB}\right)$ can be expressed as a combination of metal absorption lines and noise fluctuations. Consequently, the calculation of the metal side-bands bispectrum can be written as:
\begin{align}
    \bk_{SB}(q_{0},q_{1}) =  \left\langle \frac{B^{\mathrm{raw}}_{q_{0},q_{1}}|_{SB} - P^{n}(q)|_{SB}\cdot\left[\sum_{i=0}^{2}P^{\mathrm{raw}}(q_{i})|_{SB}\right]+2P_{n}^{2}(q)|_{SB}}{W(q_{0},R,\Delta v)W(q_{1},R,\Delta v)W(q_{0}+q_{1},R,\Delta v)} \right \rangle. \label{eq10}
\end{align}
This uncorrelated background cannot be directly estimated through bispectrum measurements in the \lya wavelength region. Therefore, to address this issue, we perform bispectrum measurements in the side-bands and subtract them from the \lya bispectrum measured over the same gas redshift range. This approach has been extensively examined in \cite{ravoux2023dark}, \cite{Palanque-Delabrouille:2013gaa}, and \cite{Chabanier_2019}. However, in our study, we apply it to $\bk$ estimation instead of $\pk$. This method is purely statistical, utilizing a distinct sample of quasars to compute both the \lya and metal bispectra within a specific redshift range. In the case of SB1, the bispectrum includes contributions from all metals with wavelengths greater than 1380~\AA, with significant contributions from $\mathrm{Si~{\textsc{iv}}}$ and $\mathrm{C~{\textsc{iv}}}$ absorption lines. Conversely, SB2 contains only $\mathrm{C~{\textsc{iv}}}$ absorptions. Therefore, we subtract only the SB1 signal from the \lya bispectrum, while using SB2 primarily for consistency checks. We estimate the metal bispectrum over the same observed wavelength range as the \lya bispectrum, meaning we use a quasar sample with a lower redshift range than the sample used to estimate the \lya bispectrum. For example, in the first redshift bin, $2.1 < z_{Ly\alpha} < 2.3$, we measure the bispectrum in SB1 (corresponding to 3770~\AA~ $<\lambda_{obs}<$ 4012~\AA) using quasars with $z_{qso}\sim 2.0$, while for the \lya bispectrum we use quasars with $z_{qso}\sim 2.2$. The measurement for isosceles triangles is shown in Fig. \ref{fig:plot05}. A total of 256,751 chunks were analyzed, undergoing the same cuts as applied to the quasar spectra in the \lya forest analysis. The top panel shows the $\bk_{SB1}$ stacking across all redshift bins (black dots with error bars), while the red dots represent the bispectrum signal in SB2, from a sample of 237,985 chunks. As expected, the amplitude of the $\bk_{SB2}$ signal is smaller than that of $\bk_{SB1}$, since SB2 excludes the $\mathrm{Si~{\textsc{iv}}}$ absorptions. The bottom panel shows $\bk_{SB1}$ for different redshift bins.
Fig. \ref{fig:plot06} illustrates the results for the squeezed limit bispectrum. Similar to the isosceles triangles, the signal amplitude of $\bksq|_{SB2}$ is smaller than that of $\bk_{SB1}$. This configuration clearly displays the oscillations caused by the $\mathrm{C~{\textsc{iv}}}$ doublet absorptions on the $\mathrm{Si~{\textsc{iv}}}$ forest. Once again, the black dots with error bars in the upper panel represent the stack of all redshift bins of $\bk_{SB1}$, while the red dots show the $\bksq|_{SB2}$ signal. The bottom panel illustrates $\bk_{SB1}$ for different redshift bins.

The solid lines represent the fit to the bispectrum shape of the side-bands. The fitting procedure we employed is inspired by \cite{ravoux2023dark} and detailed in \cite{10.1093/mnras/stad1363}. However, it was originally intended for measuring the side-bands power spectrum, and we have adapted this approach to analyze the side-bands bispectrum. Similar to the previous work, we account for the absorptions of the $\mathrm{Si~{\textsc{iv}}}$ and $\mathrm{C~{\textsc{iv}}}$ doublets. According to NIST data (\cite{kramida_nist_2009}), the $\mathrm{Si~{\textsc{iv}}}$ doublets are located at $\lambda^{\mathrm{Si~{\textsc{iv}}_{a}}}_{\mathrm{R.F.}} = 1,393.76$~\AA\ and $\lambda^{\mathrm{Si~{\textsc{iv}}_{b}}}_{\mathrm{R.F.}} = 1,402.77$~\AA, while the $\mathrm{C~{\textsc{iv}}}$ doublets are at $\lambda^{\mathrm{C~{\textsc{iv}}_{a}}}_{\mathrm{R.F.}} = 1,548.202$~\AA\ and $\lambda^{\mathrm{C~{\textsc{iv}}_{b}}}_{\mathrm{R.F.}} = 1,550.774$~\AA. The $\mathrm{C~{\textsc{iv}}}$ doublet has a separation of $\mathrm{r}=\Delta v_{\mathrm{C~{\textsc{iv}}_{a,b}}}=499~km/s$, while the $\mathrm{Si~{\textsc{iv}}}$ doublet has a separation of $\mathrm{s}=\Delta v_{\mathrm{Si~{\textsc{iv}}_{a,b}}}=1933~km/s$.
The presence of an absorption doublet in the side-bands induces oscillations in the bispectrum signal. The periodicity of these oscillations depends on the separation of the doublets. When two absorption doublets are present in the same side-band, the periodicity of the oscillations also depends on both the sum and difference of the separations of the doublets. In the top panels of Figs. \ref{fig:plot05} and \ref{fig:plot06}, both side-bands show significant oscillations caused by the $\mathrm{C~{\textsc{iv}}}$ doublet, along with large-scale oscillations due to the difference in the doublet spacing. As expected, the SB1 signal shows additional oscillations due to the $\mathrm{Si~{\textsc{iv}}}$ absorptions. The model for $\bk_{SQ}$ consists of a sum of a power law and oscillating functions:
\begin{equation}
    \bk_{\mathrm{SB1, model}} = A\cdot \left(\frac{q}{q_{2}}\right)^{-\epsilon} + \sum_{i}B_{\mathrm{doublet,i}}(q,q_{i},A_{i},a_{i},\psi_{i}).
    \label{eq11}
\end{equation}

In the isosceles configuration, where there is a $2\mathrm{q}$ dependence, the oscillations induced by a doublet have a frequency characterized by the wavenumber $2q = 2\pi / \Delta v_{i}$. For example, in the top panel of Fig. \ref{fig:plot05}, the first bump of $\bk_{SB1}$ corresponds to $2q = 2\pi / \mathrm{s} = 0.0016~(km/s)^{-1}$, and the second bump to $2q = 2\pi / \mathrm{r} = 0.0062~(km/s)^{-1}$. In the case of the squeezed limit configuration, while there is also a $2\mathrm{q}$ dependence, it does not affect the signal because $q$ is a minimum value, behaving like a constant. In this case, the frequency of the oscillations is given by $q = 2\pi / \Delta v_{i}$. In this configuration, the oscillation due to the difference in the velocities of the doublets becomes more evident, as seen in the first bump of the $\bksq|_{SB2}$ signal at $2q = \frac{2\pi}{\mathrm{s-r}} = 0.0021~(km/s)^{-1}$. We model the doublet oscillations using damped sinusoidal functions as follows:
\begin{equation}
    B_{\mathrm{doublet,i}}(q,q_{i},A_{i},a_{i},\psi_{i}) = \sum_{i}A_{i}e^{-a_{i}\left(\frac{q}{q_{i}}\right)}\sin{\left(2\pi\left(\frac{q}{q_{i}}\right)+\psi_{i}\right)}.
    \label{eq12}
\end{equation}
In the case of isosceles triangles, $q_{i} = \left(\frac{\pi}{s}, \frac{\pi}{r}, \frac{\pi}{s+r}, \frac{\pi}{s-r}\right)$. For squeezed limit triangles, $q_{i} = \left(\frac{2\pi}{s}, \frac{2\pi}{r}, \frac{\pi}{s+r}, \frac{\pi}{s-r}\right)$. The terms $(\mathrm{s+r})$ and $(\mathrm{s-r})$ are derived from a simple model that approximates the transmitted flux contrast ($\delta_{F}$) in the presence of two doublets with velocity separations $s$ and $r$, i.e., $\delta_{F} \sim \delta(v) + a\delta(v+\mathrm{s})+b\delta(v+\mathrm{r})$, where the factors $a$ and $b$ indicate the relative strengths of the absorptions of the doublets compared to the background absorption. When calculating the bispectrum of this transmitted flux contrast, oscillating functions arise that vary as $(\mathrm{s+r})$ and $(\mathrm{s-r})$.

The SB1 fitted function is then used to derive the redshift dependence of the side-band bispectrum, shown in the bottom panels of Figs. \ref{fig:plot05} and \ref{fig:plot06}. For each redshift bin, we fit the product of the global SB1 fitted function and a first-order polynomial. This fit should not be used for rigorous scientific conclusions, as the $\chi_{\nu}^{2}$ values indicate that the model based on the SB1 fitting function does not fully represent the observed data. However, it is sufficient for providing a baseline estimate. We performed the fit independently for each configuration. The fitted parameters are generally consistent between the two configurations, except for the parameters $A$ and $A_{i}$, which primarily affect the amplitude of the doublet oscillations.

\subsection{1D power spectrum assessment} \label{subsec035}
As shown in Eq.~(\ref{eq08}), calculating the bispectrum requires estimating the power spectrum, which includes both the noise power spectrum and the side-band power spectrum (SB1 only), as indicated in Eq.~(\ref{eq02}). The power spectrum allows us to assess the quality of our data analysis and the efficiency of our pipeline for calculating the bispectrum ($\bk$). This is done by comparing our $\pk$ estimation with previous measurements. While numerous $\pk$ measurements have been made in the \lya forest, some of which were mentioned in Sec. \ref{Intro}, we opted to compare our results with previous measurements from the eBOSS DR14 data (\cite{Chabanier_2019}) and the most recent measurements from the DESI early data release (\cite{ravoux2023dark}), which draws from large quasar surveys. The FFT estimator for the 1D power spectrum is calculated as an average over all available \lya forests in the measurement sample. From Eq.~(\ref{eq02}), the $\pk$ estimator is defined as:
\begin{equation}
    P^{1D}(k) = \Big \langle \frac{P^{\mathrm{raw}}(k) - P^{n}(k)}{W^{2}(k,R,\Delta \mathrm{v})} \Big \rangle - P^{\mathrm{metals}}(k). \label{eq13}
\end{equation}

\begin{figure*}
\centering
\subfigure{\includegraphics[width=7.5cm,height=9cm]{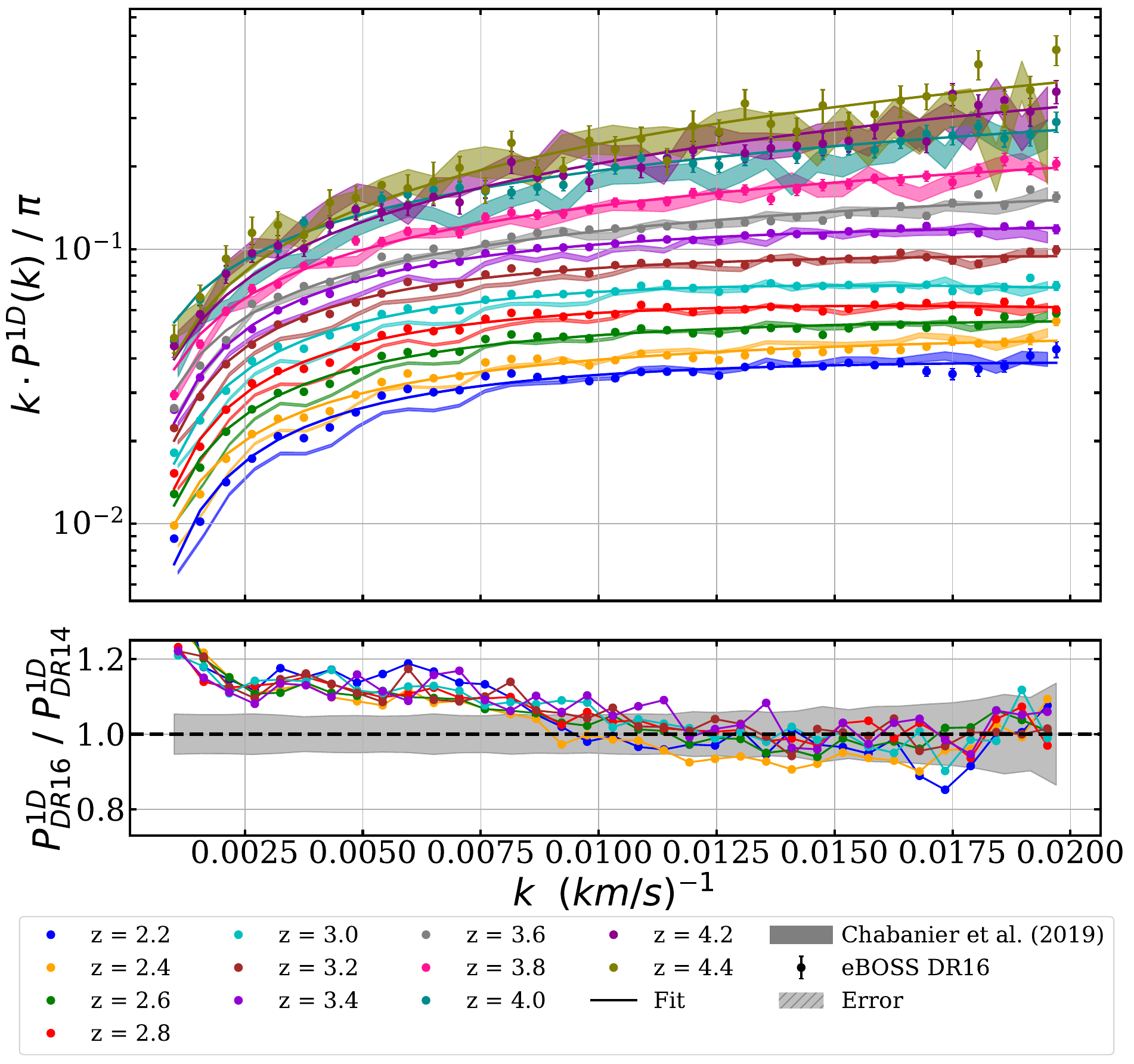}}
\subfigure{\includegraphics[width=7.5cm,height=9cm]{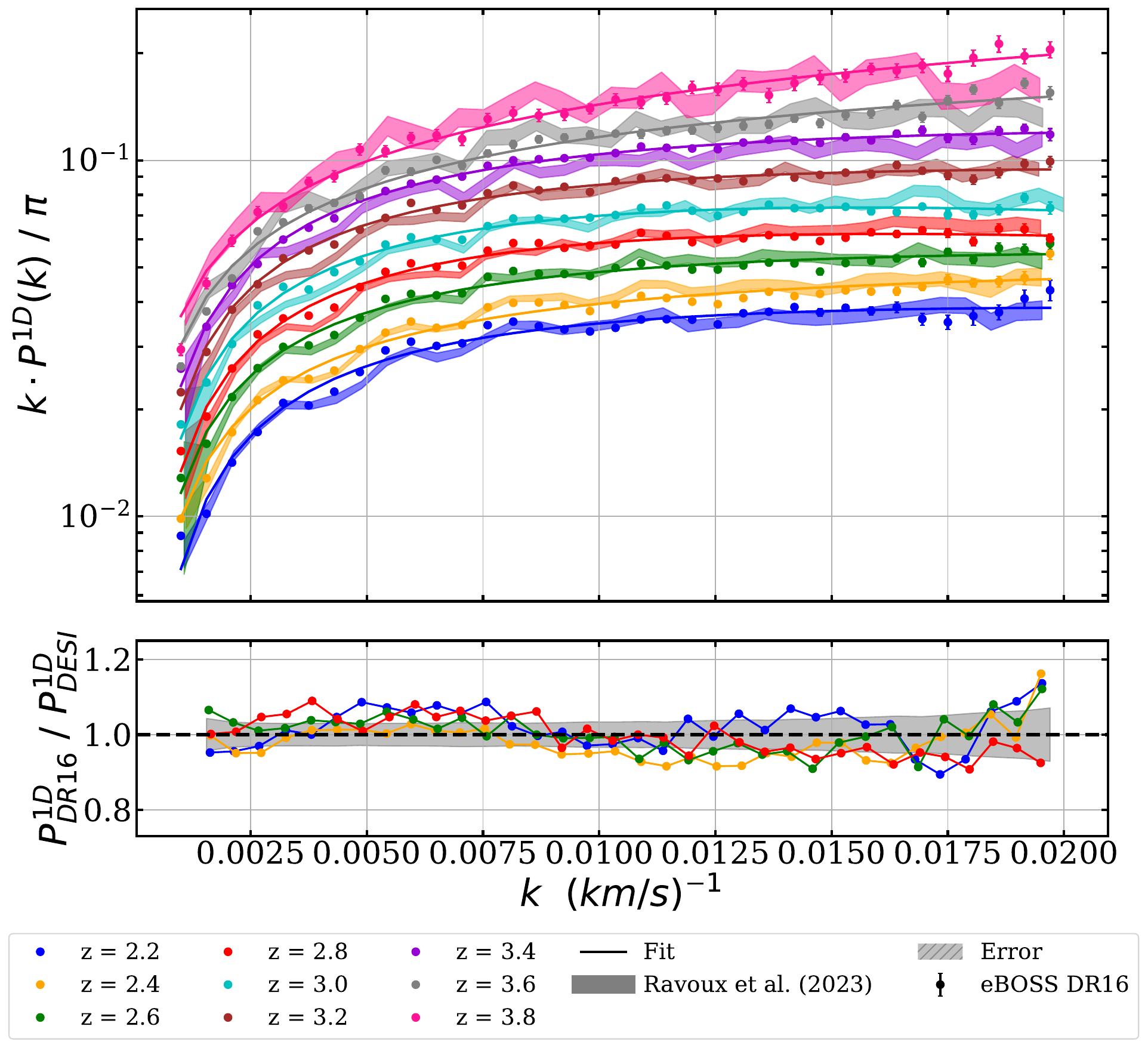}}
1D ﬂux power spectrum in eBOSS DR16 data for twelve redshift bins.\caption{1D flux power spectrum in eBOSS DR16 data for twelve redshift bins. \textit{Left:} Comparison between the measurement performed in this work and that of the eBOSS DR14 data \citep{Chabanier_2019}. Our measurement with points and error bars, the DR14 measurement with shaded colored areas. The ratio between DR16 and DR14 measurements is showed in the bottom panel. The striped grey area in the bottom panel represent the centered error bar (added in quadrature) of the ratio averaged over all shown redshift bins. We found the same discrepancies reported in \citep{ravoux2023dark} at large scales with respect to DR14 data, but we are unable to determinate the source of this difference. \textit{Right:} Same comparison, but this time between the measurement performed in this work and that of the DESI data \citep{ravoux2023dark}. Both signals show a good agreement for all redshift bins considered. Solid lines in both panels represent the fitted model given by Eq.~(\ref{eq14}). The error bars present in our measurements come from the diagonal of the bootstrap covariance matrix.}
\label{fig:plot07}
\end{figure*}

Fig. \ref{fig:plot07} (left) shows the comparison between our $\pk$ estimate (represented by dots with error bars) and the eBOSS DR14 data (shaded lines) for different redshift bins. The percentage difference between them is approximately 15\% at large scales ($k < 0.01~km/s$). A similar discrepancy was reported in \cite{ravoux2023dark}, though their comparison was between DESI and eBOSS DR14 data. Like them, we are unable to determine the source of this discrepancy. In our case, the continuum fitting provided by PCA also affects the $\pk$ signal at large scales, but its effect is insufficient to account for the magnitude of this difference. However, at small scales, both measurements are consistent within the error bars, suggesting that we have accurately estimated the noise power spectrum and corrected for the spectrograph resolution present in the window function, as these are the main factors influencing small scales.

We also compare our measurement with the DESI early data release in Fig. \ref{fig:plot07} (right). Overall, both measurements are in good agreement across the $k$ range when considering the error bars, with a percentage difference of around 5\%. The largest difference is observed around $k \sim 0.013~km/s$, likely due to the oscillation of $\pk_{SB1}$ in that region compared to the eBOSS data.

We made a fit to the $\pk$ using the fiducial power spectrum function proposed in \cite{10.1093/mnras/staa2331} given by:
\begin{equation}
    \frac{kP(k,z)}{\pi} = A \frac{(k/k_{0})^{3+n+\alpha\ln{k/k_{0}}}}{1 + (k/k_{1})^{2}}\left(\frac{1 + z}{1 + z_{0}}\right)^{B+\beta\ln{k/k_{0}}},\label{eq14}
\end{equation}
where $k_{0} = 0.009 (km/s)^{-1}$ and $z_{0} = 3.0$. The continuous thin colored lines in both plots of Fig. \ref{fig:plot07} show the fitting. 

\section{Synthetic data correction}\label{Sec04}
In this section, we examine the biases introduced at each step of the data analysis and assess their impact using simulated spectra. Our goal is to characterize their effect on the $\bk$, considering continuum fitting, masking of pixels affected by sky lines, and absorption from DLAs as well as noise and spectrograph resolution.

\subsection{Mocks}\label{subsec041}
CoLoRe (Cosmological Lofty Realisation) \cite{Ram_rez_P_rez_2022} is a package designed to produce various cosmological tracers of the same underlying matter fluctuations, such as spectroscopic galaxies, weak lensing, and CMB lensing. Additionally, it can generate \lya absorption in the spectra of high-redshift quasars, which is of particular interest for this work. Although CoLoRe can produce skewers of transmitted flux fraction, its "raw" output requires significant post-processing before it can be considered a realistic representation of the \lya forest. For this purpose, we use LyaCoLoRe \href{https://github.com/igmhub/LyaCoLoRe}{\faGithub}\footnote{Public software available at \url{https://github.com/igmhub/LyaCoLoRe}} \cite{Farr_2020}, which is capable of generating realistic skewers of transmitted flux fraction. CoLoRe generates the density field from an initial power spectrum using two approaches. The first is based on an initial Gaussian field, where a lognormal approximation is employed to model the physical density, which can help understand the linear evolution of the field. The second approach uses a formalism called Lagrangian Perturbation Theory (LPT) \cite{Bernardeau_2002}, which extends to mildly non-linear scales. In CoLoRe, both first- and second-order LPT (2LPT) methods are available. We estimated the bispectrum using both approaches, and our results indicate that the signals agree within the error bars, with deviations of less than 10\% at large scales in both configurations. This suggests that the noiseless bispectrum is dominated by the non-linear mapping from the mocks to the skewers' flux. Therefore, we have chosen the simplest and least computationally expensive method, the lognormal approach, to present the results.

The output skewers from LyaCoLoRe require the addition of instrumental noise and combination with a QSO continuum before they can be considered realistic spectra. This can be carried out using the \textit{desisim} package \href{https://github.com/desihub/desisim}{\faGithub}\footnote{Public software available at \url{https://github.com/desihub/desisim}}. \textit{Desisim} produces QSO continuum using the \textit{SIMQSO} package \href{https://github.com/desihub/simqso}{\faGithub}\footnote{Public software available at \url{https://github.com/desihub/simqso}}, where the continuum is generated by applying a broken power law and adding emission lines, modeled by Gaussian distributions. Absorption lines in the \lya quasar spectra are added by \textit{quickquasars} \href{https://github.com/desihub/desisim/blob/main/py/desisim/scripts/quickquasars.py}{\faGithub}\footnote{Public software available at \url{https://github.com/desihub/desisim/blob/main/py/desisim/scripts/quickquasars.py}} (a DESI code within \textit{desisim}), by multiplying the continuum templates with the transmitted flux of the raw mocks. Additionally, \textit{quickquasars} can introduce noise to the spectra, metal absorption lines, DLAs, BALs, and tune the pixel width of the wavelength in the spectra. For more details, see \cite{Herrera-Alcantar:2024its}.

To understand the impact of systematic errors, such as sky lines, DLA contamination, and the continuum fitting procedure, we generated mock noiseless spectra that replicate the numerical density of \lya forest chunks per redshift bin. DLA contamination is introduced using a random catalog. We created five independent realizations with different initial conditions. For each realization, we generated two sets of spectra: one including DLA information and one without it.

\subsection{Continuum-fitting correction}\label{subsec042}
As mentioned in \ref{SubSec1-2}, we decided to estimate the continuum-fitting using PCA. Since we only perform PCA on the blue-side of the \lya forest, what we are actually estimating is $C(\lambda)\overline{F}(\lambda)$. PCA produces a series of eigenvectors ordered by the fraction of the sample variance they explain. In principle, each eigenvector is a linear combinations of the input fluxes. The quasar continuum is produced through the following equation:
\begin{equation}
    C\overline{F}_{i,m}(\lambda_{\mathrm{R.F.}}) = \overline{f}(\lambda_{\mathrm{R.F.}}) + \sum_{j=1}^{m}\alpha_{ij}\xi_{j}(\lambda_{\mathrm{R.F.}}). \label{eq15}
\end{equation}

where $\overline{f}(\lambda_{\mathrm{R.F.}})$ is the mean observed flux, and $\xi_i(\lambda_{\mathrm{R.F.}})$ represent the eigenvectors. We construct the continuum using three principal components\footnote{We initially decompose into 10 eigenvectors, but the continuum is construct using only 3, as this performs better than starting with only 3 eigenvectors.}. The continuum-fitting procedure systematically distorts the measured $C(\lambda)\overline{F}(\lambda)$ by suppressing large-scale modes in the $\bk$ measurement. We observed that suppression of the $\bk$ signal at small wavenumbers increases as the number of eigenvectors is increased. We tested up to 5 eigenvectors and found that the signal suppression exceeded 25\%. Thus, we decided to use only 3 eigenvectors. This choice reduces the impact of the continuum fitting and ensures that we can adequately capture the diversity of the quasar sample, as explained in detail in \cite{Rodrigo:2023inprep}. 

We define the bias induced by the continuum-fitting as the ratio of the bispectrum computed with the true continuum (TRUECONT) to the bispectrum computed using our standard PCA-based continuum-fitting procedure:
\begin{equation}
    b_{\mathrm{cont}}(q,z) = \frac{\bk_{\mathrm{TRUECONT}}(q,z)}{\bk_{\mathrm{PCA}}(q,z)}. \label{eq16}
\end{equation}

\begin{figure}
\centering
\subfigure{\includegraphics[width=7.5cm,height=5.5cm]{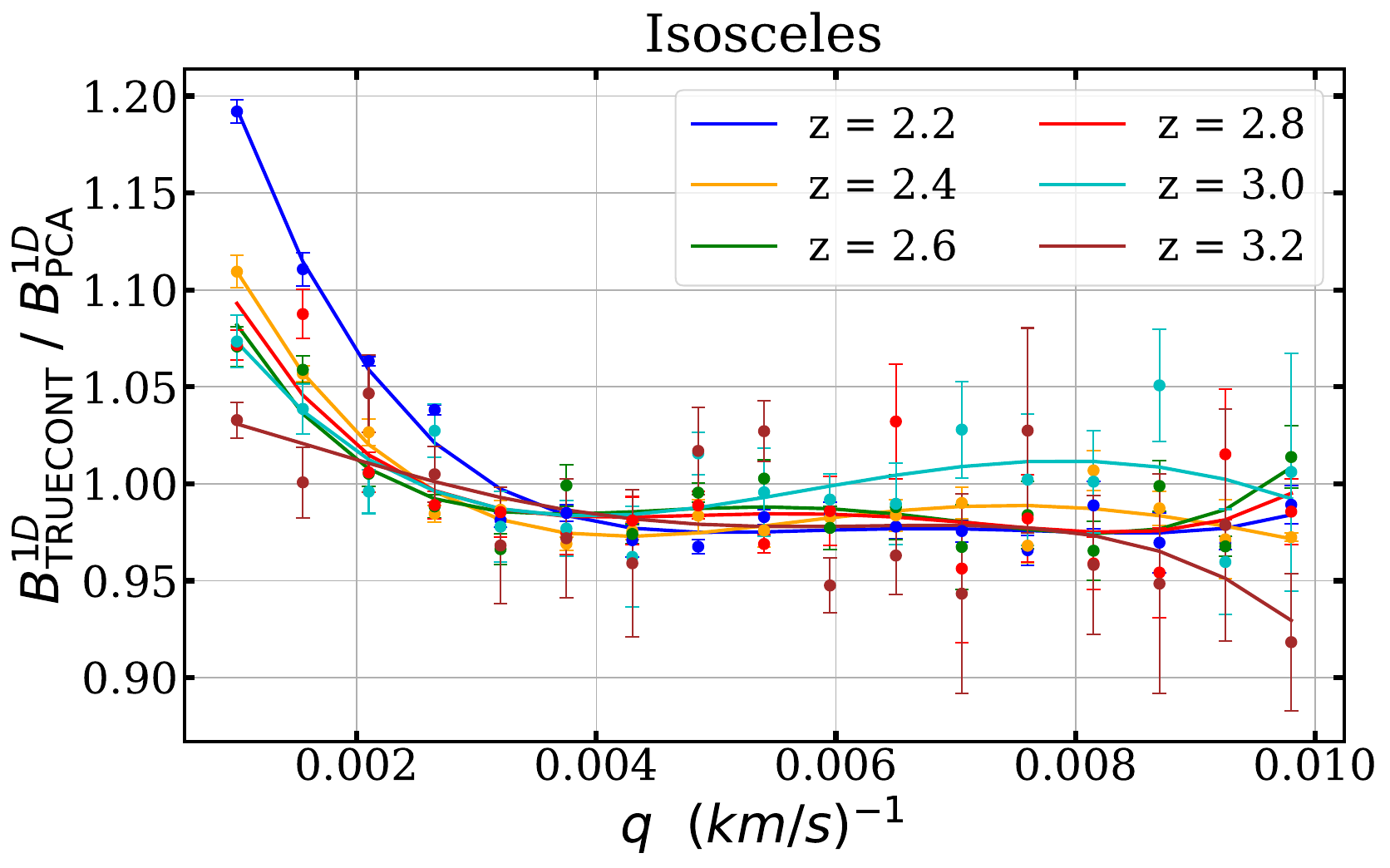}}
\subfigure{\includegraphics[width=7.5cm,height=5.5cm]{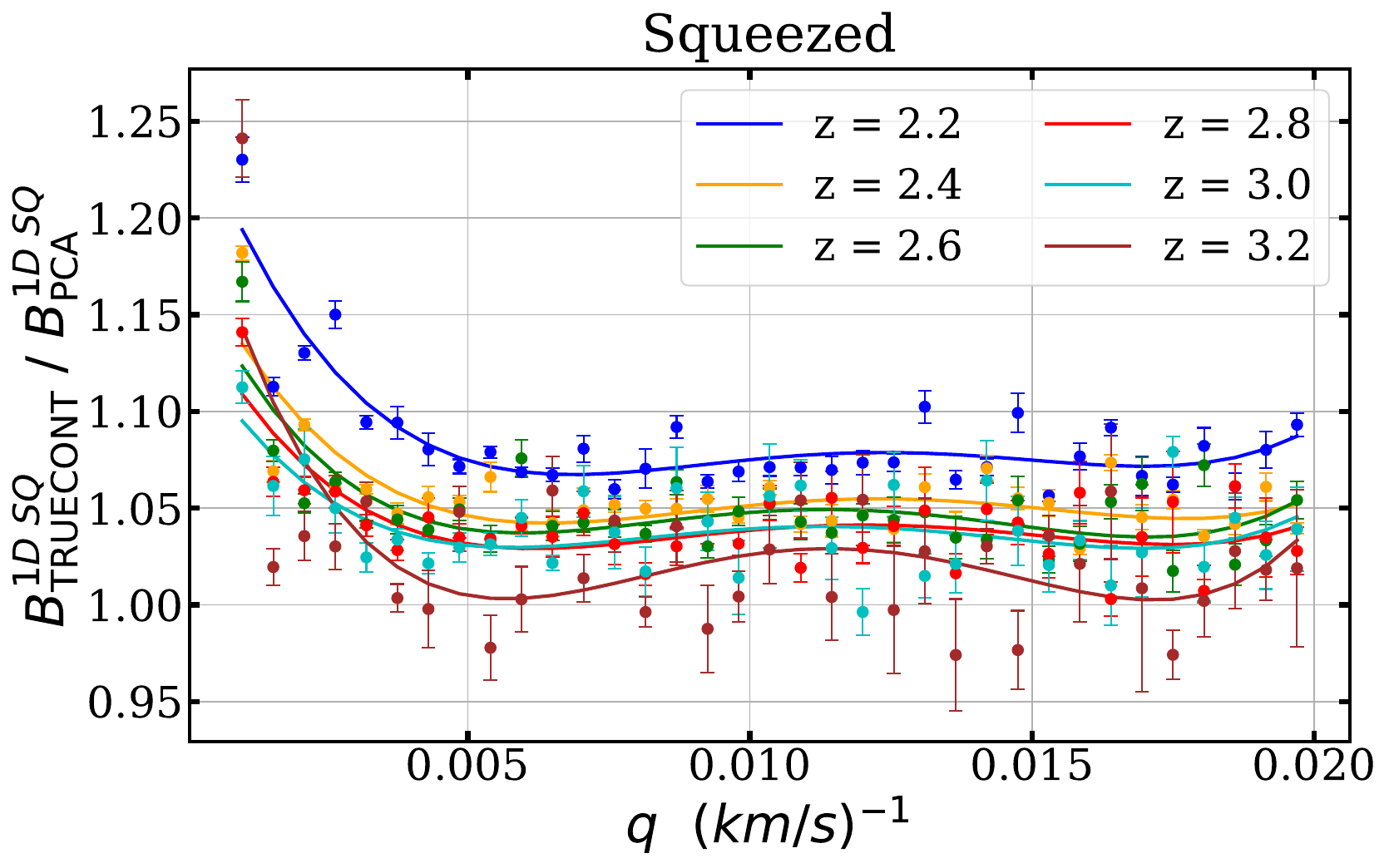}}
\caption{Ratios between the bispectrum obtained using true continuum (TRUECONT) and the one derived with our PCA continuum fitting (PCA) on the combination of 5 mocks, for both triangular configurations considered. Fitting functions are represented by continuous lines and used to correct the bispectrum measurement. The correction is fitted with a four-degree polynomial. The error bars come from to the diagonal of the bootstrap covariance matrix between the five realizations. 
}
\label{fig:plot08}
\end{figure}

Fig. \ref{fig:plot08} illustrates the measured bias. The top panel shows the isosceles configurations, where suppression of the $\bk$ signal is evident at small wavenumbers, with slightly excessive power across the remaining wavenumbers. The bottom panel represents the squeezed-limit configurations, where signal suppression is observed across all wavenumbers, with additional suppression particularly noticeable at small wavenumbers. The correction is fitted with a fourth-order polynomial dependence for $2.2 < z < 3.2$.
Overall, the impact of the continuum-fitting become more significant in the estimation of the bispectrum compared to its effect on the power spectrum, where the error is approximately 6\%.

\subsection{Spectrum pixel masking correction}\label{subsec043}
In our pipeline for estimating the $\bk$ signal, we masked pixels affected by sky lines or DLAs present in the \lya forest. We set the affected pixels to a flux value of zero and maintained this value throughout the transmission delta field. This pixel masking introduces a $q$-dependent bias, which must be quantified.

We begin by estimating the correction due to the masking of the sky lines. We compare the $\bk$ measured from mocks with sky-line masking (SKYm) to mocks without masking (PCA). The bias parameter used for the masking correction is defined as the ratio between the unmasked and masked bispectrum:
\begin{equation}
    b_{\mathrm{SKYm}}(q,z) = \frac{\bk_{\mathrm{PCA}}(q,z)}{\bk_{\mathrm{SKYm}}(q,z)}. \label{eq17}
\end{equation}

For the two chosen triangle configurations, the masking effect is most pronounced at large scales and decreases as it approaches smaller scales. Only the redshift bins at $z=2.2$, $2.4$, and $2.6$ are affected by sky lines, while the redshift bins at $z=2.8$, $3.0$, and $3.2$ remain unbiased. These unbiased redshifts are consistent with the findings of \cite{Chabanier_2019}, although their analysis focused on the $\pk$. To model this bias, we employ a fourth-order polynomial, which is used for the final $\bk$ correction.
\begin{figure}
\centering
\subfigure{\includegraphics[width=7.5cm,height=5.5cm]{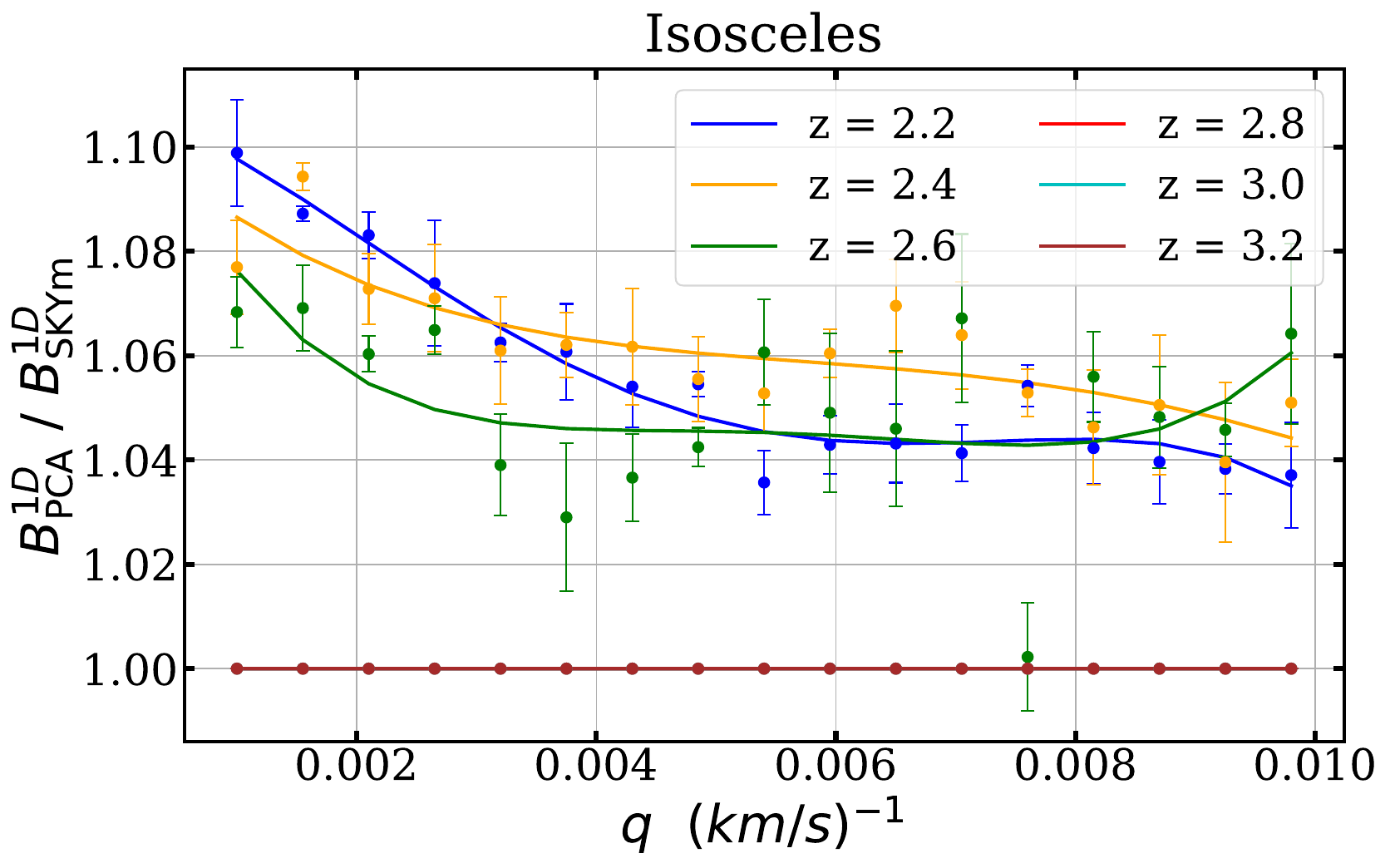}}
\subfigure{\includegraphics[width=7.5cm,height=5.5cm]{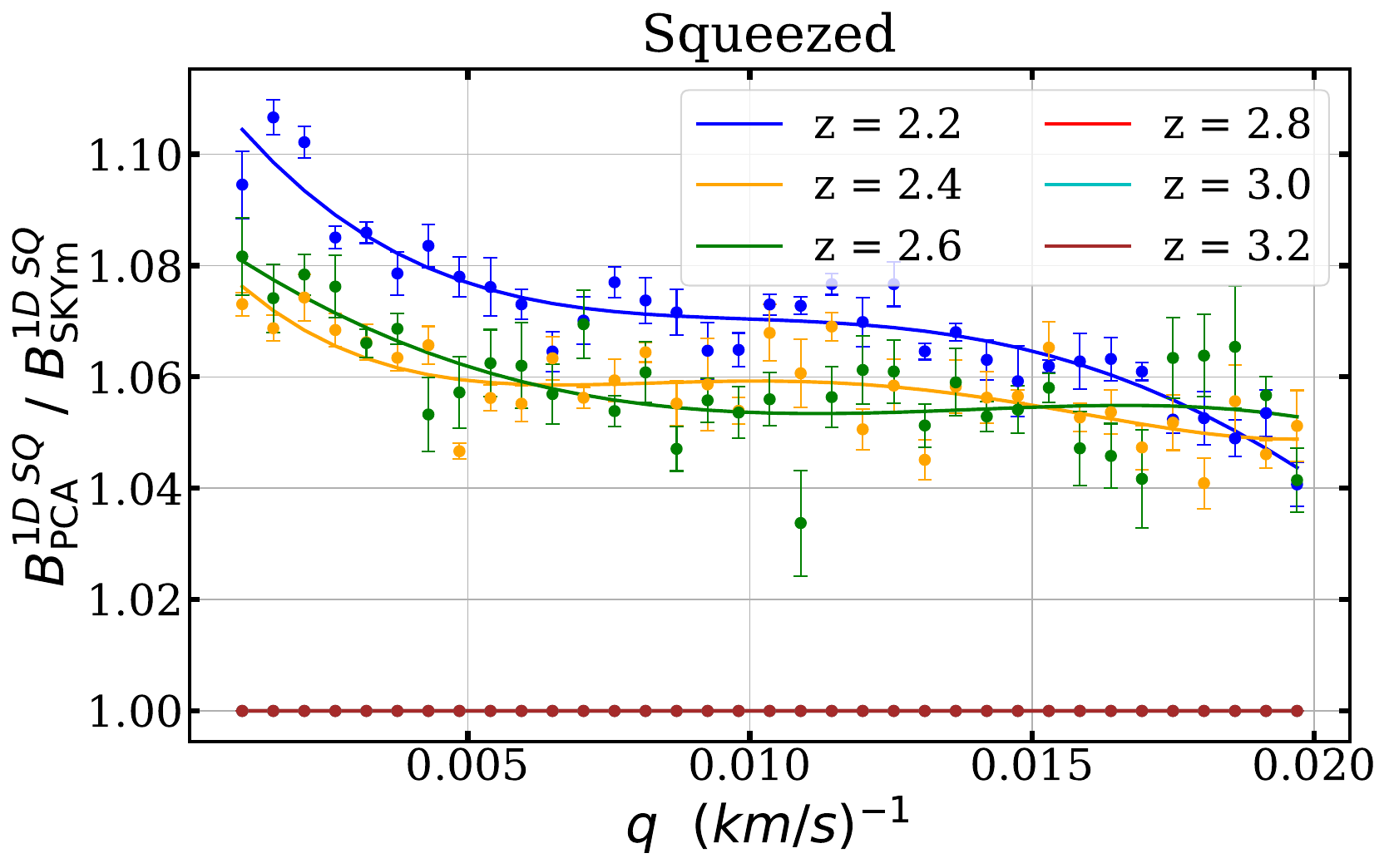}}
\caption{Ratio between the bispectrum obtained using unmasked sky lines pixels (PCA) and the one with masked pixels (SKYm) on the combination of the 5 mocks, for both triangular configurations considered. Four-order polynomial function are employed to fit the corrections in each redshift bin. The error bars come from to the diagonal of the bootstrap covariance matrix between the five realizations.}
\label{fig:plot09}
\end{figure}
Fig. \ref{fig:plot09} shows the $b_{\mathrm{SKYm}}(k,z)$ effect for both configurations. The top panel corresponds to isosceles triangles, while the bottom panel shows the squeezed-limit triangle configuration.

DLAs are added at random locations in the \lya forest during the creation of the mocks. For these initial results on the $\bk$, we are not focused on characterizing the completeness of the DLA finder applied to the data. Instead, we use a `truth' DLA catalog for masking, making it sufficient to use a random distribution to understand this effect. We mask the `truth' catalog using the same parameters as those applied to the observed DLA data catalog. We then compare the $\bk$ measured from mocks with DLA masking (DLAm) to mocks without masking (PCA). The bias parameter used for the masking correction is defined as the ratio between the unmasked and masked bispectrum:
\begin{equation}
    b_{\mathrm{DLAm}}(q,z) = \frac{\bk_{\mathrm{PCA}}(q,z)}{\bk_{\mathrm{DLAm}}(q,z)}. \label{eq18}
\end{equation}
\begin{figure}
\centering
\subfigure{\includegraphics[width=7.5cm,height=5.5cm]{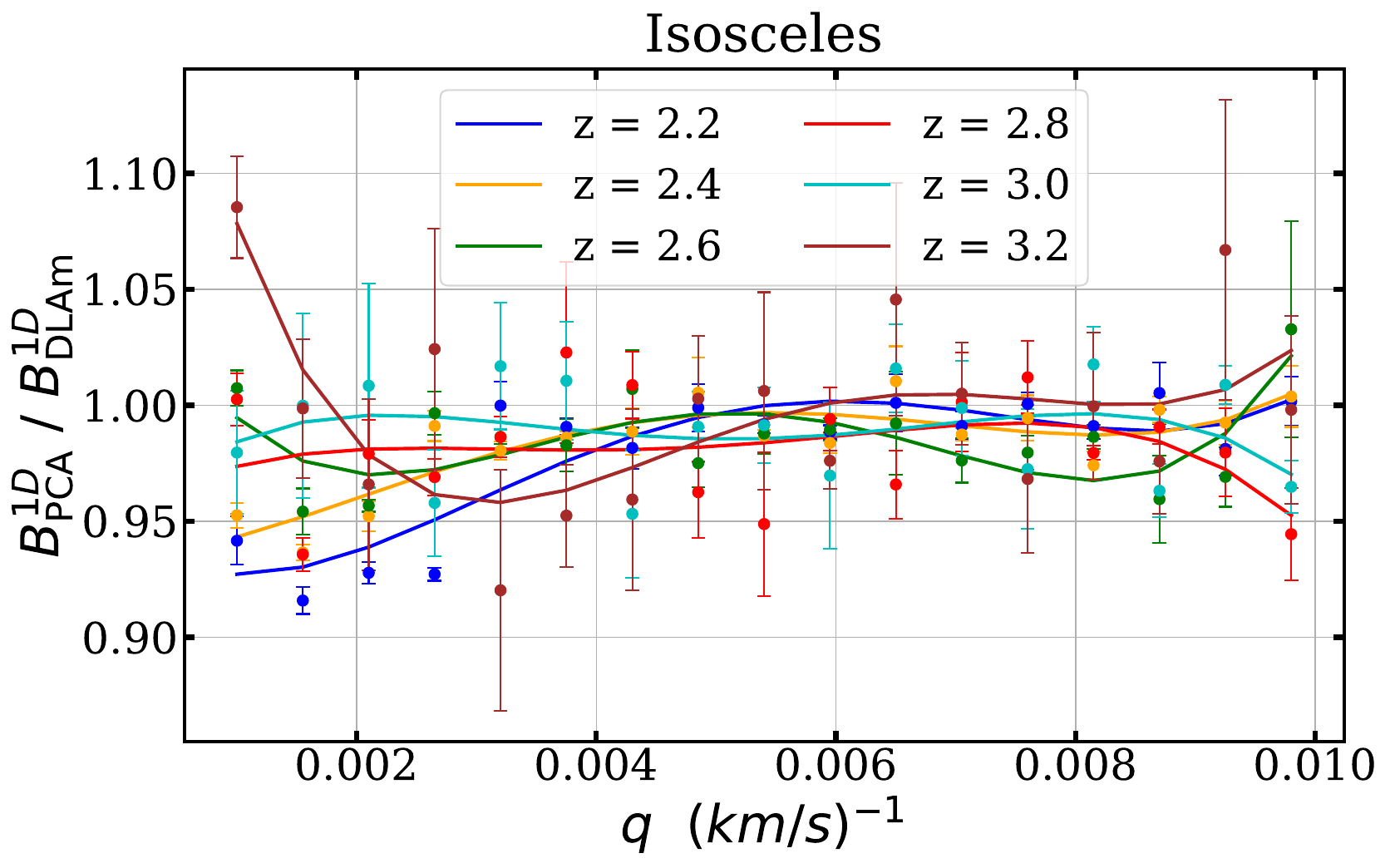}}
\subfigure{\includegraphics[width=7.5cm,height=5.5cm]{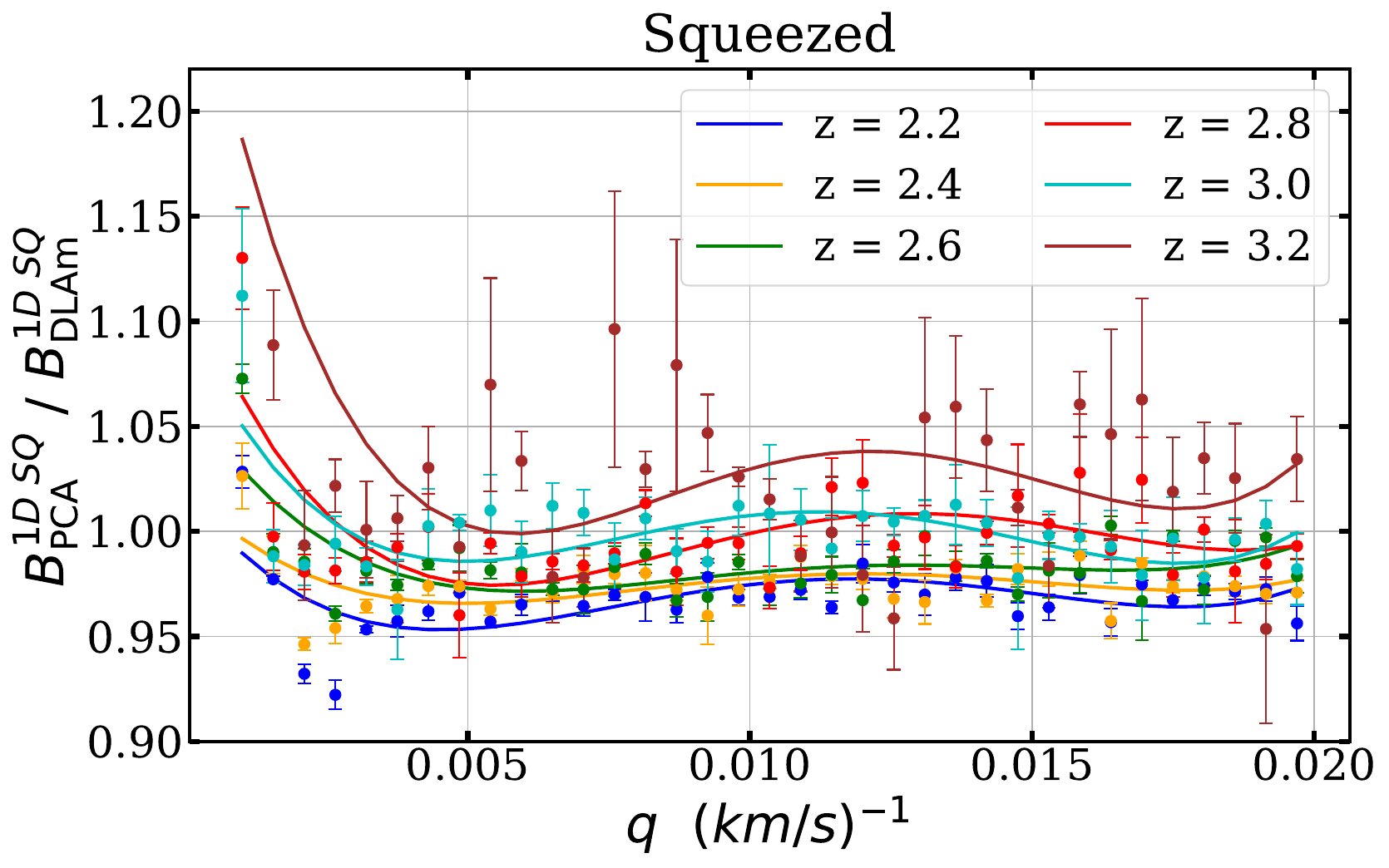}}
\caption{Ratio between the bispectrum obtained using unmasked DLAs pixels (PCA) and the one with masked pixels (DLAm) on the combination of the 5 mocks, for both triangular configurations considered. Four-order polynomial function are employed to fit the corrections in each redshift bin. The error bars come from to the diagonal of the bootstrap covariance matrix between the five realizations.}
\label{fig:plot10}
\end{figure}
Fig. \ref{fig:plot10} shows the $b_{\mathrm{DLAm}}(k,z)$ effect for both configurations. The top panel corresponds to isosceles triangles, while the bottom panel represents the squeezed-limit triangle configuration. In both configurations, there is a noticeable effect at small wavenumbers, which increases with redshift. This is consistent with the expectation that DLA contamination in the \lya forest increases with redshift. The impact of DLA masking is smaller compared to the effect of sky-line masking. These results are in good agreement with the $\pk$ measurements from eBOSS DR14 \cite{Chabanier_2019}. To model this bias, we employ a fourth-order polynomial, which is used for the final $\bk$ correction.

\section{Systematic uncertainties estimation}\label{Sec05}
We estimate the statistical uncertainty ($\sigma_{\mathrm{stat}}$) of our averaged $\bk$ measurement using the diagonal of the bootstrap covariance matrix\footnote{We compute the bootstrap covariance matrix for both configurations independently, rather than across the full 2D signal of the bispectrum.}. For our dataset of $N$ chunks, we form a bootstrap dataset by randomly selecting $N$ chunks with replacement. The covariance matrix is then computed as follows:
\begin{align}
    C^{\bk}_{ij} =  \dfrac{1}{N_{\mathrm{chunks}}}\sum_{k=1}^{N_{\mathrm{chunks}}}\left[\bk_{k}(q_{i}) - \langle \bk(q_{i}) \rangle \right]\cdot \left[\bk_{k}(q_{j}) - \langle \bk(q_{j}) \rangle \right], \label{eqcov}
\end{align}
$\langle\rangle$ means average over bootstrap realizations. In the Eq.~(\ref{eqcov}) the $\bk$ term can to be replaced by $\bksq$ in the case of the squeezed limit.
\begin{figure}
\centering
\subfigure{\includegraphics[width=7.5cm,height=5.5cm]{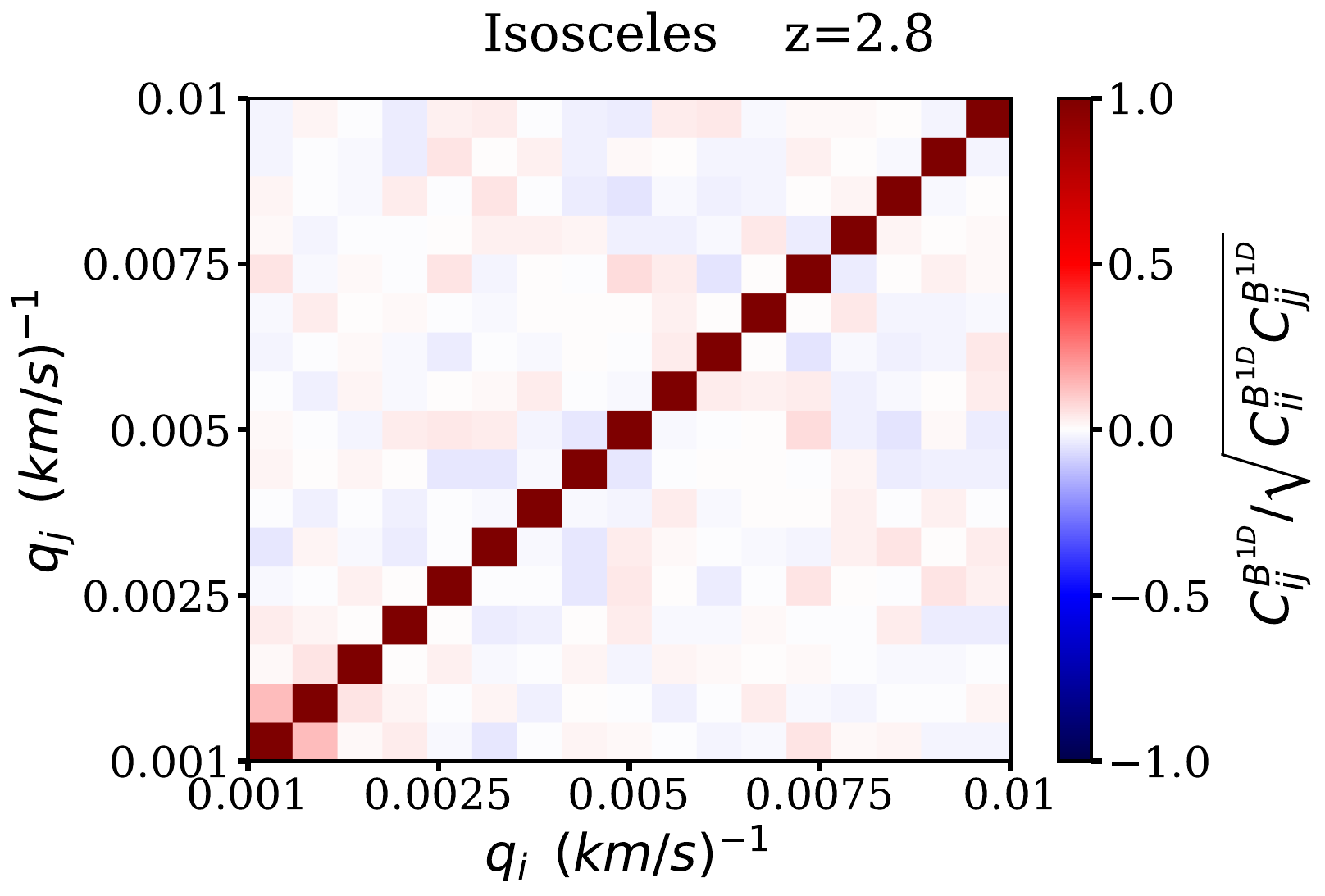}}
\subfigure{\includegraphics[width=7.5cm,height=5.5cm]{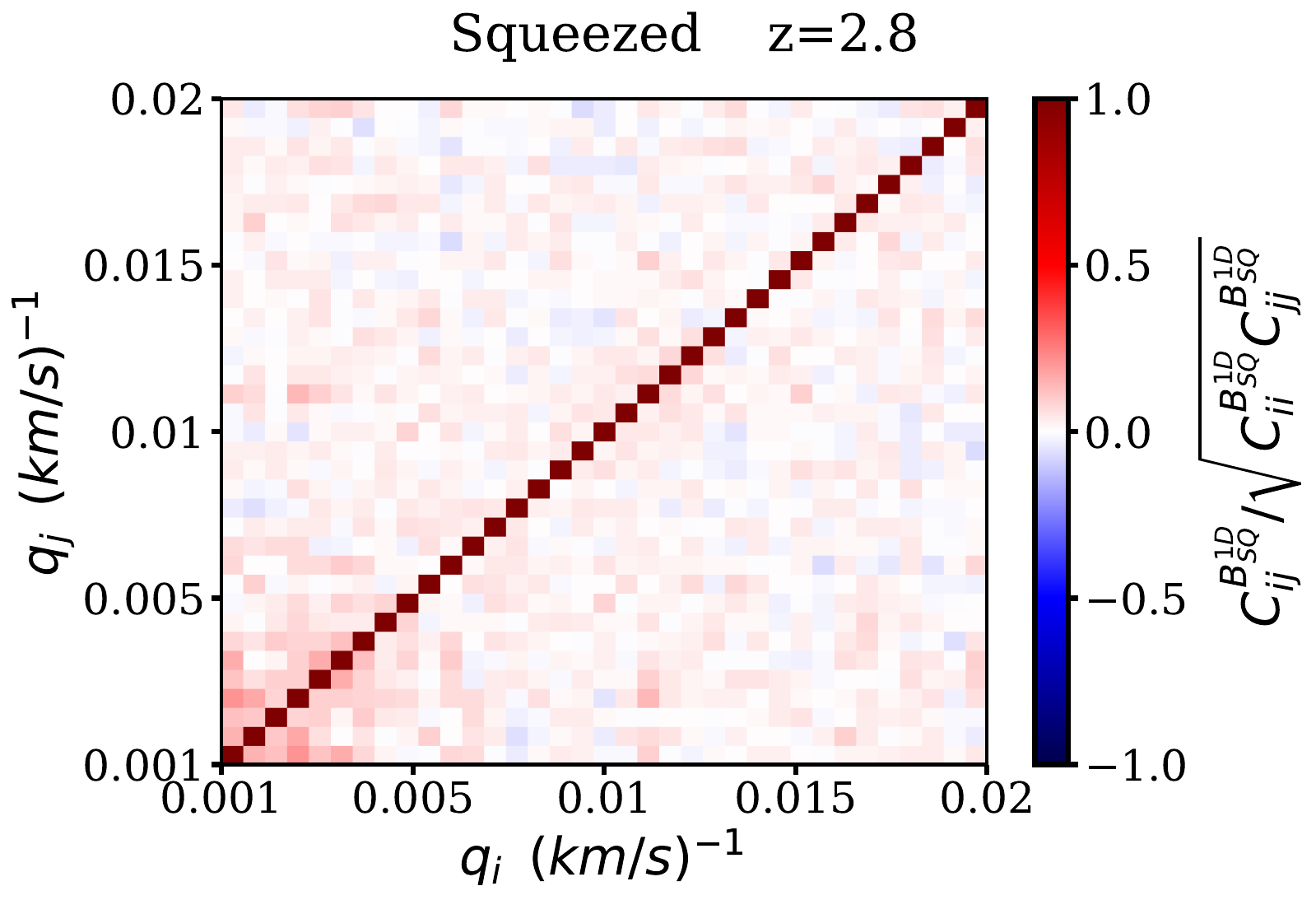}}
\caption{Bootstrap covariance matrices for both triangular configurations considered over eBOSS DR16 data set. We generate 2000 bootstrap samples. There is not correlation between adjacent q-modes, only slightly for large scales.}
\label{fig:plot11}
\end{figure}
We generate 2000 bootstrap samples of the input dataset and calculate the corresponding bootstrap covariance matrix. Fig. \ref{fig:plot11} shows the normalized bootstrap covariance matrix $C_{ij}/\sqrt{C_{ii}C_{jj}}$, where we observe zero correlation between adjacent wavenumbers. A slight correlation appears to grow from large to small scales, but this is only seen at small wavenumbers, likely due to the systematics affecting the initial $q$-modes. Convergence is achieved rapidly, and it may even be sufficient to use 200 bootstrap samples.
\begin{figure}
\centering
\subfigure{\includegraphics[width=7.5cm,height=5.5cm]{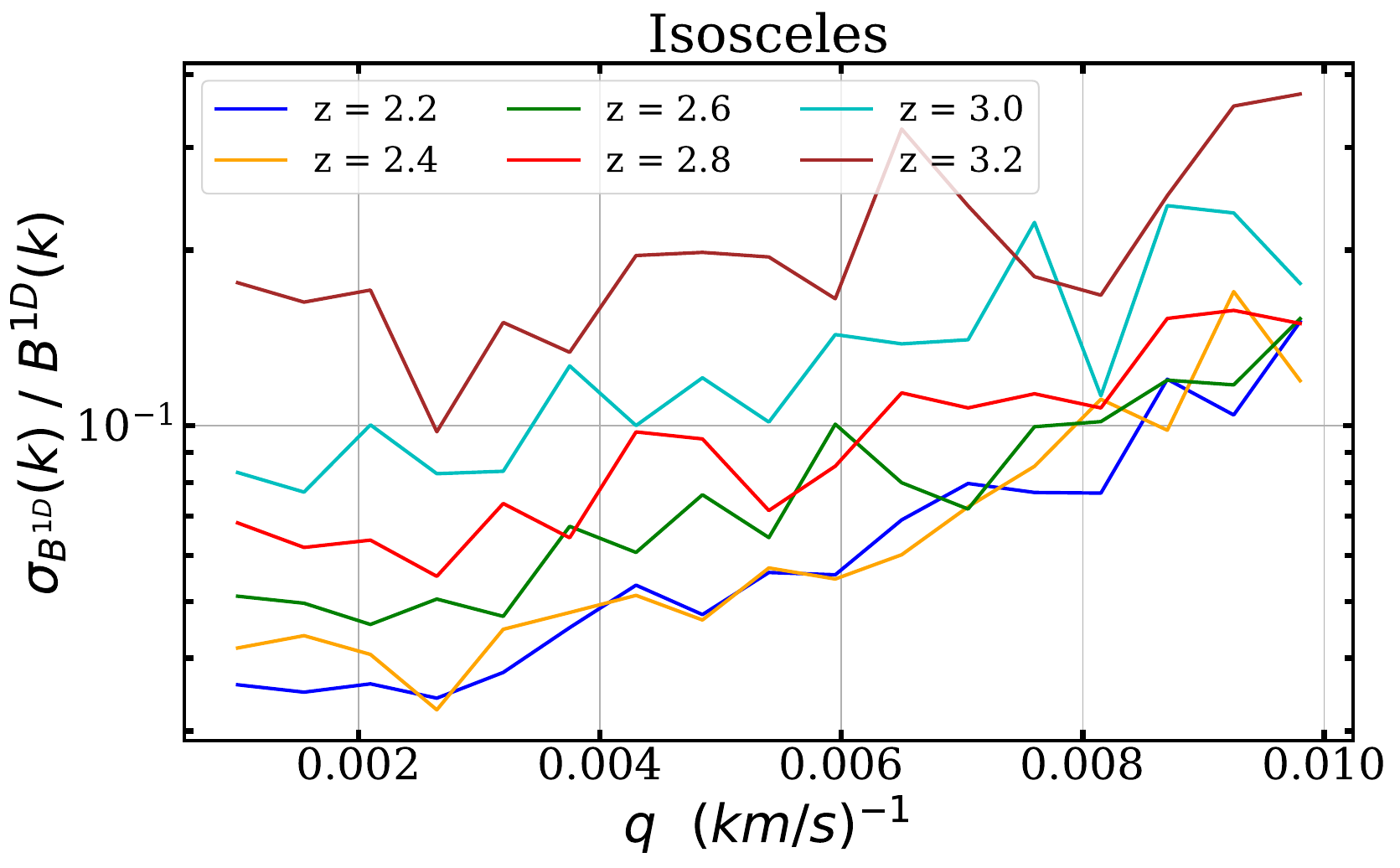}}
\subfigure{\includegraphics[width=7.5cm,height=5.5cm]{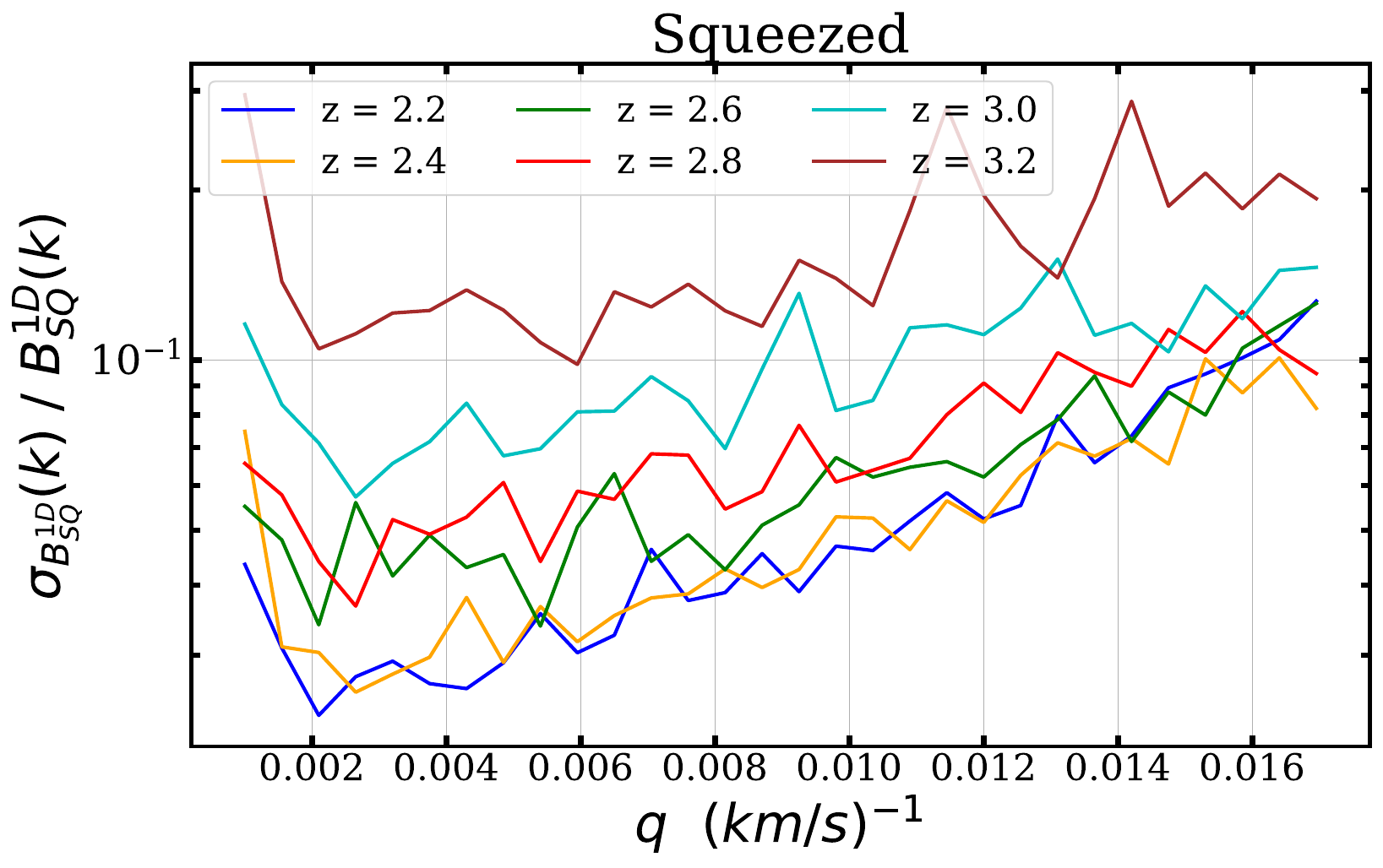}}
\caption{The estimation of the relative statistical uncertainties on the flux bispectrum measurement, for both triangular configurations considered.} 
\label{fig:plot12}
\end{figure}
Fig. \ref{fig:plot12} illustrates the dependence of the error bars on the bispectrum power. The low redshift bins are not clearly separated, and at redshifts $z = 2.2$ and $z = 2.4$, statistical uncertainties intersect at small scales. This intersection is due to a significant increase in noise in the blue spectral band. The variation in the number of chunks present in each redshift bin leads $\sigma(q)/\bk$ to increase as a function of redshift. Furthermore, the substantial change in amplitude at small scales compared to large scales is attributed to the effect of the window function, where the spectrograph's resolution directly impacts the results.

\begin{figure*}
\centering
\subfigure{\includegraphics[width=7.5cm]{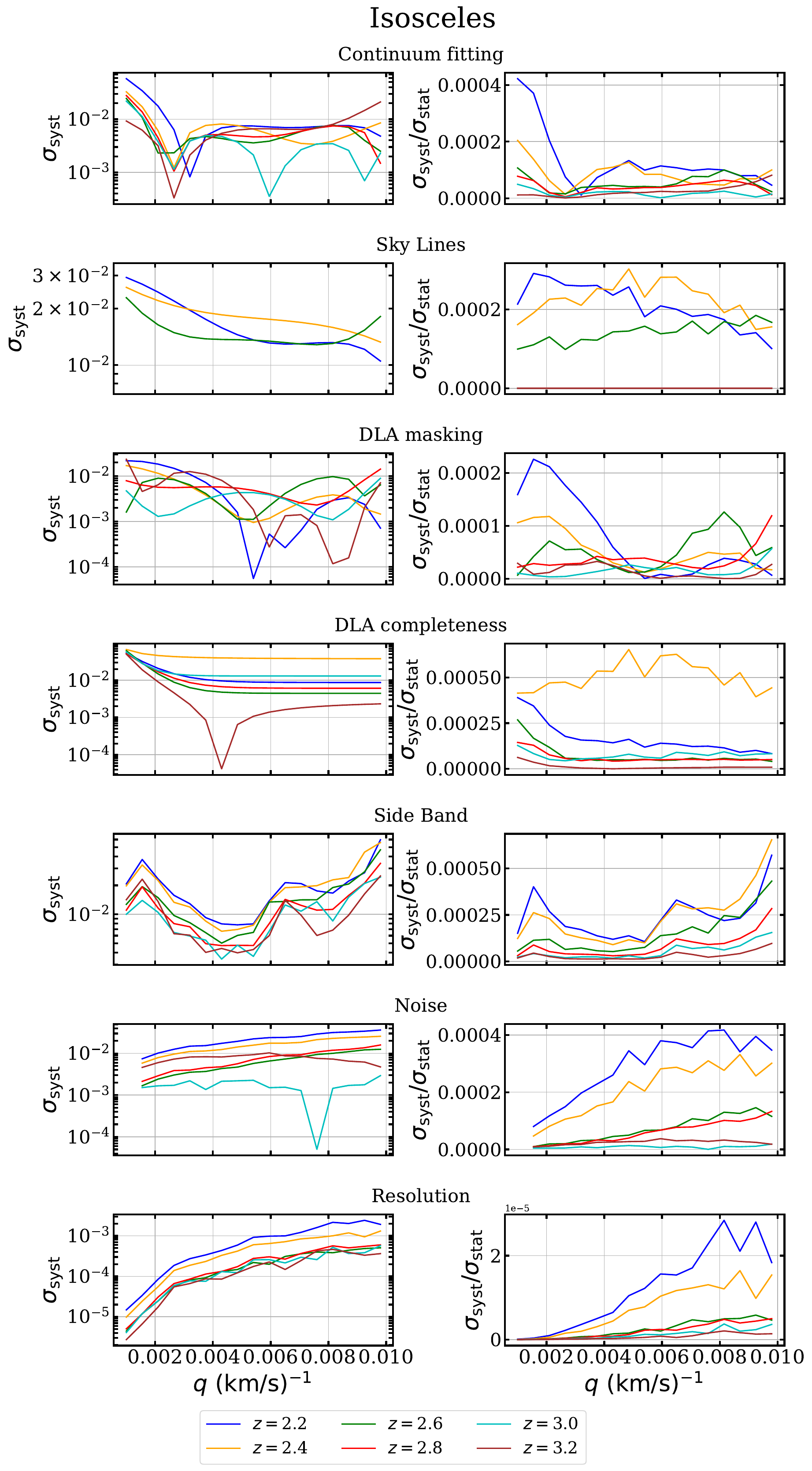}}
\subfigure{\includegraphics[width=7.5cm]{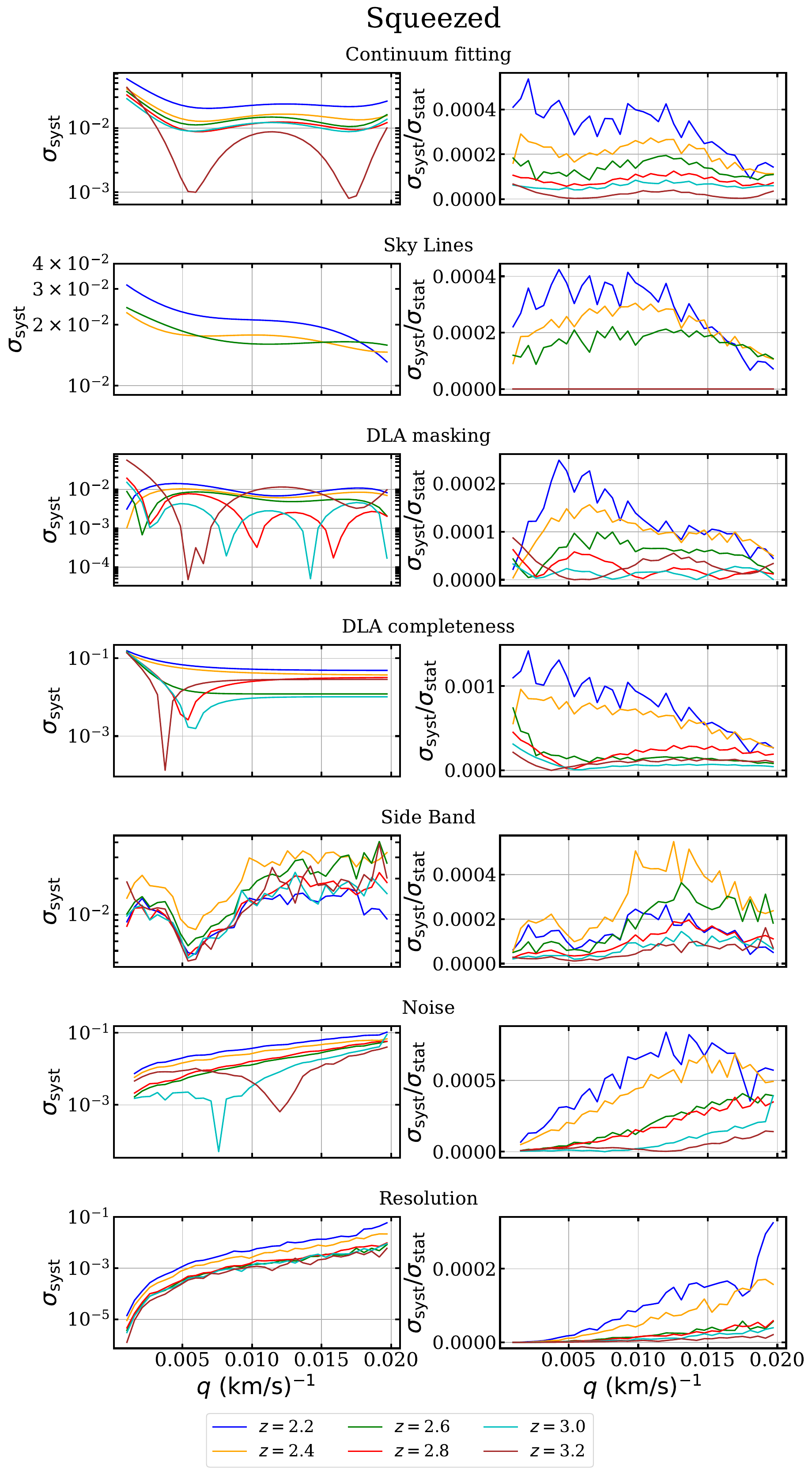}}
\caption{
Estimation of the systematic uncertainties $\sigma_{\mathrm{syst}}$ for different redshift bins in the two triangular configurations considered. Each line sub-panel is associated to a systematic considered in this work. The left panels of each triangular configuration show the absolute uncertainties, while the right panels show the relative value between systematic uncertainties and statistical uncertainties $\sigma_{\mathrm{stat}}$.
}
\label{fig:plot13}
\end{figure*}

As explained in the previous sections, starting from Eq.~(\ref{eq07}), the measurement of $\bk$ requires characterizing the impact of various instrumental and astrophysical contaminants. Therefore, we need to associate systematic errors ($\sigma_{\mathrm{syst}}$) with our $\bk$ estimate. The left panel of Fig. \ref{fig:plot13} shows the systematic uncertainties for isosceles triangles across different redshift bins and their relative values compared to statistical errors. The right panel of Fig. \ref{fig:plot13} presents the systematic uncertainties for squeezed limit triangles in different redshift bins and their relative values with respect to statistical errors. We identify seven systematic uncertainties and made a conservative choice in defining these uncertainties:
\begin{itemize}
    \item \textit{Continuum fitting}: We assign a systematic error of 30 per cent times the $b_{\mathrm{cont}}(k,z)$ correction computed in \ref{subsec042}.
    \item \textit{Sky lines masking}: The effect of the sky emission lines masking on the $\bk$ measurement was determined with synthetic data in \ref{subsec043}. We define the systematic error associated with each masking as 30 per cent of this correction. 
    \item \textit{DLA masking}:The effect of the DLAs masking on the $\bk$ measurement was also determined using synthetic data in \ref{subsec043}. We define the systematic error associated with each masking as 30 per cent of this correction. 
    \item \textit{DLA completeness}: We utilize the synthetic mock data described in Sec. \ref{Sec04} to investigate the impact of DLAs on the 1D bispectrum. We calculate the ratio between mocks with DLAs and those without DLAs. This ratio shows a similar effect to the one observed in the 1D power spectrum signal, with small wavenumbers being the most affected. To address this, we employ a model presented by \cite{10.1093/mnras/stx2942} to fit this ratio. Based on the DLA catalog provided by \cite{Chabanier:2021dis}, the authors report over 90\% efficiency using their CNN finder. However, as a conservative measure, we associate a 10\% uncertainty with the total impact of DLAs on our $\bk$ signal.
    \item \textit{Side-band}: We subtract the fitted side-bands bispectrum $\bk_{\mathrm{SB1, model}}$, modeled and computed in Sec. \ref{subsec034}, from the measured $\bk$ in the \lya forest to account for metal absorption. While one could use statistical uncertainty as the systematic uncertainty, the bootstrap error estimates might not be accurate, potentially biasing the result. Moreover, because the $\bk_{\mathrm{SB1, model}}$ signal depends on the number of side-band quasars in the relevant wavelength range, the statistical estimate of metal power might be incorrect. Projects like DESI will significantly improve the statistics in this area. Therefore, we calculate the ratio between the bispectrum before and after removing the side-band contribution. Allowing for the blending of metal lines as discussed in \cite{10.1093/mnras/stz2214} and \cite{10.1093/mnras/stab3201}, we assign a 10 percent error.
    \item \textit{Noise estimation}: As discussed throughout this text, the noise level in the bispectrum signal depends on both the raw power spectrum and the noise power spectrum, which amplifies the $P^{\mathrm{raw}}(k)$ signal. A straightforward method to characterize the noise effect on the $\bk$ signal involves multiplying the noise power spectrum by the raw power spectrum. The procedure for determining the noise level in the quasar spectrum is described in Sec. \ref{subsec033}. The noise level is corrected using the $\beta$ term, which varies for each redshift bin and dataset. We assign a systematic uncertainty to the resulting noise power spectrum, equal to 30\% of the (1-$\beta$) term times $q\cdot P(q)$. The maximum value occurs at $z=2.2$, as this is closer to the edge of the CCD.
    \item \textit{Resolution}: To investigate the impact of the resolution on the $\bk$ measurement, we conducted a study similar to the one proposed in \cite{Chabanier_2019}, focusing on the bispectrum rather than the power spectrum. The window function $W$ requires knowledge of the spectral resolution $R$ of the spectrograph, see Eq.~(\ref{window}). To incorporate its systematic uncertainty, we calculated the average resolution $\langle R\rangle$ across all the skewers contributing to each redshift bin. The systematic uncertainty on the $\bk$ signal is then given by $\left(2q^{3}R\Delta R\right)\cdot \bk$. The cubic q-term makes the large q-modes more affected by this uncertainty. The average resolution $\langle R\rangle$ also varies with respect to the redshift, from 79 $km/s$ to 68 $km/s$ with larger value for lower redshifts. The low-$z$ bins are more affected by this systematic. 
\end{itemize}

The choice of a 30 percent impact for systematic uncertainty comes from considering a random shift ranging between `100\% of the correction' and `no correction,' which we describe using a uniform distribution between 0 and 1. The standard deviation of this distribution, equal to $1/\sqrt{12} \sim 0.30$, quantifies the spread among the possible values, resulting in a systematic uncertainty equal to 30\% of the correction.

We acknowledge that we have incorporated several ideas previously used to estimate systematic uncertainty in the 1D power spectrum signal. While we have adapted these ideas for the 1D bispectrum signal, we recognize there is still room for improvement in our systematic uncertainty estimation. Continuum fitting could be enhanced either by using the method employed by \texttt{picca} or by developing a new approach that introduces less bias into the bispectrum signal. New galaxy surveys, such as DESI, will further reduce the effect of sky line masking by minimizing the number of masked pixels (see \cite{ravoux2023dark}). Similarly, improvements in the number of masked pixels for DLAs are expected. Significant progress is also anticipated in the study of quasar identification with DLAs, leading to improved DLA completeness. Additionally, the influence of other high-column density absorber systems, such as sub-DLAs, small-DLAs, and Lyman Limit systems, should be considered. These systems have been studied by \cite{10.1093/mnras/stx2942} in the context of the power spectrum, and we have shown that Rogers' modeling can also be applied to the bispectrum, suggesting it will work for these other systems as well. As the statistics of quasars with side-bands increase, the associated systematic error can be better characterized. We expect significant improvements in noise estimation for the bispectrum signal, as new methods are developed to quantify its effect. Regarding resolution, DESI is expected to yield considerable improvements. Since $\sigma_{\mathrm{stat}}$ is quite large, the ratio between $\sigma_{\mathrm{stat}}$ and $\sigma_{\mathrm{syst}}$ is small for all systematics. 

In future work, we plan to explore new ideas and methods to better quantify systematic effects. Specifically, we aim to investigate the impact of systematics using large samples of more realistic synthetic data, which will provide us with a deeper understanding of their effects on the bispectrum measurement.

\section{Results: first measurement of the \texorpdfstring{$\mathrm{Ly}\alpha$}{Lya} \texorpdfstring{$\bk$}{b1d} in eBOSS} \label{Sec06}
The $\bk$ measurement was performed using the pipeline described in Sec. \ref{Sec02}. Considering all the corrections outlined in previous sections, the final estimation of the 1D bispectrum incorporate the 2D signal from eqs. \ref{eq16}, \ref{eq17} and \ref{eq18} into the full bispectrum signal from Eq.~(\ref{eq08}). Our results will be presented for the two triangular configurations defined in Sec. \ref{Sec02}. While this is primarily for visualization purposes, it is important to emphasize that the analysis was conducted using the 2D bispectrum signal, encompassing all triangular configurations.

\subsection{Theoretical description of \texorpdfstring{$\bk$}{b1d}} \label{subsec061}
For the modelling, we use the theoretical expressions of  \cite{10.1111/j.1365-2966.2004.07404.x}, which relate the flux power spectrum and bispectrum using second-order perturbation theory (2OPT). In the Fluctuating Gunn-Peterson Approximation (FGPA), the flux can be related to the density field as 
\begin{equation}
F = \exp\{-A(1 + \delta_{IGM})^{\beta}\},  \label{eq19}  
\end{equation}
where $A$ and $\beta$ depend on redshift. $\beta = 2-0.7(\gamma-1) $, and $\gamma$ is the power-law index of the gas temperature-density relation. This model assumes that redshift space distortions are not important and neglects thermal broadening and instrumental noise. In futures works, we hope to overcome this assumption using \cite{Arinyo-i-Prats:2015vqa}, \cite{Givans:2020sez} and \cite{Verde:1998zr}. Furthermore, the model assume that at the scale of interest $\delta_{IGM} = \delta_{DM} = \delta$, expanding the density field to second order as $\delta(x) = \delta^{(1)}(x) + \delta^{(2)}(x)$, and using FPGA, we get $\delta_{F} \approx b_{1}[\delta^{(1)}(x) + \delta^{(2)}(x)] + \frac{b_{2}}{2}\delta^{(1) 2}(x)$, with $b_{1} = -A\beta$ and $b_{2} = -A\beta(\beta - 1 - A\beta)$. The expression for the Ly$\alpha$ bispectrum follows then by projecting the 3D bispectrum (\cite{Matarrese:1997sk}) along the line-of-sight, namely
\begin{align}
B(q_{0},q_{1},q_{2}) =&\enspace \Big(\frac{12}{7}c_{1}+c_{2}\Big)p(q_{0})p(q_{1})+\enspace c_{1}\Big[\Big(q_{0}q_{1}-\frac{2}{7}q^{2}_{0}\Big)p^{(-1)}(q_{0})p(q_{1}) \label{eq20} \\+&\enspace \Big(q_{1}q_{0}-\frac{2}{7}q^{2}_{1}\Big)p^{(-1)}(q_{1})p(q_{0})+\enspace \frac{6}{7}q^{2}_{0}q^{2}_{1}p^{(-1)}(q_{0})p^{(-1)}(q_{1})\Big]+cyc.(0,1,2),  \nonumber
\end{align}
where $c_{1} = 1/b_{1}$, $c_{2}=b_{2}/b^{2}_{1}$ and $p(q)$ is the 1D-Power Spectrum. 
The spectral moments $p^{(-1)}(q)$ are given by
\begin{equation}
p^{(\boldsymbol\ell)}(q) = |q|^{2\boldsymbol\ell}p(q)+2\boldsymbol\ell \int_{|q|}^{\infty} k^{-2\boldsymbol\ell -1} p(k)\enspace dk, \label{eq21}
\end{equation}
in this case $\boldsymbol\ell=-1$.  

We develop Eq.~(\ref{eq20}) for both \emph{Configuration 1} and \emph{Configuration 2}. It is essential to consider the sign in the above equations for all $q$ values and account for the symmetry between $p(q_{i})$ and $p^{(-1)}_{i}(q)$. \cite{10.1111/j.1365-2966.2004.07404.x} demonstrated that the optimal values that match with synthetic spectra are $A = 0.48$ and $\gamma = 1.5$. In our results, we fit the values of $c_{1}$ and $c_{2}$ for the full 2D $\bk$ signal, considering all triangle configurations.

\subsection{\texorpdfstring{$\bk$}{b1d} measurement}
\begin{figure*}
\centering
\includegraphics[width=15cm]{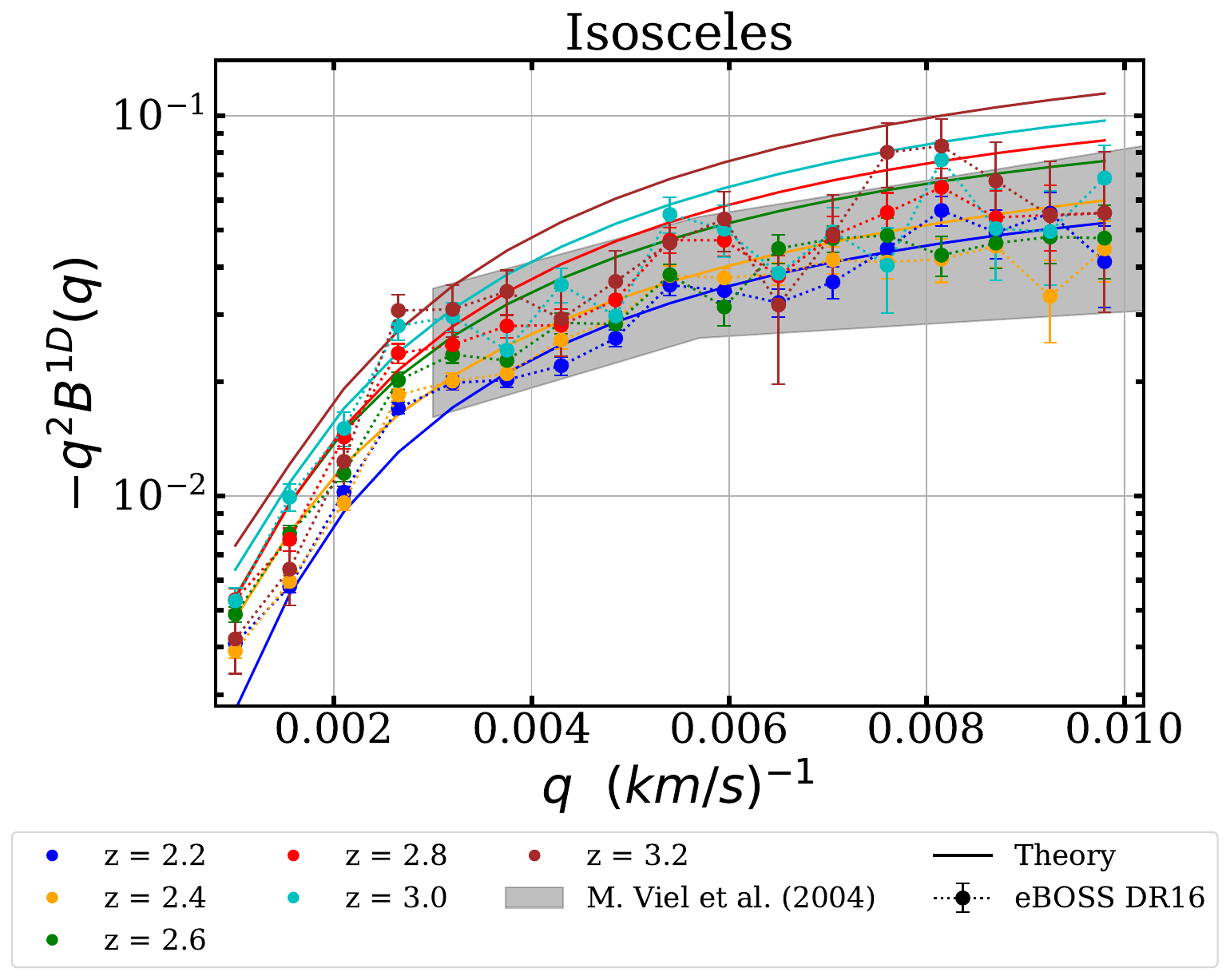}
\caption{One-dimensional \lya bispectrum for isosceles triangles using eBOSS DR16 data, for redshift bins from $z=2.2$ to $z=3.2$ (point with error bars). Solid lines represent the fit modeled theoretically by Eq.~(\ref{eq21}). The shaded gray area is the previous measurement by \citep{10.1111/j.1365-2966.2004.07404.x}. The error bars present in our measurements come from the diagonal of the bootstrap covariance matrix and systematic uncertainties added in quadrature.
}
\label{fig:plot15}
\end{figure*}

\begin{figure}
\centering
\includegraphics[width=11.5cm]{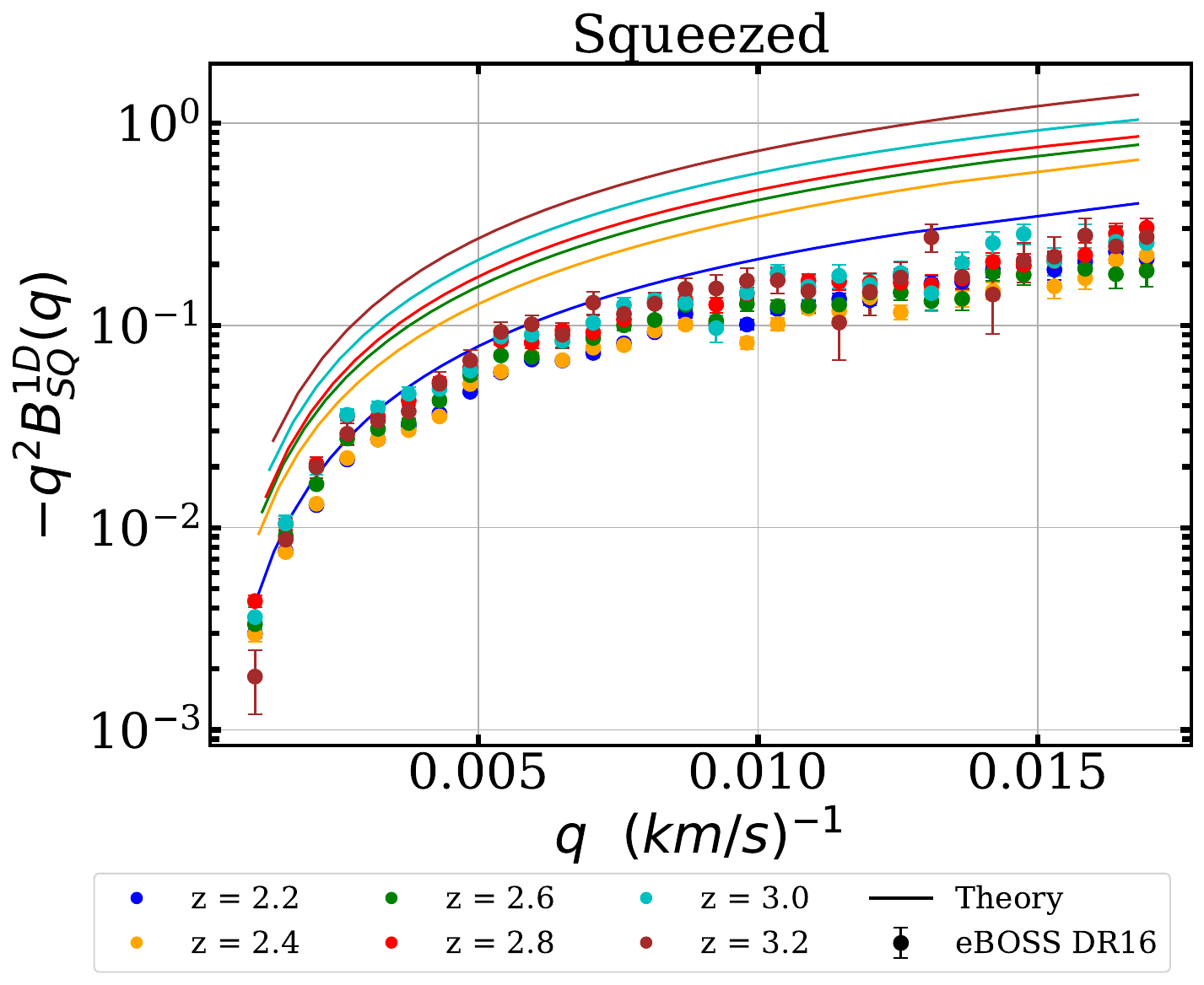}
\caption{
One-dimensional \lya bispectrum for squeezed limit triangles using eBOSS DR16 data (points with error bars). Solid lines represent the fit modeled theoretically by Eq.~(\ref{eq21}). The error bars present in our measurements come from the diagonal of the bootstrap covariance matrix and systematic uncertainties added in quadrature.
}
\label{fig:plot16}
\end{figure}

Using the procedure described in the previous sections, we estimate the 1D bispectrum over 12 redshift bins, from $z_{Ly\alpha} = 2.2$ to 4.4, and over 17 Fourier modes, ranging from $q = 1 \times 10^{-3}$ to 0.01 $(\mathrm{km/s})^{-1}$, for the isosceles triangle configuration. For the squeezed-limit triangle configuration, we calculate 30 modes, ranging from $q = 1 \times 10^{-3}$ to 0.017 $(\mathrm{km/s})^{-1}$, within the same redshift range. We chose these $q$-limits because the effect of the window function exceeds 80\%. We focus on the low redshift bins from $z_{Ly\alpha} = 2.2$ to 3.2 for both configurations, as the higher redshift bins exhibit a noisier bispectrum signal. Fig. \ref{fig:plot15} displays the 1D bispectrum measurement for the isosceles triangles at low redshift. The error bars shown correspond to the diagonal elements of the bootstrap covariance matrix, with systematic uncertainties added in quadrature. Our measurements appear to be consistent with the previous results of \cite{10.1111/j.1365-2966.2004.07404.x}, which used a redshift bin from $z_{Ly\alpha} = 2.0$ to 2.4, with a signal characterized by large error bars, as shown by the gray shaded line. The colored solid lines, corresponding to the different redshift bins, represent the fit of the theoretical 2OPT model to the observed data. Considering that we fit the 2D bispectrum signal (see the left panel of Fig. \ref{fig:plot_t0}) and the number of free parameters involved, the $\chi^{2}$ value between the model and the data indicates a good fit.
However, the squeezed-limit configuration shows a poor fit. Fig. \ref{fig:plot16} displays the $\bksq$ signal for the different redshift bins. The poor fit is likely due to the fact that the squeezed-limit configuration depends on the delta field in the minimum Fourier mode (the first Fourier mode, excluding the $q=0$ mode). This mode is significantly influenced by several systematics, as discussed in Sec. \ref{Sec05}, such as continuum fitting and pixel masking due to DLAs or sky lines. The value of $q_{\mathrm{min}}$ in the $\bksq$ signal varies for each redshift bin, increasing with redshift and ranging from $q_{\mathrm{min}} = 3.3 \times 10^{-4}$ to $5.6 \times 10^{-4}$ $(\mathrm{km/s})^{-1}$.

We also theoretically modeled the bispectrum signal using mock data, but the results were unsatisfactory. Given the error bars associated with the bispectrum signal, the $\chi^{2}$ value indicated a poor fit for both configurations, despite the $A$ and $\gamma$ values being similar to those of the observed data. Notably, the theoretical modeling of the squeezed-limit signal produced results similar to those of the observed signal. his suggests that the poor fit in this configuration may also be due to the limitations of the theoretical model. As noted in \cite{Chiang_2017}, the squeezed-limit bispectrum encodes the impact of large-scale fluctuations on the small-scale power spectrum, representing how the small-scale power spectrum "responds" to large-scale fluctuations. We find that large-scale fluctuations are strongly affected by systematics, while the 2OPT-based theoretical model may inadequately describe small scales. In future work, we will explore the theoretical modeling provided by 2OPT in more detail. It may be necessary to employ a different model or extend the model proposed by \cite{10.1111/j.1365-2966.2004.07404.x} by accounting for factors such as peculiar velocities and thermal broadening.

\begin{figure}
\centering
\subfigure{\includegraphics[width=7.5cm,height=5.5cm]{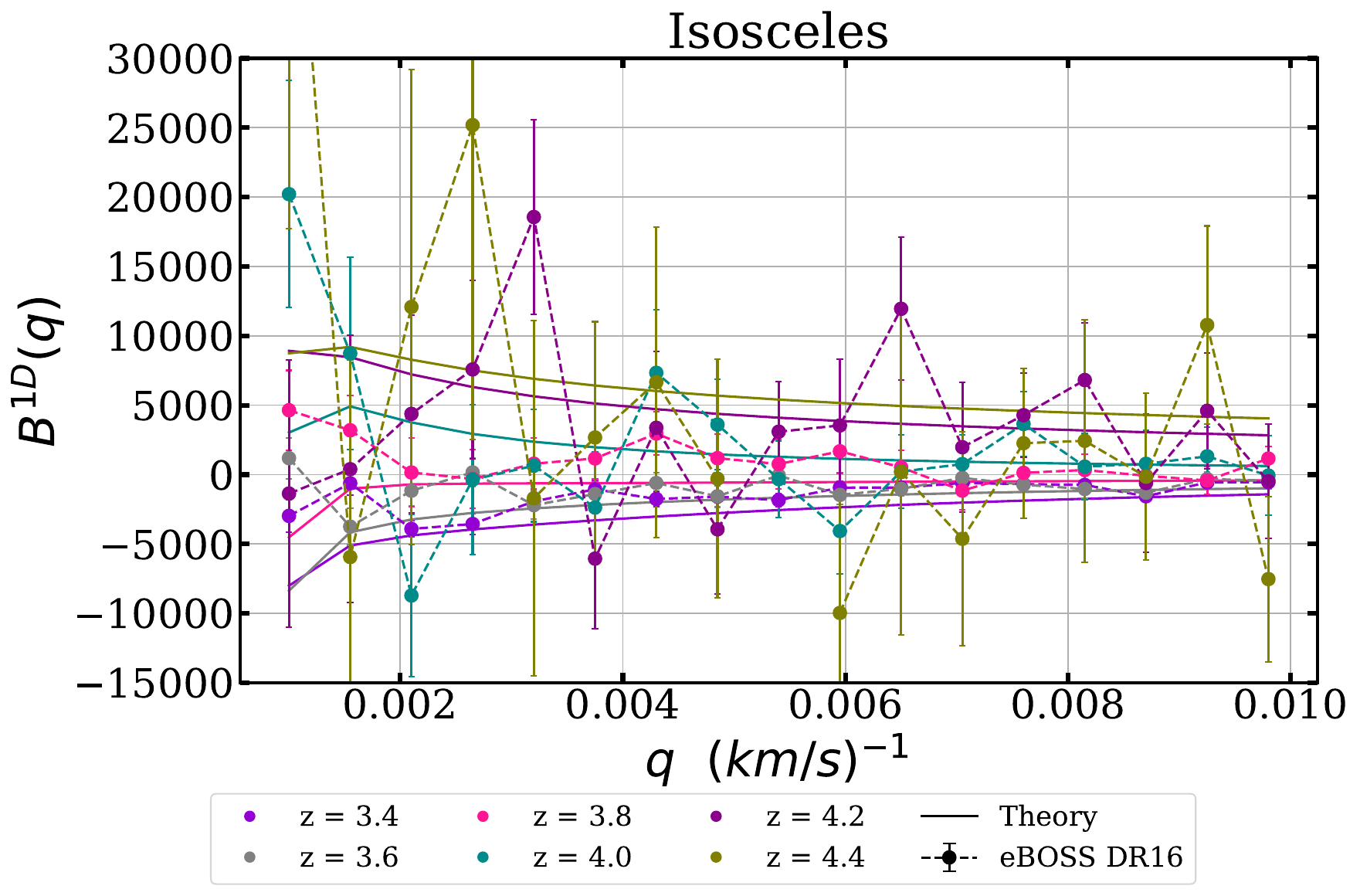}}
\subfigure{\includegraphics[width=7.5cm,height=5.5cm]{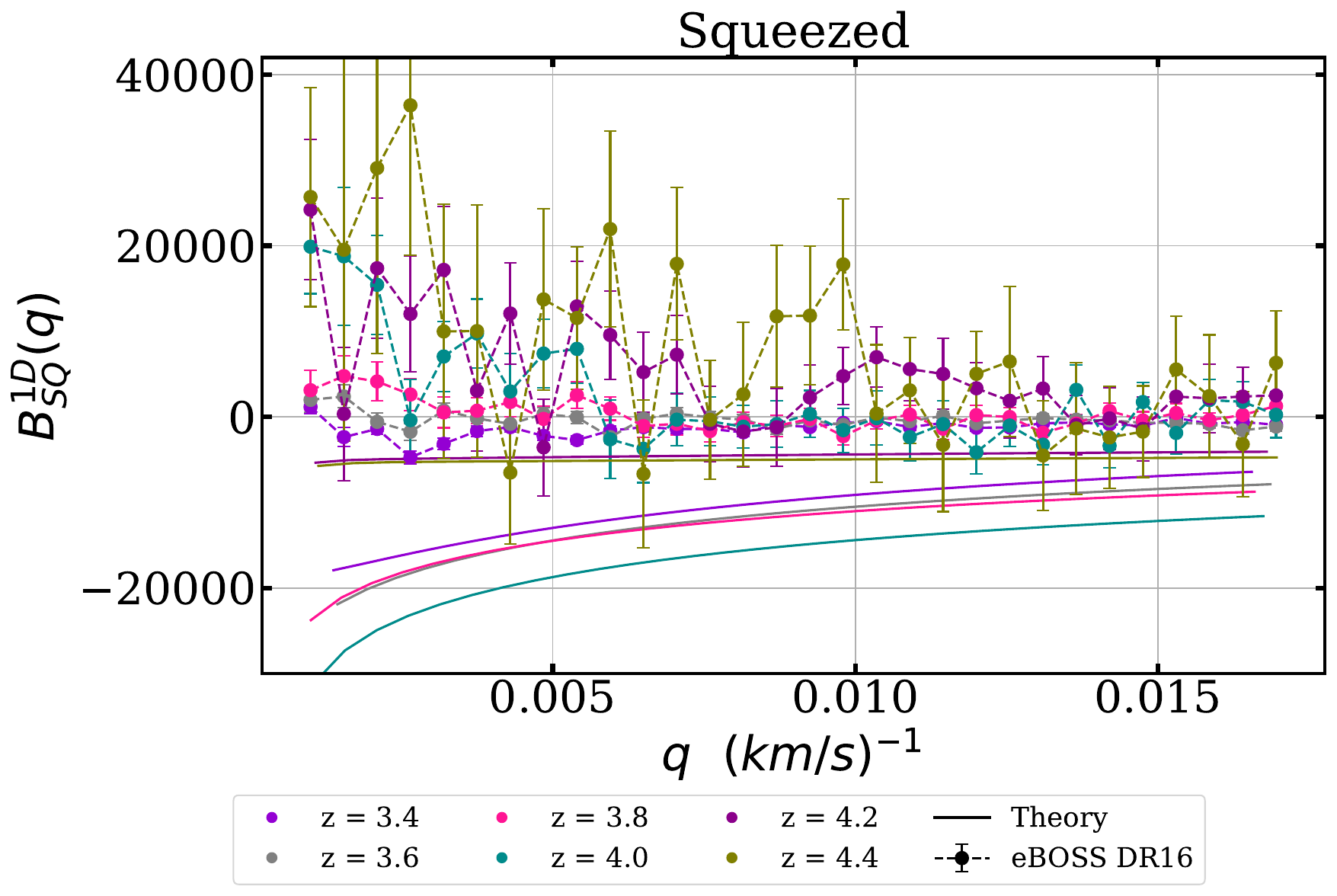}}
\caption{
One-dimensional \lya bispectrum for redshift bins from $z=3.4$ to $z=4.4$ in the two triangular configurations considered. Solid lines represent the fit theoretically modeled by Eq.~(\ref{eq21}). The error bars present in our measurements come from the diagonal of the bootstrap covariance matrix and systematic uncertainties added in quadrature.
}
\label{fig:plot17}
\end{figure}

We also measured the bispectrum for the high redshift bins ($3.4 < z_{Ly\alpha} < 4.4$); however, the signal was not significantly detected. The bispectrum signal for these redshift bins exhibits large fluctuations around zero, making it challenging to obtain a reliable estimate (see the upper panel of Fig. \ref{fig:plot17}). In the case of the squeezed triangles (lower panel of Fig. \ref{fig:plot17}), the signal appears to shift toward positive values. This shift is more noticeable in the last two redshift bins ($z_{Ly\alpha} = 4.2$ and 4.4). Nonetheless, due to the large error bars resulting from the low number of chunks, it is difficult to confidently assert this observation.

The fit to the 2D bispectrum signal depends on the parameters $c_{1}$ and $c_{2}$, which can be expressed as functions of $A$ and $\beta$ in Eq.~(\ref{eq19}). We present the results of the fit in terms of the $A$ values. Instead of using $\beta$, we opted to use $\gamma$, the power-law index of the gas temperature-density relation. The results are shown in Table \ref{tab:table1}. Our results for $z = 2.4$ are in agreement with previous measurements of the $\gamma$ parameter reported by \cite{McDonald:2000nn}, \cite{Rudie_2012}, and \cite{Telikova_2019}, with differences of less than 3\%. \cite{Telikova_2019} used high-resolution quasar spectra from the KODIAQ dataset to estimate the $\gamma$ parameter for redshift bins similar to those we have considered. Overall, our results are comparable, with differences of less than 5\%, except for the redshift bins at $z = 2.6$ and $z = 3.2$, where the discrepancy is as high as 7\%.
\begin{table}
    \centering
    \begin{tabular}{ |c|c|c|c|c| }
         \hline
         \rule[-0.8em]{0em}{2em} Data & z-bin & $A$ & $\gamma$ & $N_{\mathrm{chunks}}$\\
         \hline
         \multirow[c]{10.8}{4em}{eBOSS} & \rule[-0.8em]{0em}{2.5em} $2.2$ & $0.39\pm 0.06$ & $1.49\pm 0.50$ & 39972
         \\
         \cline{2-5}
         &\rule[-0.8em]{0em}{2.5em} $2.4$ & $0.47\pm 0.08$ & $1.48\pm 0.49$ & 34511 \\
         \cline{2-5}
         &\rule[-0.8em]{0em}{2.5em}$2.6$ & $0.53\pm 0.07$ & $1.47\pm 0.43$ & 25882 \\
         \cline{2-5}
         &\rule[-0.8em]{0em}{2.5em}$2.8$ & $0.62\pm 0.07$ & $1.50\pm 0.32$ & 19537 \\
         \cline{2-5}
         &\rule[-0.8em]{0em}{2.5em}$3.0$ & $0.69\pm 0.07$ & $1.31\pm 0.32$ & 13402 \\
         \cline{2-5}
         &\rule[-0.8em]{0em}{2.5em}$3.2$ & $0.76\pm 0.12$ & $1.38\pm 0.48$ & 8335 \\
         \hline
    \end{tabular}
    \caption{Parameter's fitting to observed 1D bispectrum for all triple q-configurations for different redshift bins. The $\gamma$ value (power-law index of the gas temperature-density relation) is estimated from the $\beta$ parameter in eq \ref{eq19}.
    }
    \label{tab:table1}
\end{table}

\subsection{\texorpdfstring{$\siiii$}{SIII} cross-correlation}
The correlated background due to absorption by \lya and $\siiii$ from the same gas cloud along the quasar line of sight has been studied in \cite{McDonald_2006} and \cite{Palanque-Delabrouille:2013gaa}. They explain that this correlation can be estimated directly from the power spectrum signal. This suggests that we can extend this idea to the bispectrum case as well. Consider two absorption lines, one due to \lya and one due to $\siiii$. The $\siiii$ absorption line is weaker than the \lya absorption by a factor of $a$ and has a separation of $\mu = 2271~km/s$. We can write the total transmission fluctuation field as:
\begin{equation}
    \delta_{F}(v) = \delta(v) + a\cdot \delta(v+\mu), \label{eq22}
\end{equation}
where $\delta(v)$ represent the \lya component. The Fourier transform of this total transmission field is given by 
\begin{equation}
    \widetilde{\delta}_{F}(k) = \widetilde{\delta}(k)\cdot\left(1 + a\cdot \mathrm{e}^{ik\mu}\right). \label{eq23}
\end{equation}
The power spectrum is then $P \propto |\widetilde{\delta}_{F}|^{2}$,
\begin{equation}
    P_{F}(k) = P(k)\left(1+a^{2}+2a\cos{(k\mu)} \right), \label{eq24}
\end{equation}
while the bispectrum is $B \propto Re[\widetilde{\delta}_{F}(q_{0})\widetilde{\delta}_{F}(q_{1})\widetilde{\delta}_{F}(-q_{0}-q_{1})]$,
\begin{align}\label{eq25}
    B_{F}(q_{0},q_{1}) = B(q_{0},q_{1})\Bigl[& 1+a^{3}+a(1+a)\Bigl(\cos{(q_{0}\mu)}+\cos{(q_{1}\mu)}+\cos{\bigl((q_{0}+q_{1})\mu\bigr)}\Bigr)\Bigr].
\end{align}

Eq.~(\ref{eq25}) describes the oscillations due to the \lya/$\siiii$ cross-correlation for all triangle configurations along the line of sight. This expression helps us describe the dark areas in the left panel of Fig. \ref{fig:plot_t0}. However, we need to model the signal originating solely from the \lya forest. Although this modeling could theoretically be based on the approach explained in Sec. \ref{subsec061}, the results may be insufficient, as a good fit is not achieved for all triangle configurations, as previously shown. For this reason, we decided to model the $\bk$ signal independently for the two chosen triangle configurations. As a first step, let us derive the appropriate expressions for these configurations from Eq.~(\ref{eq25}).

Implementing the isosceles triangles configuration defined as $q=q_{1}=q_{0}$, 
\begin{align}\label{eq26}
    \bk(q) = \bk_{Ly\alpha}(q)\Bigl[1+a^{3}+a(1+a)\Bigl(2\cos{(q\mu)}+\cos{(2q\mu})\Bigr)\Bigr],
\end{align}
now over the squeezed limit triangle configuration given by $q_{0}=q - q_{\mathrm{min}}$ and $q_{1}=-q - q_{\mathrm{min}}$,
\begin{align}
    \bksq(q) = \bk_{SQ~Ly\alpha}(q)\Bigl[1+a^{3}+a(1+a)\Bigl(&2\cos{(q\mu)}\cos{(q_{\mathrm{min}}\mu)} +\cos{(2q_{\mathrm{min}}\mu})\Bigr)\Bigr]. \label{eq27}
\end{align}

The bispectrum generated solely by \lya absorption lines is modeled using Eq.~(\ref{eq14}). Although this equation was originally designed to model $k\pk(k)/\pi$, it provides a good fit to the quantity $-q^{2}\bk(q)$ for either configuration. We independently fit $\bk_{Ly\alpha}$ and $\bk_{SQ~Ly\alpha}$, and then fit the oscillating components of eqs. \ref{eq26} and \ref{eq27}, respectively. Table \ref{tab:table2} presents the parameter results from Eq.~(\ref{eq14}) for the $\pk$ and the two bispectrum configurations at $z=2.4$.
\begin{table}
    \centering
    \begin{tabular}{ |c|c|c|c|c|c|c| }
         \hline
         Signal & $k_{1}$ & $A$ & $n$ & $\alpha$ & $B$ & $\beta$\\
         \hline
          $\bk$ & 0.009 & 0.57 & -3.52 & -0.25 & 11.99 & -9.69
         \\
         \hline
          $\bksq$ & 0.89 & 0.46 & -3.60 & -0.10 & 9.79 & -10.28 
          \\
         \hline
         $\pk$ & 0.9 & 0.31 & -4.22 & -0.11 & 13.24 & -9.73
         \\
         \hline
    \end{tabular}
    \caption{Parameter's fitting to observed 1D bispectrum oscillations for both q-configurations at z=2.4, as well as to power spectrum.
    }
    \label{tab:table2}
\end{table}

\begin{figure}
\centering
\includegraphics[width=12.5cm]{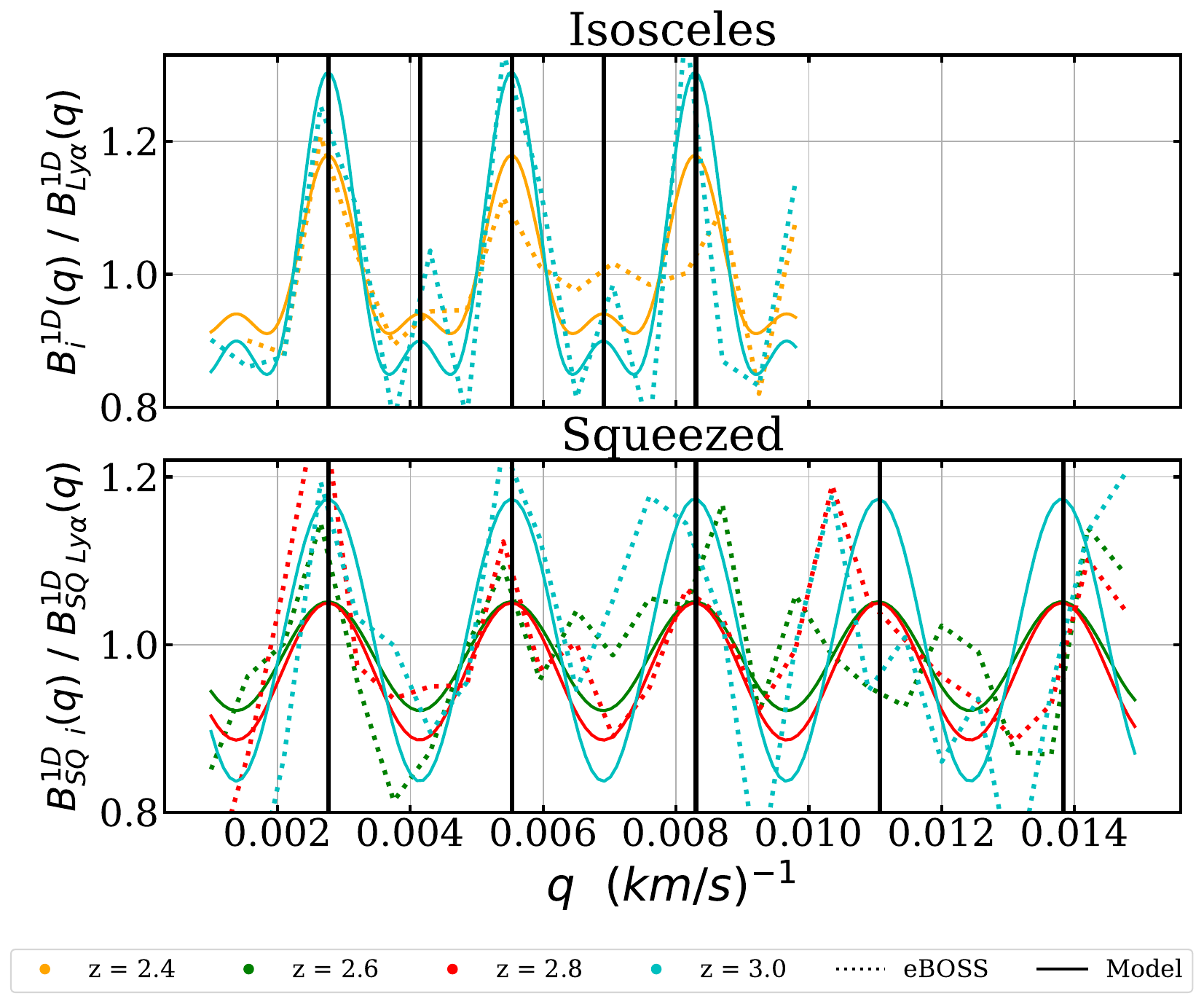}
\caption{
Oscillations present in the 1D bispectrum measurement for the two triangle configuration considered. The oscillations are induced by \lya/$\siiii$ cross-correlation. The dotted lines correspond to eBOSS DR16 data, while the solid lines show our best fit modelled by Eqs.~(\ref{eq26}) and (\ref{eq27}). The bispectrum signal without oscillation is computed using \ref{eq14}, but fitting to the quantity $-q^{2}\bk(q)$.
}
\label{fig:plot18}
\end{figure}

We clearly detect, the oscillation pattern in the bispectrum due the \lya/$\siiii$ cross-correlation, see Fig. \ref{fig:plot18}. For our first fit to $\bk$ accounting for $\siiii$, we use $a$ as a extra free parameter of the fit. We found a remarkable improvement in $\chi^{2}$, from 38.35 to 8.26 at z=3.0, with similar results for other redshift bins. As in \cite{McDonald_2006}, after a very simple fit we find that $a\sim 0.04$, of course this value depends on the redshift.

The top panel of Fig. \ref{fig:plot18} shows the oscillations of isosceles triangles configuration as described by Eq.~(\ref{eq26}). The larges oscillations are produced by the term `$2\cos{(q\mu)}$', with a frequency of $q=2\pi/\mu$. The term `$\cos{(2q\mu)}$' induces smaller oscillations in the $\bk$ signal, with a frequency of $q=\pi/\mu$. The vertical lines indicate the peaks of both induced oscillations. The figure clearly shows the smaller amplitude oscillation at $q=4.1\times10^{-3}$ and $6.9\times10^{-3}~(km/s)^{-1}$ in the estimated $\bk$ signal. The lower panel corresponds to the oscillations present in the $\bksq$ signal, modeled by Eq.~(\ref{eq27}), where the oscillation frequency is $q=2\pi/\mu$. Similarly, the vertical lines indicate the oscillation frequency.

\section{Conclusions}\label{Sec07}
The Lyman alpha forest (\lya) is a unique high-redshift, multi-scale probe of the matter distribution in the universe. It can be used to characterise the expansion history of the universe on the largest scales, while also providing insights into the growth of structure, the nature of dark matter, neutrino masses, and primordial features when focusing on the smallest scales. A common approach to extract the small-scale information is to use the two-point correlation function in Fourier space along the line of sight, known as the 1D power spectrum ($\pk$). However, additional information can be obtained by using higher order correlation functions, whose signal decays as the order increases. The next estimator to the $\pk$ in this signal-to-noise hierarchy is the 1D bispectrum ($\bk$), which correlates three pixels along the line of sight. Thus, an important question is whether we can measure the $\bk$ for the Lyman alpha forest using the current high-precision data from spectroscopic surveys.

In this context, we perform the first measurement of the Lyman alpha forest 1D bispectrum using the eBOSS DR16 data in the redshift range $2.2 < z < 4.4$. A statistically significant bispectrum signal is observed in the range $2.2 < z < 3.2$, where the signal is negative, smooth, and exhibits a clear redshift dependence. Moreover, the signal clearly reveals oscillations due to the metal (particularly $\siiii$) absorption lines in the \lya forest region, that we effectively account for with a simple model. Our measurement encompasses all possible triangle configurations up to a given maximal frequency, as shown in Fig.~\ref{fig:plot_t0}. However, we focus on two specific triangle configurations: isosceles and squeezed limit. Figures  \ref{fig:plot15} and \ref{fig:plot16} display the bispectrum signal for these two configurations, respectively. 

To assess the robustness of the $\bk$ signal, we conduct a thorough investigation of the systematic uncertainties affecting the measurement. We use adapted synthetic data to correct for the impact of masking pixels affected by DLAs and sky lines, as well as to account for the continuum fitting procedure that we choose. These mocks were processed using the same pipeline as the observed data. The major source of uncertainty at small scales arises from the spectrograph's resolution and noise power estimation, with the latter being dominant at low redshifts. In contrast, at large scales, the incompleteness of the DLA catalog is the primary source of uncertainty. There may also be uncertainty due to the incompleteness of the BAL catalog, but we leave a detailed study to a future publication. Another systematic effect concerns metal absorption on the redder side of the \lya peak (known as side-bands). We perform the first measurement of the 1D bispectrum for these side-bands, with that defined closer to the \lya peak showing the highest power, as in the $\pk$ case. We characterise this side-band bispectrum using oscillatory functions of the $\mathrm{Si~{\textsc{iv}}}$ and $\mathrm{C~{\textsc{iv}}}$ doublets, and use it to remove the background power present in the bispectrum signal of the Lyman-alpha forest.

To understand the shape and redshift scaling of the full 1D bispectrum signal, we use a simple model based on second-order perturbation theory, which relates the power spectrum to the bispectrum using also a simple bias model. We find that this analytic description can reproduce the slope of the estimated bispectrum for both the isosceles triangle configuration and the squeezed limit. However, it cannot accurately reproduce the amplitude of the signal, especially in the squeezed limit configuration where we expect non-linear physics to play a major role. An interesting future avenue would be to include EFT corrections and a more complete bias model (see for example \cite{Aviles:2023fqx}) to achieve similar results as those in the recent studies of the power spectrum found in \cite{Ivanov:2023yla,Ivanov:2024jtl}. However, with this simple modeling we are able to estimate the power-law index $\gamma$ of the gas temperature-density relation for different redshift bins. These values are in agreement with previous measurements reported in \cite{McDonald:2000nn}, \cite{Rudie_2012}, and \cite{Telikova_2019}.

Although we measure the bispectrum for the high redshift bins $3.4<z<4.4$, the signal is quite noisy and exhibits a tendency to change sign. Upcoming spectroscopic surveys, such as DESI, will help clarify whether these high-reshift signals are statistical anomalies or related to the evolution of mean transmission with redshift. We plan to refine our analysis using DESI data, as improvements in noise calibration and spectral resolution, as shown by \cite{ravoux2023dark}, should enhance the signal quality. Furthermore, extending the analysis to more sophisticated analytic models, such as those in \cite{Ivanov:2023yla}, or emulators based on hydrodynamic simulations (e.g. \cite{DESI:2024hqh}), in combination with the clean bispectrum measurements obtained with DESI data using the pipeline presented here, could allow the use of the $\bk$ to place additional constraints on cosmological and intergalactic medium parameters, beyond those provided by the power spectrum alone.

\section{Acknowledgements}
We would like to thank Andreu Font-Ribera and Alma X. Gonzalez Morales for advise and early discussions on the project, and for providing some of the mock catalogues. We also thank the DESI \lya WG members who have contributed indirectly to the infrastructure of the pipeline analysis use here. R.D.L.C. and G.N. acknowledge the computational resources of the DCI-UG DataLab, and the financial support of CONAHCYT, through the graduate fellowship programs and grants "Ciencia de Frontera 2019" No. 102958 and "Ciencia Básica y de Frontera" No. CBF2023-2024-162; the DAIP-UG and the Instituto Avanzado de Cosmologia. V.I. is supported by the Kavli Foundation. C.R. Acknowledge the funding from Excellence Initiative of Aix-Marseille University - A*MIDEX, a French ``Investissements d'Avenir'' program (AMX-20-CE-02 - DARKUNI).




 
\appendix

\section{Maximal \texorpdfstring{$q$}{q} scale and signal for different triangle configurations}\label{appendix:apex1}

\begin{figure}
\centering
\includegraphics[width=12.5cm]{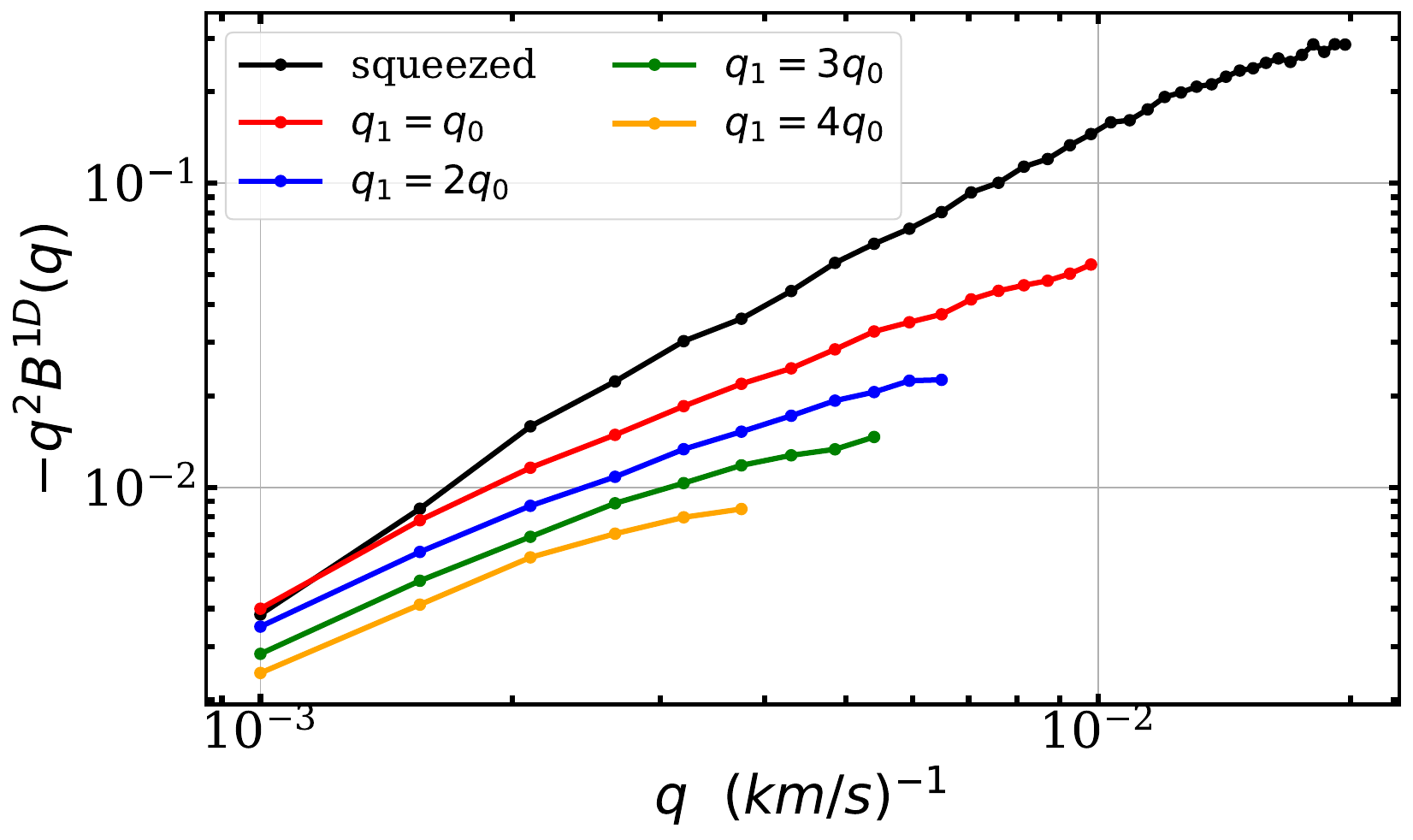}
\caption{One-dimensional bispectrum for different triangles configurations. The $q_{max}$ value for the different configurations depends on the size of the forest.}
\label{fig:apex01}
\end{figure}

The bispectrum is constructed using triangles whose vertices are at modes $q_{0}$, $q_{1}$ and $-(q_{0}+q_{1})$. Among all possible triangle configurations, some exhibit higher power, particularly those that share a common q-mode. This phenomenon is explained in \cite{Matarrese:1997sk} and \cite{Gil-Marin:2014sta}. We verify this by extracting the signal for different triangle configurations from the complete bispectrum signal (see Fig. \ref{fig:plot_t0}). Fig. \ref{fig:apex01} shows the $\bk$ for different common q-vectors ($q_{1}/q_{0}=1,2,3,4$) and the special case of the squeezed limit (Eq.~(\ref{configs}). Since the shot-noise in the bispectrum signal is derived from the power spectrum signal, the value of $q_{\mathrm{max}}$ for different triangle configurations corresponds to the $k_{\mathrm{max}}$ value in the $\pk$ signal. This $k_{\mathrm{max}}$ value is determined by the size of the forest, the spectrograph resolution and the Nyqvist-Shannon limit ($k_{\mathrm{Nyqvist}}=\pi/\Delta v$). In our $\pk$ estimate, $k_{\mathrm{max}}=0.02 ~ (\mathrm{km/s})^{-1}$. According to Eq.~(\ref{eq08}), the power spectrum includes a term that depends on $(q_{0}+q_{1})$. Since we have chosen triangles of the form $q_{1}/q_{0}=n$, where $n$ is an integer, it follows that $q_{0}+q_{1}=nq_{0}=k_{\mathrm{max}}$. Therefore, $q_{0}=q_{\mathrm{max}}=k_{\mathrm{max}}/n$.

 \bibliographystyle{JHEP}  
\bibliography{bibliografia} 

\end{document}